\documentstyle[preprint,prd,aps]{revtex}
\begin{document}
\draft
\title{
Self-energy-part resummed quark and gluon propagators in a 
spin-polarized quark matter and generalized Boltzmann 
equations} 
\author{A. Ni\'{e}gawa\thanks{Electronic address: 
niegawa@sci.osaka-cu.ac.jp}
} 
\address{Graduate School of Science, Osaka City University, 
Sumiyoshi-ku, Osaka 558-8585, JAPAN}
\date{Received today}
\maketitle
\begin{abstract}
We construct perturbative frameworks for studying nonequilibrium 
spin-polarized quark matter. We employ the closed-time-path 
formalism and use the gradient approximation in the derivative 
expansion. After constructing self-energy-part resummed quark and 
gluon propagators, we formulate two kind of mutually equivalent 
perturbative frameworks: The first one is formulated on the basis of 
the initial-particle distribution function, and the second one is 
formulated on the basis of \lq\lq physical''-particle distribution 
function. In the course of construction of the second framework, the 
generalized Boltzmann equations and their relatives {\em directly} 
come out, which describe the evolution of the system. The 
frameworks are relevant to the study of a magnetic character of 
quark matters, e.g., possible quark stars. 
\hspace*{1ex} 
\end{abstract} 
\narrowtext 
\setcounter{equation}{0}
\setcounter{section}{0}
\section{Introduction}
\def\theequation{\mbox{\arabic{section}.\arabic{equation}}}
Recent possible discovery of a quark star \cite{1,2} renews our 
interest in the study of quark matters. It has been pointed out 
\cite{3} the possibly of existing quark liquid in a ferro-magnetic 
phase. For analyzing the magnetic property of quark matters in a 
consistent manner \cite{4,5,6}, it is necessary to construct a 
self-energy-part resummed quark and gluon propagators in a 
spin-polarized quark matter, and thereby frame a perturbation 
theory. 

The spin-polarized quark matter is, in general, out of equilibrium. 
For dealing with such systems, we employ the closed-time-path 
formalism \cite{4,5}. In this formalism, propagators, vertices, and 
self-energy parts enjoy ($2 \times 2$)-matrix forms, denoted \lq\lq 
$\; \hat{} \;$''. Let $\hat{G} (x, y)$ be a generic two-point 
function. Fourier transforming with respect to $x - y$ (Wigner 
transformation), we have $\hat{G} (P, X)$ with $X = (x + y) / 2$. We 
assume that $\hat{G} (P, X)$ depends weakly on $X$. Then, as usual, 
employing a derivative expansion (DEX), we use the gradient 
approximation: 
\[ 
\hat{G} (P; X) \simeq \hat{G} (P; Y) + (X - Y)^\mu \partial_{Y^\mu} 
\hat{G} (P; Y) \, . 
\] 
We refer to the first term on the right-had side (RHS) as the 
leading part (term) and to the second term as the gradient part 
(term). Throughout this paper, we assume that the density matrix is 
color singlet, so that the quark and gluon propagators are diagonal 
in color space and independent of color index. Then, we drop the 
color index throughout. 

The plan of the paper is as follows. In Sec. II, the leading term 
in the DEX of the self-energy-part resummed (SEPR) quark propagator 
is constructed. In Sec. III, we construct the leading term of the 
SEPR gluon propagator in a Coulomb gauge. In Secs. II and III, the 
argument $X$ is dropped throughout. In Sec. IV, we present the 
gradient terms of the quark and gluon propagators. Then, we frame 
two mutually equivalent perturbative frameworks. One framework is 
constructed in terms of the \lq\lq bare''-number-density function 
(and its relative), and the other, which we call physical-$N$ 
scheme, is constructed in terms of the \lq\lq 
renormalized''-number-density function (and its relative). The 
latter scheme accompanies the generalized Boltzmann equation for 
the \lq\lq renormalized''-number-density function and its 
relatives. The form for the leading part of the SEPR gluon 
propagator in a covariant gauge is given in Appendix D. 
\setcounter{equation}{0}
\setcounter{section}{1}
\section{Quark propagator} 
\def\theequation{\mbox{\arabic{section}.\arabic{equation}}}
\subsection{Preliminaries} 
\subsubsection*{Spin-polarization vector} 
We define a spin-polarization vector ${\cal S} (P)$ as follows. For 
a timelike ($P^2 = p_0^2 - \vec{p}^{\, 2} > 0$) mode, we choose 
${\cal S}^\mu = (0, \vec{\zeta})$ $(\equiv \zeta^\mu)$ 
[$\vec{\zeta}^{\, 2} = 1$] in the rest frame, where $P^\mu = 
(\epsilon (p_0) \sqrt{P^2}, \vec{0})$. Similarly, for a spacelike 
($P^2 < 0$) mode, we choose ${\cal S}^\mu = (0, \vec{\zeta})$ in the 
\lq\lq $p_0 = 0$ frame,'' where $P^\mu = (0, \sqrt{- P^2} \, 
\vec{\xi})$ $( \equiv \sqrt{ - P^2} \, \xi^\mu)$ [$\vec{\xi}^{\, 2} 
= 1, \; \vec{\xi} \cdot \vec{\zeta} = 0$]. ${\cal S} (P)$ in any 
frame, where $P^\mu = (p_0, \vec{p})$, is obtained through a Lorentz 
transformation: 
\widetext 
\begin{eqnarray}
{\cal S}^\mu (P) & = & \theta (P^2) \left\{ \frac{\vec{p} \cdot 
\vec{\zeta} \, \left[ P^\mu + \epsilon (p_0) \sqrt{P^2} \, n^\mu 
\right]}{\sqrt{P^2} \, [\sqrt{P^2} + |p_0|]} + \zeta^\mu \right\} 
\nonumber \\ 
& & + \theta (- P^2) \left\{ - \frac{\vec{p} \cdot \vec{\zeta} \, 
\left[ P^\mu + \epsilon (\vec{p} \cdot \vec{\xi}) \sqrt{- P^2} \, 
\xi^\mu \right]}{\sqrt{- P^2} \, [\sqrt{- P^2} + |\vec{p} \cdot 
\vec{\xi}|]} + \zeta^\mu \right\} \, , \nonumber \\ 
\label{pol} \\ 
{\cal S} \cdot P & = & 0 \, , \;\;\;\;\; {\cal S}^2 = - 1 \, , 
\label{seisitsu} \\ 
n^\mu & = & (1, \vec{0}) \, . 
\nonumber 
\end{eqnarray} 
\narrowtext 
\noindent 
When a magnetic field is applied along the $\vec{\zeta}$ direction, 
$p_0 > 0$ modes with positive (negative) charge go to the state 
${\cal S}^\mu (P)$ ($- {\cal S}^\mu (P))$, while their \lq\lq 
antiparticle'' counterparts ($p_0 < 0$ modes) go to the state $- 
{\cal S}^\mu (P)$ (${\cal S}^\mu (P)$). In what follows, the 
concrete form (\ref{pol}) is not used, but only the properties 
(\ref{seisitsu}) will be used. 

The projection operators ${\cal P}_\pm (P)$ onto the states of 
definite polarization ($\pm$) reads 
\[
{\cal P}_\rho (P) = \frac{1 + \rho \epsilon (p_0) \gamma_5 
{\cal S}\kern-0.1em\raise0.3ex\llap{/}\kern0.15em\relax (P)}{2} \, . 
\] 
\subsubsection*{Orthogonal basis in Minkowski space and the standard 
form}
As an orthogonal basis in Minkowski space, we choose 
\[
\begin{array}{ll} 
P^\mu \, ,  & \;\;\; {\cal S}^\mu \, , \\ 
N^\mu = n^\mu - \frac{p_0}{P^2} P^\mu+ {\cal S}_0 {\cal S}^\mu \, , 
& \;\;\; (N^2 = {\cal S}_0^2 - \vec{p}^{\, 2} / P^2 ) \, , \\ 
e_\perp^\mu = i \epsilon^{\mu \nu \rho \sigma} P_\nu N_\rho {\cal 
S}_\sigma \, , & \;\;\; (e_\perp^2 = - P^2 N^2) \, . 
\end{array} 
\]
A generic $(4 \times 4)$-matrix function $A (P, N, {\cal S}, 
e_\perp)$ in a Dirac-matrix space is written in the from, 
\begin{eqnarray}
A & = & A_1' + A_2' \gamma_5 + A_3' 
P\kern-0.1em\raise0.3ex\llap{/}\kern0.15em\relax + A_4' 
N\kern-0.16em\raise0.3ex\llap{/}\kern0.09em\relax + A_5' {\cal 
S}\kern-0.1em\raise0.3ex\llap{/}\kern0.15em\relax + A_6' 
e_\perp\kern-0.62em\raise0.1ex\llap{/}\kern-0.17em\relax 
\mbox{\hspace*{0.55em}}\nonumber \\ 
&& + A_7' \gamma_5 P\kern-0.1em\raise0.3ex\llap{/}\kern0.15em\relax 
+ A_8' \gamma_5 N\kern-0.16em\raise0.3ex\llap{/}\kern0.09em\relax + 
A_9' \gamma_5 {\cal 
S}\kern-0.1em\raise0.3ex\llap{/}\kern0.15em\relax + A_{10}' \gamma_5 
e_\perp\kern-0.62em\raise0.1ex\llap{/}\kern-0.17em\relax 
\mbox{\hspace*{0.55em}} \nonumber \\ 
&& + A_{11}' P\kern-0.1em\raise0.3ex\llap{/}\kern0.15em\relax 
N\kern-0.16em\raise0.3ex\llap{/}\kern0.09em\relax + A_{12}' 
P\kern-0.1em\raise0.3ex\llap{/}\kern0.15em\relax {\cal 
S}\kern-0.1em\raise0.3ex\llap{/}\kern0.15em\relax + A_{13}' 
P\kern-0.1em\raise0.3ex\llap{/}\kern0.15em\relax 
e_\perp\kern-0.62em\raise0.1ex\llap{/}\kern-0.17em\relax 
\mbox{\hspace*{0.55em}} + A_{14}' 
N\kern-0.16em\raise0.3ex\llap{/}\kern0.09em\relax {\cal 
S}\kern-0.1em\raise0.3ex\llap{/}\kern0.15em\relax \nonumber \\ 
&& + A_{15}' N\kern-0.16em\raise0.3ex\llap{/}\kern0.09em\relax 
e_\perp\kern-0.62em\raise0.1ex\llap{/}\kern-0.17em\relax 
\mbox{\hspace*{0.55em}} + A_{16}' {\cal 
S}\kern-0.1em\raise0.3ex\llap{/}\kern0.15em\relax 
e_\perp\kern-0.62em\raise0.1ex\llap{/}\kern-0.17em\relax 
\mbox{\hspace*{0.55em}} \, . 
\label{gene}
\end{eqnarray}
We decompose $A$ into four parts, 
\begin{equation} 
A = \sum_{\rho, \, \sigma = \pm} {\cal P}_\rho A {\cal P}_\sigma 
\equiv \sum_{\rho, \, \sigma = \pm} {\cal P}_\rho A^{\rho \sigma} 
{\cal P}_\sigma \, , 
\label{14-0} 
\end{equation} 
and write $A^{\rho \sigma}$ in the form, 
\begin{eqnarray}
A^{\rho \rho} & = & A_1^{\rho \rho} + A_2^{\rho \rho} \, 
P\kern-0.1em\raise0.3ex\llap{/}\kern0.15em\relax + A_3^{\rho \rho} 
\, N\kern-0.16em\raise0.3ex\llap{/}\kern0.09em\relax + A_4^{\rho 
\rho} \, P\kern-0.1em\raise0.3ex\llap{/}\kern0.15em\relax 
N\kern-0.16em\raise0.3ex\llap{/}\kern0.09em\relax \, , \nonumber \\ 
A^{\rho - \rho} & = & \gamma_5 \left[ A_1^{\rho - \rho} + A_2^{\rho 
- \rho} \, P\kern-0.1em\raise0.3ex\llap{/}\kern0.15em\relax + 
A_3^{\rho - \rho} \, 
N\kern-0.16em\raise0.3ex\llap{/}\kern0.09em\relax + A_4^{\rho - 
\rho} \, P\kern-0.1em\raise0.3ex\llap{/}\kern0.15em\relax 
N\kern-0.16em\raise0.3ex\llap{/}\kern0.09em\relax \right] \, . 
\nonumber \\ 
&& 
\label{P-P} 
\end{eqnarray}
It is a straightforward task to obtain 
\begin{equation}
\begin{array}{ll}
A_1^{\rho \rho} = A_1' + \rho \epsilon (p_0) A_9' \, , & \; 
A_2^{\rho \rho} = A_3' - \rho \epsilon (p_0) N^2 A_{15}' \, , \\ 
A_3^{\rho \rho} = A_4' + \rho \epsilon (p_0) P^2 A_{13}' \, , & \; 
A_4^{\rho \rho} = A_{11}' + \rho \epsilon (p_0) A_{6}' \, , \\ 
A_1^{\rho - \rho} = A_2' - \rho \epsilon (p_0) A_5' \, , & \; 
A_2^{\rho - \rho} = A_7' + \rho \epsilon (p_0) A_{12}' \, , \\ 
A_3^{\rho - \rho} = A_8' + \rho \epsilon (p_0) A_{14}' \, , & \; 
A_4^{\rho - \rho} = A_{16}' - \rho \epsilon (p_0) A_{10}' \, . 
\end{array}
\label{1.61}
\end{equation}
We refer Eq. (\ref{P-P}) to as the standard form (SF) and $A^{\rho 
\sigma}$ or $A_j^{\rho \sigma}$ to as an SF-element of $A$. It is to 
be understood that the (bare and self-energy-part resummed) 
propagators and the self-energy part, which appear in the following, 
are to be written in the SF. 
\subsection{Bare propagator}
First of all, we note that the bare propagator matrix $\hat{S} (P)$ 
and the self-energy-part resummed propagator matrix $\hat{G} (P)$ 
enjoy the \lq\lq symmetry'' property, 
\begin{equation} 
\hat{S}^\dagger (P) = - \hat{\tau}_1 \gamma^0 \, \displaystyle{ 
\raisebox{1.8ex}{\scriptsize{t}}} \mbox{\hspace{-0.33ex}} \hat{S} 
(P) \gamma^0  \hat{\tau}_1 \, , \;\; \hat{G}^\dagger (P) = - 
\hat{\tau}_1 \gamma^0 \, \displaystyle{ 
\raisebox{1.8ex}{\scriptsize{t}}} \mbox{\hspace{-0.33ex}} \hat{G} 
(P) \gamma^0  \hat{\tau}_1 \, , 
\label{seisei3} 
\end{equation} 
which results from the hermiticity of the density matrix. Here, 
$\hat{\tau}_1$ is the first Pauli matrix, $\dagger$ acts on Dirac 
gamma matrix function, e.g., $(A 
P\kern-0.1em\raise0.3ex\llap{/}\kern0.15em\relax)^\dagger = A^* 
P^\mu \gamma^\dagger_\mu$, and $\displaystyle{ 
\raisebox{1.8ex}{\scriptsize{t}}} \mbox{\hspace{-0.33ex}} \hat{S} 
(P)$ denotes the transpose of the $(2 \times 2)$ matrix function 
$\hat{S} (P)$, etc.  

The bare propagator $\hat{S} (P)$ is an inverse of $\hat{S}^{-1} (P) 
= (P\kern-0.1em\raise0.3ex\llap{/}\kern0.15em\relax - m) 
\hat{\tau}_3$. A general solution to $\hat{S}^{-1} \hat{S} = \hat{S} 
\hat{S}^{-1} = 1$ is 
\begin{eqnarray}
\hat{S} (P) &=& \hat{S}^{(0)} (P) + S_K (P) \hat{M}_+ \, , 
\label{333} \\ 
\hat{S}^{(0)} (P) & = & \sum_{\rho = \pm} {\cal P}_\rho \left[ 
\hat{S}_{RA} (P) - f_\rho (S_R - S_A) \hat{M}_+ \right] \, , \\ 
S_K (P) & = & - \sum_{\rho = \pm} C_{\rho - \rho} (P) \left( 
\Delta_R (P) - \Delta_A (P) \right) \nonumber \\ 
&& \times {\cal P}_\rho \gamma_5 
(P\kern-0.1em\raise0.3ex\llap{/}\kern0.15em\relax - m) \, 
N\kern-0.16em\raise0.3ex\llap{/}\kern0.09em\relax {\cal P}_{- \rho} 
\label{335} \, , 
\end{eqnarray}
where the suffix \lq\lq $K$'' stands for the \lq\lq Keldish 
component'' and 
\begin{eqnarray}
\hat{S}_{RA} (P) &=& 
\left( 
\begin{array}{cc}
S_R & \;\; 0 \\ 
S_R - S_A & \;\; - S_A 
\end{array}
\right) \, , \nonumber \\ 
\hat{M}_\pm & = & \left( 
\begin{array}{cc}
1 & \;\; \pm 1 \\ 
\pm 1 & \;\; 1 
\end{array}
\right) \, , \nonumber \\ 
S_{R (A)} &=& (P\kern-0.1em\raise0.3ex\llap{/}\kern0.15em\relax 
+ m) \Delta_{R (A)} (P) = 
\frac{P\kern-0.1em\raise0.3ex\llap{/}\kern0.15em\relax + m}{P^2 - 
m^2 \pm i p_0 0^+} \, , \nonumber \\ 
&& \label{2jyuu} \\ 
f_\rho (P) & = & \theta (p_0) N_\rho (|p_0|, \vec{p}) \nonumber \\ 
&& + \theta (- p_0) \left[ 1 - \bar{N}_\rho (|p_0|, - \vec{p}) 
\right] \, . 
\label{bareN} 
\end{eqnarray}
Here $S_{R (A)}$ is the retarded (advanced) propagator, and $N_\rho 
(|p_0|, \vec{p})$ [$\bar{N}_\rho (|p_0|, - \vec{p})$] ($\rho = \pm$) 
is the \lq\lq bare'' number-density function of a quark [an 
antiquark] with polarization $\rho {\cal S} (P)$, energy $|p_0|$ ($= 
\sqrt{\vec{p}^{\, 2} + m^2}$), and momentum $\vec{p}$ [$- \vec{p}$]. 
$S_K$ in Eq. (\ref{335}) connects opposite polarization states. From 
Eqs. (\ref{seisei3}), (\ref{333}), and (\ref{335}), we have 
\[
\left( C_{+ -} (P) \right)^* = C_{- +} (P) \, . 
\]

The derivative expansion is an efficient device for dealing with 
quasiuniform systems near equilibrium or nonequilibrium 
quasistationary systems. For such systems, $S_K$ is small when 
compared to $\hat{S}^{(0)}$. 
\subsection{Dyson equation} 
The self-energy-part ($\hat{\Sigma}$) resummed propagator $\hat{G}$ 
obeys the Dyson equation, 
\begin{equation}
\hat{G} (P) = \hat{S} (P) [1 + \hat{\Sigma} (P) \hat{G} (P)] =  [1 + 
\hat{G} (P) \hat{\Sigma} (P)] \hat{S} (P) \, . 
\label{SD} 
\end{equation}
We write $\hat{G}$ and $\hat{\Sigma}$ in SFs, 
\[
\hat{G} = \sum_{\rho, \, \sigma = \pm} {\cal P}_\rho \hat{G}^{\rho 
\sigma} {\cal P}_\sigma \, , \mbox{\hspace{10ex}} \hat{\Sigma} = 
\sum_{\rho, \, \sigma = \pm} {\cal P}_\rho \hat{\Sigma}^{\rho 
\sigma} {\cal P}_\sigma \, . 
\] 
It is worth mentioning that, for the system that enjoys an azimuthal 
symmetry around the $\vec{\zeta}$-direction, $\hat{\Sigma}$, and 
then also $\hat{G}$, are independent of $E_\perp^\mu$, provided that 
we choose $\vec{\xi} = \vec{p} \times \vec{\zeta} / |\vec{p} \times 
\vec{\zeta}|$. Then, from Eqs. (\ref{gene}) - (\ref{1.61}), we have 
\[
\begin{array}{ll}
\hat{\Sigma}_4^{\rho - \rho} = 0 \;\;\;\;\;\;\;\;\;\;\;\;\;\;\; & 
(\rho = \pm) \, , \\ 
\hat{\Sigma}_j^{++} = \hat{\Sigma}_j^{--} 
\;\;\;\;\;\;\;\;\;\;\;\;\;\;\; & (j = 2, 3, 4) \, . 
\end{array}
\]
Same relations hold for $\hat{G}$'s. 

Substituting the SFs for $\hat{S}$, $\hat{\Sigma}$, and $\hat{G}$ in 
Eq. (\ref{SD}), we obtain coupled equations, 
\begin{eqnarray}
\hat{G}^{\rho \sigma} & = & \hat{S}^{\rho \sigma} + \left( \hat{S} 
\hat{\Sigma} \hat{G} \right)^{\rho \sigma} = \hat{S}^{\rho \sigma} + 
\left( \hat{G} \hat{\Sigma} \hat{S} \right)^{\rho \sigma} \nonumber 
\\ 
&& \mbox{\hspace*{28ex}} (\rho, \sigma = \pm) \, , 
\label{SD1} 
\end{eqnarray}
where $\left( \hat{S} \hat{\Sigma} \hat{G} \right)^{\rho \sigma} 
\equiv \sum_{\xi, \, \zeta = \pm} \hat{S}^{\rho \xi} 
\hat{\Sigma}^{\xi \zeta} \hat{G}^{\zeta \sigma}$, etc. 
The relation that involves $(...)^{\rho \sigma}$ is to be understood 
to hold when sandwiched between projection operators ${\cal P}_\rho 
\, ... \, {\cal P}_\sigma$. We write Eq. (\ref{SD1}), with obvious 
notation, as 
\begin{equation}
\hat{\bf G} = \hat{\bf S} + \hat{\bf S} \hat{\bf \Sigma} \hat{\bf G} 
= \hat{\bf S} + \hat{\bf G} \hat{\bf \Sigma} \hat{\bf S} \, , 
\label{SD11} 
\end{equation}
where bold-face letters denote ($2 \times 2$)-matrix in a \lq\lq 
polarization space''. 

From Eq. (\ref{seisei3}), we obtain the symmetry relations for the 
SF-elements of $\hat{G}^{\rho \sigma}$ (cf. Eqs. (\ref{14-0}) and 
(\ref{P-P})), 
\begin{equation}
\left( \hat{G}_j^{\rho \sigma} (P) \right)^* = - \sigma_j^{\rho 
\sigma} \hat{\tau}_1 \, \displaystyle{ 
\raisebox{1.8ex}{\scriptsize{t}}} \mbox{\hspace{-0.33ex}} 
\hat{G}_j^{\sigma \rho} (P) \hat{\tau}_1 \, , 
\label{symm31} 
\end{equation}
where 
\[ 
\sigma_j^{\rho \sigma} = \left\{ 
\begin{array}{lcl}
+ \;\; \mbox{for} \;\; (\rho \sigma, j) & = & (\rho \rho,1) \, , \, 
(\rho \rho,2) \, , \, (\rho \rho,3) \, , \\ 
&& (\rho - \rho,2) \, , \, (\rho - \rho,3) \, , \\ 
&& (\rho - \rho,4) \, , \\ 
- \;\; \mbox{for} \;\; (\rho \sigma, j) & = & (\rho \rho,4) \, , \, 
(\rho - \rho,1) \, . 
\end{array} 
\right. 
\]
Similar relations hold for $\hat{\Sigma}_j^{\rho \sigma}$'s. 

Let us introduce ($2 \times 2$)-matrix function ${\bf f}$ in the 
polarization space, 
\[
{\bf f} = \mbox{diag} \, (f_+, f_-) \, . 
\]
Then $\hat{\bf S}$ is written as 
\begin{eqnarray}
\hat{\bf S} & = & \hat{\bf S}^{(0)} + {\bf S}_K \hat{M}_+ \, , 
\label{hadaka} \\ 
\hat{\bf S}^{(0)} &=& \hat{S}_{RA} {\bf 1} - {\bf f} (S_R - S_A) 
\hat{M}_+ \nonumber \\ 
{\bf S}_K &= & \left( S_R (P) - S_A (P) \right) \gamma_5 
N\kern-0.1em\raise0.3ex\llap{/}\kern0.15em\relax {\bf C} (P) \, , 
\label{esu} \\ 
{\bf C} (P) &=& \left( 
\begin{array}{cc} 
0 & \;\; C_{+-} (P) 
\label{esu12} \\ 
C_{-+} (P) & \;\; 0 
\end{array} 
\right) \, . \nonumber 
\end{eqnarray}

Among the components of $\hat{\bf \Sigma}$, is a relation, 
\begin{equation}
{\bf \Sigma}_{11} + {\bf \Sigma}_{12} + {\bf \Sigma}_{21} + {\bf 
\Sigma}_{22} = 0 \, . 
\label{seip}
\end{equation}
Then, $\hat{\bf \Sigma}$ is written as 
\begin{eqnarray}
\hat{\bf \Sigma} & = & \hat{\bf \Sigma}^{{(0)}} - {\bf \Sigma}_K 
\hat{M}_- \, , 
\label{siguma} \\ 
\hat{\bf \Sigma}^{{(0)}} &=& \left( 
\begin{array}{cc} 
{\bf \Sigma}_R & \;\; 0 \\ 
- {\bf \Sigma}_R + {\bf \Sigma}_A & \;\; - {\bf \Sigma}_A 
\end{array} 
\right) - ({\bf \Sigma}_R {\bf f} - {\bf f} {\bf \Sigma}_A) 
\hat{M}_- \, , \nonumber \\ 
\\ 
{\bf \Sigma}_R & = & {\bf \Sigma}_{11} + {\bf \Sigma}_{12} = - {\bf 
\Sigma}_{22} - {\bf \Sigma}_{21} \, , \\ 
{\bf \Sigma}_A & = & {\bf \Sigma}_{11} + {\bf \Sigma}_{21} = - {\bf 
\Sigma}_{22} - {\bf \Sigma}_{12} \, , \\ 
{\bf \Sigma}_K & = & {\bf f} {\bf \Sigma}_{11} - {\bf \Sigma}_{11} 
{\bf f} + {\bf f} {\bf \Sigma}_{21} + {\bf \Sigma}_{12} (1 - {\bf 
f}) \, . 
\label{sigma3} 
\end{eqnarray}
From Eq. (\ref{symm31}) with $\hat{\Sigma}_j^{\rho \sigma}$ for 
$\hat{G}_j^{\rho \sigma}$, we obtain the symmetry relations, 
\begin{eqnarray*} 
\left( \Sigma_{Rj}^{\rho \sigma} (P) \right)^* & = & \sigma_j^{\rho 
\sigma} \Sigma_{Aj}^{\sigma \rho} (P) \, , \\ 
\left( \Sigma_{Kj}^{\rho \sigma} (P) \right)^* & = & - 
\sigma_j^{\rho \sigma} \Sigma_{Kj}^{\sigma \rho} (P) \, . 
\end{eqnarray*} 

Among the components of $\hat{\bf G}$, is a relation, 
\begin{equation}
{\bf G}_{11} + {\bf G}_{22} = {\bf G}_{12} + {\bf G}_{21} \, . 
\label{seip1}
\end{equation}
Then, $\hat{\bf G}$ is written as 
\begin{eqnarray}
\hat{\bf G} & = & \hat{\bf G}^{{(0)}} + {\bf G}_K \hat{M}_+ \, , 
\label{gii} \\ 
\hat{\bf G}^{{(0)}} &=& \left( 
\begin{array}{cc}
{\bf G}_R & \;\; 0 \\ 
{\bf G}_R - {\bf G}_A & \;\; - {\bf G}_A 
\end{array} 
\right) - ({\bf G}_R {\bf f} - {\bf f} {\bf G}_A) \hat{M}_+ \, , 
\nonumber \\ 
\\ 
{\bf G}_R & = & {\bf G}_{11} - {\bf G}_{12} = - {\bf G}_{22} + {\bf 
G}_{21} \, , \\ 
{\bf G}_A & = & {\bf G}_{11} - {\bf G}_{21} = - {\bf G}_{22} + {\bf 
G}_{12} \, , \\ 
{\bf G}_K & = & {\bf G}_{11} {\bf f} - {\bf f}{\bf G}_{11} + {\bf f} 
{\bf G}_{21} + {\bf G}_{12} (1 - {\bf f}) \, . 
\label{gii3} 
\end{eqnarray}
It is worth mentioning that, for equilibrium systems, ${\bf 
\Sigma}_K = {\bf G}_K = 0$. From Eq. (\ref{symm31}), follows the 
symmetry relations, 
\begin{eqnarray} 
\left( G_{Rj}^{\rho \sigma} (P) \right)^* & = & \sigma_j^{\rho 
\sigma} G_{Aj}^{\sigma \rho} (P) \, , \nonumber \\ 
\left( G_{Kj}^{\rho \sigma} (P) \right)^* & = & - \sigma_j^{\rho 
\sigma} G_{Kj}^{\sigma \rho} (P) \, . 
\label{hikkuri3} 
\end{eqnarray} 

Substitution of Eqs. (\ref{hadaka}), (\ref{siguma}), and (\ref{gii}) 
into Eq. (\ref{SD11}) yields 
\begin{eqnarray}
\hat{\bf G}^{(0)} &=& \hat{\bf S}^{(0)} + \hat{\bf S}^{(0)} \hat{\bf 
\Sigma}^{(0)} \hat{\bf G}^{(0)} = \hat{\bf S}^{(0)} + \hat{\bf 
G}^{(0)} \hat{\bf \Sigma}^{(0)} \hat{\bf S}^{(0)} \, , 
\nonumber \\ 
\label{RRAA} \\ 
{\bf G}_K &=& {\bf S}_K + S_R {\bf \Sigma}_R {\bf G}_K + {\bf S}_K 
{\bf \Sigma}_A {\bf G}_A - S_R {\bf \Sigma}_K {\bf G}_A \nonumber \\ 
&& = {\bf S}_K + {\bf G}_R {\bf \Sigma}_R {\bf S}_K + {\bf G}_K {\bf 
\Sigma}_A S_A - {\bf G}_R {\bf \Sigma}_K S_A \, . \nonumber \\ 
\label{mei} 
\end{eqnarray}
From Eq. (\ref{RRAA}), we obtain 
\begin{eqnarray}
{\bf G}_{R (A)} = \left[ 
P\kern-0.1em\raise0.3ex\llap{/}\kern0.15em\relax - m - {\bf 
\Sigma}_{R (A)} \right]^{- 1} \, , 
\label{ohosi} 
\end{eqnarray}
where use has been made of $\left( \hat{S}^{(0) - 1} \right)^{\rho 
\sigma} = \delta^{\rho \sigma} 
(P\kern-0.1em\raise0.3ex\llap{/}\kern0.15em\relax - m) 
\hat{\tau}_3$. We get from Eq. (\ref{ohosi}), after some 
manipulation, 
\begin{eqnarray}
G_R^{\rho \rho} & = & \left[ 
P\kern-0.1em\raise0.3ex\llap{/}\kern0.15em\relax - m - 
\Sigma_R^{\rho \rho} - \Sigma_R^{\rho - \rho} \, 
G_R^{(\mbox{\scriptsize{pre}}) - \rho - \rho} \, \Sigma_R^{- 
\rho \rho} \right]^{- 1} \, , \nonumber \\ 
\label{final} \\ 
G_R^{\rho - \rho} & = & G_R^{(\mbox{\scriptsize{pre}}) \rho \rho} \, 
\Sigma_R^{\rho - \rho} \, G_R^{- \rho - \rho} = G_R^{\rho \rho} \, 
\Sigma_R^{\rho - \rho} \, G_R^{(\mbox{\scriptsize{pre}}) - \rho - 
\rho} \, , \nonumber \\ 
\label{final1}
\end{eqnarray}
where 
\begin{equation}
G_R^{(\mbox{\scriptsize{pre}}) \rho \rho} = \left[ 
P\kern-0.1em\raise0.3ex\llap{/}\kern0.15em\relax - m - 
\Sigma_R^{\rho \rho} \right]^{- 1} \, . 
\label{gyaku}
\end{equation}
As has been remarked above after Eq. (\ref{SD1}), Eq. (\ref{gyaku}) 
is to be understood to mean 
\begin{eqnarray} 
&& {\cal P}_\rho G_R^{(\mbox{\scriptsize{pre}}) \rho \rho} \left[ 
P\kern-0.1em\raise0.3ex\llap{/}\kern0.15em\relax - m - 
\Sigma_R^{\rho \rho} \right] {\cal P}_\rho \nonumber \\ 
&& \mbox{\hspace*{4ex}} = {\cal P}_\rho \left[ 
P\kern-0.1em\raise0.3ex\llap{/}\kern0.15em\relax - m - 
\Sigma_R^{\rho \rho} \right] G_R^{(\mbox{\scriptsize{pre}}) \rho 
\rho} {\cal P}_\rho = {\cal P}_\rho \, 1 \, {\cal P}_\rho = {\cal 
P}_\rho \, . \nonumber \\ 
&& 
\label{tuika}
\end{eqnarray} 
Such an understanding also applies to Eq. (\ref{final}). Concrete 
form for ${\bf G}_{R (A)}$ will be given in the next section. 

As for ${\bf G}_K$, Eq. (\ref{mei}), we show in Appendix A that 
\begin{eqnarray} 
{\bf G}_K & = & {\bf G}_K^{(1)} + {\bf G}_K^{(2)} + {\bf G}_K^{(3)} 
\, , 
\label{ST20} \\ 
{\bf G}_K^{(1)} &=& - {\bf G}_R {\bf \Sigma}_K {\bf G}_A \, , 
\label{label} \\ 
{\bf G}_K^{(2)} &=& {\bf G}_R \left[ \gamma_5 
N\kern-0.1em\raise0.3ex\llap{/}\kern0.15em\relax {\bf C} (P) 
{\bf \Sigma}_A - {\bf \Sigma}_R \gamma_5 
N\kern-0.1em\raise0.3ex\llap{/}\kern0.15em\relax {\bf C} (P) \right] 
{\bf G}_A \nonumber \\ 
& \equiv & {\bf G}_R {\bf H}_l {\bf G}_A \, , 
\label{ST22} \\ 
{\bf G}_K^{(3)} &=& {\bf G}_R \gamma_5 
N\kern-0.1em\raise0.3ex\llap{/}\kern0.15em\relax {\bf C} (P) - 
\gamma_5 N\kern-0.1em\raise0.3ex\llap{/}\kern0.15em\relax {\bf C} 
(P) {\bf G}_A \, . 
\label{2.40d} 
\end{eqnarray} 
\lq\lq $l \, $'' of ${\bf H}_l$ in Eq. (\ref{ST22}) stands for the 
\lq\lq leading part'' of the DEX. As mentioned above at the end of 
Subsec. B, for quasiuniform systems near equilibrium or 
nonequilibrium quasistationary systems, ${\bf G}_K^{(2)}$ and ${\bf 
G}_K^{(3)}$ are much smaller than ${\bf G}_K^{(1)}$. The SF for 
${\bf G}_K^{(1)}$ will be given in the next section. The standard 
forms for ${\bf H}_l$ in Eq. (\ref{ST22}) and $\gamma_5 
N\kern-0.1em\raise0.3ex\llap{/}\kern0.15em\relax {\bf C} (P)$ in Eq. 
(\ref{2.40d}) are also given in the next section. The SFs for ${\bf 
G}_K^{(2)}$ and ${\bf G}_K^{(3)}$ are obtained by repeatedly using 
the formulae in Appendix B. 
\subsection{Self-energy-part resummed propagator $\hat{\bf G}$} It 
is convenient to introduce 
\[ 
\hat{\tilde{G}}^{\rho \rho} = \left( 
\begin{array}{cc} 
G_R^{\rho \rho} & \;\; G_K^{(1) \rho \rho} \\ 
0 & \;\; - G_A^{\rho \rho} 
\end{array} 
\right) \, , \mbox{\hspace*{2ex}} \hat{\tilde{\Sigma}}^{\rho \sigma} 
= \left( 
\begin{array}{cc} 
\Sigma_R^{\rho \sigma} & \;\; \Sigma_K^{\rho \sigma} \\ 
0 & \;\; - \Sigma_A^{\rho \sigma} 
\end{array} 
\right) \, . 
\]
We observe that Eq. (\ref{final}) and $G_K^{(1) \rho \rho}$ in Eq. 
(\ref{label}) are unified to a matrix equation, 
\begin{equation}
\hat{\tilde{G}}^{\rho \rho} = \left[ 
(P\kern-0.1em\raise0.3ex\llap{/}\kern0.15em\relax - m) \hat{\tau}_3 
- \hat{\tilde{\Sigma}}^{\rho \rho} - \hat{\tilde{\Sigma}}^{\rho - 
\rho} \hat{\tilde{G}}^{(\mbox{\scriptsize{pre}}) - \rho - \rho} 
\hat{\tilde{\Sigma}}^{- \rho \rho} \right]^{- 1} \, , 
\label{ffinal} 
\end{equation}
where 
\begin{eqnarray}
\hat{\tilde{G}}^{(\mbox{\scriptsize{pre}}) \rho \rho} & = & \left( 
\begin{array}{cc}
G_R^{(\mbox{\scriptsize{pre}}) \rho \rho} & \;\; 
G_K^{(\mbox{\scriptsize{pre}}) \rho \rho} \\ 
0 & \;\; - G_A^{(\mbox{\scriptsize{pre}}) \rho \rho} 
\end{array}
\right) 
\nonumber \\ 
& = & \left[ (P\kern-0.1em\raise0.3ex\llap{/}\kern0.15em\relax - m) 
\hat{\tau}_3 - \hat{\tilde{\Sigma}}^{\rho \rho} \right]^{- 1} \, . 
\label{prere} 
\end{eqnarray}
\subsubsection*{Forms for $G^{(\mbox{\scriptsize{pre}}) \rho 
\rho}_{R (A)}$ and $G^{(\mbox{\scriptsize{pre}}) \rho \rho}_K$} The 
SF for $(P\kern-0.1em\raise0.3ex\llap{/}\kern0.15em\relax - m) 
\hat{\tau}_3 - \hat{\tilde{\Sigma}}^{\rho \rho}$ reads 
\begin{eqnarray}
&& (P\kern-0.1em\raise0.3ex\llap{/}\kern0.15em\relax - m) 
\hat{\tau}_3 - \hat{\tilde{\Sigma}}^{\rho \rho} (P) \nonumber \\ 
&& \mbox{\hspace*{5ex}} = - \left[ m \hat{\tau}_3 + 
\hat{\tilde{\Sigma}}_1^{\rho \rho} (P) \right] + \left[ \hat{\tau}_3 
- \hat{\tilde{\Sigma}}_2^{\rho \rho} (P) \right] 
P\kern-0.1em\raise0.3ex\llap{/}\kern0.15em\relax \nonumber \\ 
&& \mbox{\hspace*{8ex}} - \hat{\tilde{\Sigma}}_3^{\rho \rho} (P) 
N\kern-0.16em\raise0.3ex\llap{/}\kern0.09em\relax - 
\hat{\tilde{\Sigma}}_4^{\rho \rho} (P) 
P\kern-0.1em\raise0.3ex\llap{/}\kern0.15em\relax 
N\kern-0.16em\raise0.3ex\llap{/}\kern0.09em\relax \, . 
\label{zero}
\end{eqnarray}
One obtains the expressions for the SF-elements 
$G^{(\mbox{\scriptsize{pre}}) \rho\rho}_{Tj}$ $(T = R, A, K$ and $j 
= 1 - 4)$ through straightforward but tedious manipulation of Eq. 
(\ref{prere}), which includes Eq. (\ref{tuika}). Writing 
$\Sigma_j \equiv \Sigma_{Rj}^{\rho \rho}$ for short, we have 
\widetext 
\begin{eqnarray*}
G^{(\mbox{\scriptsize{pre}}) \rho \rho}_{Rj} & = & \sigma_j^{\rho 
\rho} \left( G^{(\mbox{\scriptsize{pre}}) \rho \rho}_{Aj} \right)^* 
 \;\;\;\;\;\; (j = 1 - 4) \, , 
\\ 
G^{(\mbox{\scriptsize{pre}}) \rho \rho}_{R1} & = & \frac{m + 
\Sigma_1}{{\cal D}_{\mbox{\scriptsize{pre}}}^{\rho \rho}} \, , 
\;\;\;\;\;\;\;\;\;\;\;\;\;\; G^{(\mbox{\scriptsize{pre}}) \rho 
\rho}_{R2} = \frac{1 - \Sigma_2}{{\cal 
D}_{\mbox{\scriptsize{pre}}}^{\rho \rho}} \, , \nonumber \\ 
G^{(\mbox{\scriptsize{pre}}) \rho \rho}_{Rl} & = & - 
\frac{\Sigma_l}{{\cal D}_{\mbox{\scriptsize{pre}}}^{\rho \rho}} 
\;\;\;\;\;\; (l = 3, 4) \, , \nonumber \\ 
G^{(\mbox{\scriptsize{pre}}) \rho \rho}_{Kj} & = & 
\frac{\sum_{l = 1}^4 {\cal N}_j^{(l)} \Sigma_{Kl}^{\rho 
\rho}}{Im \left\{ [ ( m + \Sigma_1 )^2 - N^2 (\Sigma_3)^2] 
[(\Sigma_2^* - 1)^2 - N^2 (\Sigma_4^* )^2] \right\} } \;\;\;\;\;\;\; 
(j = 1 -  4) \, , \nonumber 
\end{eqnarray*}
where
\begin{equation} 
{\cal D}_{\mbox{\scriptsize{pre}}}^{\rho \rho} = \left[ (1 - 
\Sigma_2)^2 - N^2 (\Sigma_4)^2 \right] P^2 - (m + \Sigma_1)^2 + N^2 
(\Sigma_3)^2 
\label{1.54ddd} 
\end{equation} 
and 
\[
\begin{array}{ll} 
{\cal N}_1^{(1)} = - \left[ E^{(+)}_{13} F_{24} + E^{(+)}_{24} 
F_{13} \right] \, , \, \; & {\cal N}_1^{(2)} = - 2 F_{13} Re 
F_{1234}^{(-)} \, , \nonumber \\ 
{\cal N}_1^{(3)} = 2 N^2 Re \left[ F_{13} \, H_{24} + F_{24} \, 
H_{13} \right] \, , \, \; & {\cal N}_1^{(4)} = 2 i N^2 F_{13} Im 
F_{1423}^{(+)} \, , \nonumber \\ 
{\cal N}_2^{(1)} = - 2 F_{24} Re F^{(+)}_{1234} \, , \, \; & {\cal 
N}_2^{(2)} = - \left[ E_{24}^{(-)} F_{13} + E_{13}^{(-)} F_{24} 
\right] \, , \nonumber \\ 
{\cal N}_2^{(3)} = 2 N^2 F_{24} Re F^{(+)}_{1423} \, , \, \; & {\cal 
N}_2^{(4)} = 2 i N^2 Im \left[ F_{13} \, H_{24} + F_{24} \, H_{13} 
\right] \, , \nonumber \\ 
{\cal N}_3^{(1)} = - 2 Re \left[ F_{13} \, H_{24} - F_{24} \, H_{13} 
\right] \, , \, \; & {\cal N}_3^{(2)} = - 2 F_{13} Re F_{1423}^{(-)} 
\, , \nonumber \\ 
{\cal N}_3^{(3)} = - \left[ E_{13}^{(+)} F_{24} - E_{24}^{(+)} 
F_{13} \right] \, , \, \; & {\cal N}_3^{(4)} = 2 i F_{13} Im 
F_{1234}^{(-)} \, , \nonumber \\ 
{\cal N}_4^{(1)} = - 2 i F_{24} Im F_{1423}^{(-)} \, , \, \; & {\cal 
N}_4^{(2)} = 2 i Im \left[ - F_{13} \, H_{24} + F_{24} \, H_{13} 
\right] \, , \nonumber \\ 
{\cal N}_4^{(3)} = 2 i F_{24} Im F_{1234}^{(+)} \, , \, \; & {\cal 
N}_4^{(4)} = - \left[ E_{13}^{(-)} F_{24} - E_{24}^{(-)} F_{13} 
\right] \, , \nonumber 
\end{array}
\]
with 
\[
\begin{array}{llllll} 
E_{13}^{(\pm)} &=& | m + \Sigma_1 |^2 \pm N^2 |\Sigma_3|^2 \, , 
\;\;\; & E_{24}^{(\pm)} & = & | 1 - \Sigma_2 |^2 \pm N^2 
|\Sigma_4|^2 \, , \\ 
F_{24} & = & Im \frac{(1 - \Sigma_2)^2 - N^2 (\Sigma_4)^2}{{\cal 
D}_{\mbox{\scriptsize{pre}}}^{\rho \rho}} \, , \;\;\; & \;\, 
F_{13} & = & Im \frac{(m + \Sigma_1)^2 - N^2 (\Sigma_3)^2}{{\cal 
D}_{\mbox{\scriptsize{pre}}}^{\rho \rho}} \, , \\ 
F_{1234}^{(\pm)} & = & (m + \Sigma_1) (1 - \Sigma_2^*) \pm N^2 
\Sigma_3 \Sigma_4^* \, , & & & \\ 
F_{1423}^{(\pm)} & = & (m + \Sigma_1) \Sigma_4^* \pm (1 - \Sigma_2) 
\Sigma_3^* \, , & & & \\ 
H_{13} &=& (m + \Sigma_1) \, \Sigma_3^* \, , \;\;\; & \;\, 
H_{24} & = & (1 - \Sigma_2) \, \Sigma_4^* \, . 
\end{array} 
\]
\subsubsection*{Form for $G_{R (A)}^{\rho \rho}$ and $G_K^{(1) \rho 
\rho}$, Eqs. (\ref{final}) and (\ref{label})} 
Using the definition (\ref{teigi}) in Appendix B, one can write the 
quantity in the square brackets in Eq. (\ref{ffinal}) as 
\begin{equation}
\left( P\kern-0.1em\raise0.3ex\llap{/}\kern0.15em\relax - m \right) 
\hat{\tau}_3 - \hat{\tilde{\Sigma}}^{\rho \rho} - \left[ \left[ 
\hat{\tilde{\Sigma}} \otimes 
\hat{\tilde{G}}^{(\mbox{\scriptsize{pre}})} \right] \otimes 
\hat{\tilde{\Sigma}} \right]^{\rho \rho} \, . 
\label{star} 
\end{equation}
The SF for this is obtained by the repeated use of the formulae in 
Appendix B: 
\begin{eqnarray}
\mbox{Eq. (\ref{star})} & = & - \left[ m \hat{\tau}_3 + 
\hat{\tilde{\Sigma}}_1^{\rho \rho} + \hat{A}_1^{\rho - \rho} 
\hat{\tilde{\Sigma}}_1^{- \rho \rho} - P^2 \hat{A}_2^{\rho - \rho} 
\hat{\tilde{\Sigma}}_2^{- \rho \rho} - N^2 \hat{A}_3^{\rho - \rho} 
\hat{\tilde{\Sigma}}_3^{ - \rho \rho} - P^2 N^2 \hat{A}_4^{\rho - 
\rho} \hat{\tilde{\Sigma}}_4^{- \rho \rho} \right] \nonumber \\ 
&& + \left[ \hat{\tau}_3 - \hat{\tilde{\Sigma}}_2^{\rho \rho} - 
\hat{A}_1^{\rho - \rho} \hat{\tilde{\Sigma}}_2^{- \rho \rho} + 
\hat{A}_2^{\rho - \rho} \hat{\tilde{\Sigma}}_1^{- \rho \rho} - N^2 
\hat{A}_3^{\rho - \rho} \hat{\tilde{\Sigma}}_4^{- \rho \rho} - N^2 
\hat{A}_4^{\rho - \rho} \hat{\tilde{\Sigma}}_3^{- \rho \rho} \right] 
P\kern-0.1em\raise0.3ex\llap{/}\kern0.15em\relax \nonumber \\ 
&& - \left[ \hat{\tilde{\Sigma}}_3^{\rho \rho} + \hat{A}_1^{\rho - 
\rho} \hat{\tilde{\Sigma}}_3^{- \rho \rho} - P^2 \hat{A}_2^{\rho - 
\rho} \hat{\tilde{\Sigma}}_4^{- \rho \rho} - \hat{A}_3^{\rho - \rho} 
\hat{\tilde{\Sigma}}_1^{- \rho \rho} - P^2 \hat{A}_4^{\rho - \rho} 
\hat{\tilde{\Sigma}}_2^{- \rho \rho} \right] 
N\kern-0.1em\raise0.3ex\llap{/}\kern0.15em\relax \nonumber \\ 
&& - \left[ \hat{\tilde{\Sigma}}_4^{\rho \rho} + \hat{A}_1^{\rho - 
\rho} \hat{\tilde{\Sigma}}_4^{- \rho \rho} - \hat{A}_2^{\rho - \rho} 
\hat{\tilde{\Sigma}}_3^{- \rho \rho} + \hat{A}_3^{\rho - \rho} 
\hat{\tilde{\Sigma}}_2^{- \rho \rho} + \hat{A}_4^{\rho - \rho} 
\hat{\tilde{\Sigma}}_1^{- \rho \rho} \right] 
P\kern-0.1em\raise0.3ex\llap{/}\kern0.15em\relax 
N\kern-0.1em\raise0.3ex\llap{/}\kern0.15em\relax \, , 
\label{nagai}
\end{eqnarray}
where 
\begin{eqnarray}
\hat{A}_1^{\rho - \rho} & = & \hat{\tilde{\Sigma}}_1^{\rho - \rho} 
\hat{\tilde{G}}_1^{(\mbox{\scriptsize{pre}}) - \rho - \rho} + P^2 
\hat{\tilde{\Sigma}}_2^{\rho - \rho} 
\hat{\tilde{G}}_2^{(\mbox{\scriptsize{pre}}) - \rho - \rho} 
+ N^2 \hat{\tilde{\Sigma}}_3^{\rho - \rho} 
\hat{\tilde{G}}_3^{(\mbox{\scriptsize{pre}}) - \rho - \rho} - P^2 
N^2 \hat{\tilde{\Sigma}}_4^{\rho - \rho} 
\hat{\tilde{G}}_4^{(\mbox{\scriptsize{pre}}) - \rho - \rho} \, , 
\nonumber \\ 
\hat{A}_2^{\rho - \rho} & = & \hat{\tilde{\Sigma}}_1^{\rho - \rho} 
\hat{\tilde{G}}_2^{(\mbox{\scriptsize{pre}}) - \rho - \rho} + 
\hat{\tilde{\Sigma}}_2^{\rho - \rho} 
\hat{\tilde{G}}_1^{(\mbox{\scriptsize{pre}}) - \rho - \rho} - N^2 
\hat{\tilde{\Sigma}}_3^{\rho - \rho} 
\hat{\tilde{G}}_4^{(\mbox{\scriptsize{pre}}) - \rho - \rho} + N^2 
\hat{\tilde{\Sigma}}_4^{\rho - \rho} 
\hat{\tilde{G}}_3^{(\mbox{\scriptsize{pre}}) - \rho - \rho} \, , 
\nonumber \\ 
\hat{A}_3^{\rho - \rho} & = & \hat{\tilde{\Sigma}}_1^{\rho - \rho} 
\hat{\tilde{G}}_3^{(\mbox{\scriptsize{pre}}) - \rho - \rho} + P^2 
\hat{\tilde{\Sigma}}_2^{\rho - \rho} 
\hat{\tilde{G}}_4^{(\mbox{\scriptsize{pre}}) - \rho - \rho} + 
\hat{\tilde{\Sigma}}_3^{\rho - \rho} 
\hat{\tilde{G}}_1^{(\mbox{\scriptsize{pre}}) - \rho - \rho} - P^2 
\hat{\tilde{\Sigma}}_4^{\rho - \rho} 
\hat{\tilde{G}}_2^{(\mbox{\scriptsize{pre}}) - \rho - \rho} \, , 
\nonumber \\ 
\hat{A}_4^{\rho - \rho} & = & \hat{\tilde{\Sigma}}_1^{\rho - \rho} 
\hat{\tilde{G}}_4^{(\mbox{\scriptsize{pre}}) - \rho - \rho} + 
\hat{\tilde{\Sigma}}_2^{\rho - \rho} 
\hat{\tilde{G}}_3^{(\mbox{\scriptsize{pre}}) - \rho - \rho} 
- \hat{\tilde{\Sigma}}_3^{\rho - \rho} 
\hat{\tilde{G}}_2^{(\mbox{\scriptsize{pre}}) - \rho - \rho} + 
\hat{\tilde{\Sigma}}_4^{\rho - \rho} 
\hat{\tilde{G}}_1^{(\mbox{\scriptsize{pre}}) - \rho - \rho} \, . 
\label{nagai1}
\end{eqnarray}
We observe that Eq. (\ref{nagai}) with Eq. (\ref{nagai1}) is 
obtained from Eq. (\ref{zero}) through the following substitutions 
($T$ stands for $R, A$, or $K$), 
\begin{eqnarray}
\Sigma_{T1}^{\rho \rho} & \to & \Sigma_{T1}^{\rho \rho} + 
T_{111}^{\rho \rho} + P^2 T_{221}^{\rho \rho} + N^2 T_{331}^{\rho 
\rho} - P^2 N^2 T_{441}^{\rho \rho} - P^2 \left[ T_{122}^{\rho \rho} 
+ T_{212}^{\rho \rho} - N^2 T_{342}^{\rho \rho} + N^2 T_{432}^{\rho 
\rho} \right] \nonumber \\  
&& - N^2 \left[ T_{133}^{\rho \rho} + P^2 T_{243}^{\rho \rho} + 
T_{313}^{\rho \rho} - P^2 T_{423}^{\rho \rho} \right] - P^2 N^2 
\left[ T_{144}^{\rho \rho} + T_{234}^{\rho \rho} - T_{324}^{\rho 
\rho} + T_{414}^{\rho \rho} \right] \, , 
\nonumber \\  
\Sigma_{T2}^{\rho \rho} & \to & \Sigma_{T2}^{\rho \rho} + 
T_{112}^{\rho \rho} + P^2 T_{222}^{\rho \rho} + N^2 T_{332}^{\rho 
\rho} - P^2 N^2 T_{442}^{\rho \rho} - T_{121}^{\rho \rho} - 
T_{211}^{\rho \rho} + N^2 T_{341}^{\rho \rho} - N^2  T_{431}^{\rho 
\rho} \nonumber \\  
&& + N^2 \left[ T_{134}^{\rho \rho} + P^2 T_{244}^{\rho \rho} + 
T_{314}^{\rho \rho} - P^2 T_{424}^{\rho \rho} \right] + N^2 \left[ 
T_{143}^{\rho \rho} + T_{233}^{\rho \rho} - T_{323}^{\rho \rho} + 
T_{413}^{\rho \rho} \right] \, , 
\nonumber \\ 
\Sigma_{T3}^{\rho \rho} & \to & \Sigma_{T3}^{\rho \rho} + 
T_{113}^{\rho \rho} + P^2 T_{223}^{\rho \rho} + N^2 T_{333}^{\rho 
\rho} - P^2 N^2 T_{443}^{\rho \rho} - P^2 \left[ T_{124}^{\rho \rho} 
+ T_{214}^{\rho \rho} - N^2 T_{344}^{\rho \rho} + N^2 T_{434}^{\rho 
\rho} \right] \nonumber \\  
&& - \left[ T_{131}^{\rho \rho} + P^2 T_{241}^{\rho \rho} + 
T_{311}^{\rho \rho} - P^2 T_{421}^{\rho \rho} \right] - P^2 \left[ 
T_{142}^{\rho \rho} + T_{232}^{\rho \rho} - T_{322}^{\rho \rho} + 
T_{412}^{\rho \rho} \right] \, , \nonumber 
\\  
\Sigma_{T4}^{\rho \rho} & \to & \Sigma_{T4}^{\rho \rho} + 
T_{114}^{\rho \rho} + P^2 T_{224}^{\rho \rho} + N^2 T_{334}^{\rho 
\rho} - P^2 N^2 T_{444}^{\rho \rho} - \left[ T_{123}^{\rho \rho} + 
T_{213}^{\rho \rho} - N^2 T_{343}^{\rho \rho} + N^2 T_{433}^{\rho 
\rho} \right] \nonumber \\ 
&& + \left[ T_{132}^{\rho \rho} + P^2 T_{242}^{\rho \rho} + 
T_{312}^{\rho \rho} - P^2 T_{422}^{\rho \rho} \right] + \left[ 
T_{141}^{\rho \rho} + T_{231}^{\rho \rho} - T_{321}^{\rho \rho} + 
T_{411}^{\rho \rho} \right] \, . 
\label{1.57d1}
\end{eqnarray}
Here, for $T = R$ and $A$, 
\[
R_{ijl}^{\rho \rho} = \Sigma_{Ri}^{\rho - \rho} 
G_{Rj}^{(\mbox{\scriptsize{pre}}) - \rho - \rho} \Sigma_{Rl}^{- \rho 
\rho} 
\] 
and 
\[ 
A_{ijl}^{\rho \rho} = \Sigma_{Ai}^{\rho - \rho} 
G_{Aj}^{(\mbox{\scriptsize{pre}}) - \rho - \rho} \Sigma_{Al}^{- \rho 
\rho} \, , 
\]
respectively, and, for $T = K$, 
\begin{eqnarray*} 
K_{ijl}^{\rho \rho} & = & \Sigma_{Ri}^{\rho - \rho} \left[ 
G^{(\mbox{\scriptsize{pre}}) - \rho - \rho}_{Rj} \Sigma_{Kl}^{- \rho 
\rho} - G^{(\mbox{\scriptsize{pre}}) - \rho - \rho}_{Kj} 
\Sigma_{Al}^{- \rho \rho} \right] \\ 
&& + \Sigma_{Ki}^{\rho - \rho} G^{(\mbox{\scriptsize{pre}}) - \rho - 
\rho}_{Aj} \Sigma_{Al}^{- \rho \rho} \, . 
\end{eqnarray*} 
Then, the expressions for $G_{Rj}^{\rho \rho}$, $G_{Aj}^{\rho 
\rho}$, and $G_{Kj}^{(1) \rho \rho}$ $(j = 1 - 4)$ are obtained from 
those of their counterparts, in respective order, 
$G_{Rj}^{(\mbox{\scriptsize{pre}}) \rho \rho}$, 
$G_{Aj}^{(\mbox{\scriptsize{pre}}) \rho \rho}$, and 
$G_{Kj}^{(\mbox{\scriptsize{pre}}) \rho \rho}$ with the above 
substitutions. 
\subsubsection*{The forms for $H_l^{\rho \sigma}$ in $G_K^{(2) \rho 
\sigma}$, Eq. (\ref{ST22}), and for $\gamma_5 
N\kern-0.1em\raise0.3ex\llap{/}\kern0.15em\relax C^{\rho \sigma} 
(P)$ in ${\bf G}_K^{(3)}$, Eq. (\ref{2.40d})} The form for 
$H_l^{\rho \sigma}$ is obtained by using the formulae in Appendix B: 
\begin{eqnarray}
H_l^{\rho \rho} &=& - C_{\rho - \rho} \left( N^2 \Sigma_{A3}^{- \rho 
\rho} - N^2 \Sigma_{A4}^{- \rho \rho} 
P\kern-0.1em\raise0.3ex\llap{/}\kern0.15em\relax + \Sigma_{A1}^{- 
\rho \rho} N\kern-0.153em\raise0.3ex\llap{/}\kern0.177em\relax - 
\Sigma_{A2}^{- \rho \rho} 
P\kern-0.1em\raise0.3ex\llap{/}\kern0.15em\relax 
N\kern-0.153em\raise0.3ex\llap{/}\kern0.177em\relax \right) 
\nonumber \\ 
&& + \left( N^2 \Sigma_{R3}^{\rho - \rho} - N^2 \Sigma_{R4}^{\rho - 
\rho} P\kern-0.1em\raise0.3ex\llap{/}\kern0.15em\relax - 
\Sigma_{R1}^{\rho - \rho} 
N\kern-0.153em\raise0.3ex\llap{/}\kern0.177em\relax + 
\Sigma_{R2}^{\rho - \rho} 
P\kern-0.1em\raise0.3ex\llap{/}\kern0.15em\relax 
N\kern-0.153em\raise0.3ex\llap{/}\kern0.177em\relax \right) C_{- 
\rho \rho} \, , \nonumber \\ 
H_l^{\rho - \rho} &=& \gamma_5 \left[ C_{\rho - \rho} \left( N^2 
\Sigma_{A3}^{- \rho - \rho} - N^2 \Sigma_{A4}^{- \rho - \rho} 
P\kern-0.1em\raise0.3ex\llap{/}\kern0.15em\relax + \Sigma_{A1}^{- 
\rho - \rho} N\kern-0.153em\raise0.3ex\llap{/}\kern0.177em\relax - 
\Sigma_{A2}^{- \rho - \rho} 
P\kern-0.1em\raise0.3ex\llap{/}\kern0.15em\relax 
N\kern-0.153em\raise0.3ex\llap{/}\kern0.177em\relax 
\right) \right. \nonumber \\ 
&& \left. + \left( N^2 \Sigma_{R3}^{\rho \rho} - N^2 
\Sigma_{R4}^{\rho \rho} 
P\kern-0.1em\raise0.3ex\llap{/}\kern0.15em\relax - \Sigma_{R1}^{\rho 
\rho} N\kern-0.153em\raise0.3ex\llap{/}\kern0.177em\relax + 
\Sigma_{R2}^{\rho \rho} 
P\kern-0.1em\raise0.3ex\llap{/}\kern0.15em\relax 
N\kern-0.153em\raise0.3ex\llap{/}\kern0.177em\relax \right) C_{\rho 
- \rho} \right] \, . 
\label{Hlead} 
\end{eqnarray}

The form for $\gamma_5 
N\kern-0.1em\raise0.3ex\llap{/}\kern0.15em\relax {\bf C}^{\rho 
\sigma} (P)$ is given by Eq. (\ref{Hlead}) with 
\begin{eqnarray*}
\Sigma_{A1}^{\rho \sigma} \to \delta^{\rho \sigma} \, , 
\;\;\;\;\;\;\;\; 
&& \Sigma_{Aj}^{\rho \sigma} \to 0 \;\;\;\;\; (j = 2 - 4) \, , \\ 
&& \Sigma_{Rj}^{\rho \sigma} \to 0 \;\;\;\;\; (j = 1 - 4) \, . 
\end{eqnarray*}
\narrowtext 
\subsubsection*{The form for $\hat{G}^{\rho - \rho}$} 
Having obtained the expression for $\hat{G}^{\rho \rho}$, we can get 
the expression for $\hat{G}^{\rho - \rho}$ from Eqs. Eq. 
(\ref{final1}), (\ref{ST20}) - (\ref{2.40d}), by repeatedly using 
the formulae in Appendix B. 
\setcounter{equation}{0}
\setcounter{section}{2}
\section{Gluon propagator} 
\def\theequation{\mbox{\arabic{section}.\arabic{equation}}}
\subsection{Preliminary}
We adopt a Coulomb gauge. The result for a covariant gauge is 
summarized in Appendix D. 

As an orthogonal basis in Minkowski space, we choose 
\begin{equation}
\begin{array}{ll}
\tilde{P}^\mu \equiv P_\mu - p_0 n^\mu = (0, \vec{p}) \, , 
& \; \tilde{\zeta}^\mu= \left( 0, \; \vec{\zeta} - 
(\vec{\zeta} \cdot \vec{p}) \, \vec{p} / \vec{p}^{\, 2} \right) \, , 
\nonumber \\ 
n^\mu = (1, \vec{0}) \, , & \; E_\perp^\mu = \epsilon^{\mu \nu 
\rho \sigma} \tilde{P}_\nu \tilde{\zeta}_\rho n_\sigma \, . 
\end{array}
\label{orth2}
\end{equation}
These vectors are orthogonal with each other and their norms are 
\[
\begin{array}{ll} 
\tilde{P}^2 = - \vec{p}^{\, 2} \, , \;\;\; & \tilde{\zeta}^2 = - 1 + 
(\vec{\zeta} \cdot \vec{p})^2 / \vec{p}^{\, 2} \, , \\ 
n^2 = 1 \, , \;\;\; & E_\perp^2 = \vec{p}^{\, 2} \tilde{\zeta}^2 
\, . 
\end{array} 
\]
Incidentally, $\epsilon^{\mu \nu \rho \sigma} \tilde{P}_\rho 
\tilde{\zeta}_\sigma$, $\epsilon^{\mu \nu \rho \sigma} 
\tilde{P}_\rho n_\sigma$, and $\epsilon^{\mu \nu \rho \sigma} 
\tilde{\zeta}_\rho n_\sigma$ are not independent but are constructed 
out of the above four vectors, e.g., $\epsilon^{\mu \nu \rho \sigma} 
\tilde{P}_\rho \tilde{\zeta}_\sigma = (E_\perp^\mu n^\nu - n^\mu 
E_\perp^\nu)$, etc. 

We define the projection operators, 
\begin{eqnarray}
&& {\cal P}_T^{\mu \nu} (P) = g^{\mu \nu} - \frac{n^\mu n^\nu}{n^2} 
- \frac{\tilde{P}^\mu \tilde{P}^\nu}{\tilde{P}^2} \, , 
\label{Pro1} \\ 
&& {\cal P}_L^{\mu \nu} (P) = \frac{n^\mu n^\nu}{n^2} \, , 
\label{proL} \\ 
&& {\cal P}_G^{\mu \nu} (P) = \frac{\tilde{P}^\mu 
\tilde{P}^\nu}{\tilde{P}^2} \, . 
\label{Pro3} 
\end{eqnarray}
Although, $n^2 = 1$, we have written $n^2$ explicitly for later 
convenience. In the above definitions, \lq $T$', \lq $L$', and \lq 
$G$' stand, in respective order, for transverse, longitudinal, and 
gauge fixing. (Following tradition, we call $n^\mu n^\nu / n^2$ in 
Eq. (\ref{proL}) the \lq\lq longitudinal projection operator''.) 
From Eqs. (\ref{orth2}) - (\ref{Pro3}), one can show that 
\begin{eqnarray}
\tilde{P}_\mu {\cal P}_U^{\mu \nu} & = & {\cal P}_U^{\nu \mu} 
\tilde{P}_\mu = \delta_{U G} \tilde{P}^\nu \, , \nonumber \\ 
n_\mu {\cal P}_U^{\mu \nu} & = & {\cal P}_U^{\nu \mu} n_\mu = 
\delta_{U L} n^\nu \, , \nonumber \\ 
\tilde{\zeta}_\mu {\cal P}_U^{\mu \nu} & = & {\cal P}_U^{\nu \mu} 
\tilde{\zeta}_\mu = \delta_{U T} \tilde{\zeta}^\nu \, , \nonumber 
\\ 
(E_\perp)_\mu {\cal P}_U^{\mu \nu} & = & {\cal P}_U^{\nu \mu} 
(E_\perp)_\mu = \delta_{U T} E_\perp^\nu \, . 
\label{ara} 
\end{eqnarray}

Let ${\bf A}$ be a generic second-rank tensor in Minkowski space, 
whose $(\mu \nu)$-component is $({\bf A})^{\mu \nu} = A^{\mu \nu}$. 
$A^{\mu \nu}$ is decomposed as 
\begin{eqnarray} 
A^{\mu \nu} (P) & = & \sum_{U, \, V = T, L, G} {\cal P}_U^{\mu \rho} 
\left( A_{U V} \right)_{\rho \sigma} {\cal P}_V^{\sigma \nu} 
\nonumber \\ 
& \equiv & \sum_{U, \, V = T, L, G} \left( {\cal P}_U \cdot A_{U V} 
\cdot {\cal P}_V \right)^{\mu \nu} \, , 
\label{decomp1} \\ 
A_{TT}^{\mu \nu} &=& A_1^{TT} {\cal P}^{\mu \nu}_T + A_2^{TT} 
\tilde{\zeta}^\mu \tilde{\zeta}^\nu - A_3^{T T'} \tilde{\zeta}^\mu 
E_\perp^\nu \nonumber \\ 
&& + A_3^{T' T} E_\perp^\mu \tilde{\zeta}^\nu \, , \nonumber \\ 
A_{LL}^{\mu \nu} &=& A_1^{LL} {\cal P}^{\mu \nu}_L \, , \nonumber \\ 
A_{GG}^{\mu \nu} &=& A_1^{GG} {\cal P}^{\mu \nu}_G \, , \nonumber \\ 
A_{TL}^{\mu \nu} &=& A_1^{TL} \tilde{\zeta}^\mu n^\nu + A_2^{TL} 
E_\perp^\mu n^\nu \, , \nonumber \\ 
A_{LT}^{\mu \nu} &=& A_1^{LT} n^\mu \tilde{\zeta}^\nu - A_2^{LT} 
n^\mu E_\perp^\nu \, , \nonumber \\ 
A_{TG}^{\mu \nu} &=& A_1^{TG} E_\perp^\mu \tilde{P}^\nu + A_2^{TG} 
\tilde{\zeta}^\mu \tilde{P}^\nu \, , \nonumber \\ 
A_{GT}^{\mu \nu} &=& A_1^{GT} \tilde{P}^\mu E_\perp^\nu - A_2^{GT} 
\tilde{P}^\mu \tilde{\zeta}^\nu \, , \nonumber \\ 
A_{LG}^{\mu \nu} &=& A_1^{LG} n^\mu \tilde{P}^\nu \, , \nonumber \\ 
A_{GL}^{\mu \nu} &=& - A_1^{GL} \tilde{P}^\mu n^\nu \, . 
\label{standardg} 
\end{eqnarray}
From Eq. (\ref{ara}), follows $\left( {\cal P}_U \cdot A_{U V} \cdot 
{\cal P}_V \right)^{\mu \nu} = A_{U V}^{\mu \nu}$ $(U, V = T, L, 
G)$. We call Eqs. (\ref{decomp1}) and (\ref{standardg}) the standard 
form (SF) and refer $A_{UV}^{\mu \nu}$ ($U, V = T, L, G$) or 
$A_j^{UV}$ ($U, V = T, T', L, G$) to as an SF-element of $A^{\mu 
\nu}$. It is to be understood that the (bare and self-energy-part 
resummed) propagators and the self-energy part, which appear in the 
following, are to be written in the SF. 
\subsection{Bare propagators}
\subsubsection*{Bare gluon propagator}
First of all, we note that the bare propagator matrix $\hat{\bf D} 
(P)$ and the self-energy-part resummed propagator matrix $\hat{\bf 
G} (P)$ enjoy the \lq\lq symmetry'' property, 
\begin{eqnarray} 
\left( \hat{D}^{\mu \nu} (P) \right)^* & = & - \hat{\tau}_1 
\displaystyle{ \raisebox{1.8ex}{\scriptsize{t}}} 
\mbox{\hspace{-0.33ex}} \hat{D}^{\nu \mu} (P) \hat{\tau}_1 \, , 
\nonumber \\ 
\left( \hat{G}^{\mu \nu} (P) \right)^* & = & - \hat{\tau}_1 
\displaystyle{ \raisebox{1.8ex}{\scriptsize{t}}} 
\mbox{\hspace{-0.33ex}} \hat{G}^{\nu \mu} (P) \hat{\tau}_1 \, , 
\nonumber \\ 
\hat{D}^{\mu \nu} (P) & = & \displaystyle{ 
\raisebox{1.8ex}{\scriptsize{t}}} \mbox{\hspace{-0.33ex}} 
\hat{D}^{\nu \mu} (- P) \, , \nonumber \\ 
\hat{G}^{\mu \nu} (P) & = & \displaystyle{ 
\raisebox{1.8ex}{\scriptsize{t}}} \mbox{\hspace{-0.33ex}} 
\hat{G}^{\nu \mu} (- P) \, . 
\label{seisei} 
\end{eqnarray} 
The first two equations results from the hermiticity of the density 
matrix. 

$\hat{D}^{\mu \nu} (P)$ is an inverse of 
\begin{equation}
(\hat{D}^{-1} (P))^{\mu \nu} = - \left[ P^2 g^{\mu \nu} - P^\mu 
P^\nu + \frac{1}{\lambda} \tilde{P}^\mu \tilde{P}^\nu \right] 
\hat{\tau}_3 
\label{2.8d} 
\end{equation} 
with $\lambda$ a gauge parameter. A general solution to $\left( 
{\bf D}^{-1} {\bf D} \right)^{\mu \nu} = g^{\mu \nu}$ is written as 
\begin{eqnarray} 
\hat{\bf D} &=& \hat{\bf D}^{(0)} + {\bf D}_K \hat{M}_+ \, , 
\label{bare0g} \\ 
\hat{\bf D}^{(0)} &=& \hat{\bf D}_{RA} + \tilde{f} ({\bf D}_R - {\bf 
D}_A) \hat{M}_+ \, , \\ 
\hat{\bf D}_{RA} &=& \left( 
\begin{array}{cc} 
{\bf D}_R & \;\; 0 \\ 
{\bf D}_R - {\bf D}_A & \;\; - {\bf D}_A 
\end{array} 
\right) \, , \\ 
D_K^{\mu \nu} & = & \left( D_K \right)_{TT}^{\mu \nu} = - 
\tilde{C}^{\mu \nu} \left( \Delta_R (P) - \Delta_A (P) \right) \, , 
\nonumber \\ 
&& \label{Kdesu} \\ 
\tilde{C}^{\mu \nu} &=& C_2^{TT} (P) \tilde{\zeta}^\mu 
\tilde{\zeta}^\nu - C_3^{T T'} (P) \tilde{\zeta}^\mu E_\perp^\nu 
\nonumber \\ 
&& + C_3^{T' T} (P) E_\perp^\mu \tilde{\zeta}^\nu \, , 
\label{3.13d} 
\end{eqnarray} 
where 
\begin{eqnarray}
\tilde{f} (P) & = & \theta (p_0) N (|p_0|, \vec{p}) - \theta (- p_0) 
[1 + N (|p_0|, - \vec{p}) ] \, , 
\nonumber \\ 
D_R^{\mu \nu} & = & \left( D_A^{\mu \nu} \right)^* = - \Delta_R 
{\cal P}_T^{\mu \nu} - \frac{1}{\tilde{P}^2} \left[ 1 + \lambda 
\frac{p_0^2}{\tilde{P}^2} \right] {\cal P}_L^{\mu \nu} \nonumber \\ 
&& \mbox{\hspace*{9.8ex}} - \frac{\lambda}{\tilde{P}^2} {\cal 
P}_G^{\mu \nu} - \lambda \frac{p_0}{\tilde{P}^4} (\tilde{P}^\mu 
n^\nu + n^\mu \tilde{P}^\nu) \, . \nonumber \\ 
&& 
\label{Ryodesu} 
\end{eqnarray}
Here $N$ is the number density of the transverse gluon and 
$\Delta_{R (A)}$ is as in Eq. (\ref{2jyuu}). From Eqs. 
(\ref{seisei}), (\ref{bare0g}), and (\ref{Kdesu}), we have 
\begin{equation}
\left( C_2^{TT} \right)^* = C_2^{TT} \, , \;\;\;\;\;\;\;\;\;\;\; 
\left( C_3^{TT'} \right)^* = - C_3^{T'T} \,  
\label{mujina} 
\end{equation}

Note that, for the quasiuniform systems near equilibrium, 
$C_2^{TT}$, $C_3^{TT'}$, and $C_3^{T'T}$ are small when compared to 
$\tilde{f}$. 
\subsubsection*{Bare ghost propagator} 
A bare Fadeev-Popov (FP) ghost propagator $\hat{\tilde{D}}$ is 
\begin{equation}
\hat{\tilde{D}} = \frac{1}{\tilde{P}^2} \, \hat{\tau}_3 \, . 
\label{ghostp} 
\end{equation}
\subsection{Dyson equation} 
\subsubsection*{Gluon sector} 
The self-energy-part ($\hat{\bf \Pi}$) resummed propagator $\hat{\bf 
G}$ obeys 
\begin{equation} 
\hat{\bf G} (P) = \hat{\bf D} (P) - \hat{\bf D} (P) \hat{\bf \Pi} 
(P) \hat{\bf G} (P) \, . 
\label{SDg} 
\end{equation} 
From Eq. (\ref{seisei}), we obtain the symmetry relations, for the 
SF-elements of $\hat{G}^{\mu \nu}$, 
\begin{eqnarray}
\left( \hat{G}_j^{UV} (P) \right)^* &=& - \sigma_j^{UV} \hat{\tau}_1 
\displaystyle{ \raisebox{1.8ex}{\scriptsize{t}}} 
\mbox{\hspace{-0.33ex}} \hat{G}_j^{VU} (P) \hat{\tau}_1 \, , 
\nonumber \\ 
\hat{G}_j^{UV} (P) & = & \displaystyle{ 
\raisebox{1.8ex}{\scriptsize{t}}} \mbox{\hspace{-0.33ex}} 
\hat{G}_j^{VU} (- P) \;\;\;\;\;\; (U, V = T, T', L, G) \, , 
\nonumber \\ 
&& 
\label{symm3} 
\end{eqnarray}
where $\sigma_j^{UV} = \sigma_j^{VU}$ with $\sigma_j^{UU} = +$ ($U = 
T, L, G$) and 
\[
\sigma_j^{UV} = \left\{ 
\begin{array}{lcl}
+ \;\; \mbox{for} \;\; (UV, j) & = & (TL,1) \, , \, (TG,1) \\ 
- \;\; \mbox{for} \;\; (UV, j) & = & (TT',3) \, , \, (TL,2) \, , \\ 
&& (TG,2) \, , \, (LG,1) \, . 
\end{array} 
\right. 
\]
Similar relations hold for $\left( \hat{\Pi}_j \right)_{UV}$'s. 

Components of $\hat{\bf \Pi}$ follow the same relation as 
(\ref{seip}) and then $\hat{\bf \Pi}$ is written as 
\begin{eqnarray} 
\hat{\bf \Pi} &=& \hat{\bf \Pi}^{(0)} - {\bf \Pi}_K \hat{M}_- \, , 
\label{hossh} \\ 
\hat{\bf \Pi}^{(0)} &=& 
\left( 
\begin{array}{cc} 
{\bf \Pi}_R & \;\; 0 \\ 
- {\bf \Pi}_R + {\bf \Pi}_A & \;\; - {\bf \Pi}_A 
\end{array} 
\right) 
+ \tilde{f} ({\bf \Pi}_R - {\bf \Pi}_A ) \hat{M}_- \, , \nonumber \\ 
\\ 
{\bf \Pi}_R & = & {\bf \Pi}_{11} + {\bf \Pi}_{12} \, , \\ 
{\bf \Pi}_A & = & {\bf \Pi}_{11} + {\bf \Pi}_{21} \, , \\ 
{\bf \Pi}_K &=& (1 + \tilde{f}) {\bf \Pi}_{12} - \tilde{f} {\bf 
\Pi}_{21} \, . 
\label{owani1}
\end{eqnarray} 
From Eq. (\ref{symm3}) with $\hat{\Pi}_j$ for $\hat{G}_j$, we obtain 
the symmetry relations $(U, V, = T, T', L, G)$, 
\begin{eqnarray} 
\left( \Pi_{Aj}^{UV} (P) \right)^* & = & \sigma_j^{UV} \Pi_{Rj}^{VU} 
(P) \, , 
\label{mak} \\ 
\left( \Pi_{Kj}^{UV} (P) \right)^* & = & - \sigma_j^{UV} 
\Pi_{Kj}^{VU} (P) \, , 
\label{maki} \\ 
\Pi_{Rj}^{UV} (P) & = & \Pi_{Aj}^{VU} (- P) = \sigma_j^{UV} 
(\Pi_{Rj}^{UV} (- P))^*  \, , \nonumber \\ 
\Pi_{Kj}^{UV} (P) & = & \Pi_{Kj}^{VU} (- P) \nonumber \\ 
&& - \epsilon (p_0) \left[ N (|p_0|, \vec{p}) - N (|p_0|, - \vec{p}) 
\right] \nonumber \\ 
&& \times \left( \Pi_{Rj}^{VU} (- P) - \Pi_{Aj}^{VU} (- P) \right) 
\, . \nonumber 
\end{eqnarray} 

Components of ${\bf G}$ follow the same relation as (\ref{seip1}) 
and then $\hat{\bf G}$ is written as 
\begin{eqnarray} 
\hat{\bf G} &=& \hat{\bf G}^{(0)} + {\bf G}_K \hat{M}_+ \, , 
\label{hossh1} \\ 
\hat{\bf G}^{(0)} &=& \left( 
\begin{array}{cc} 
{\bf G}_R & \;\; 0 \\ 
{\bf G}_R - {\bf G}_A & \;\; - {\bf G}_A 
\end{array} 
\right) 
+ \tilde{f} ({\bf G}_R - {\bf G}_A ) \hat{M}_+ \, , \nonumber \\ 
\\ 
{\bf G}_R & = & {\bf G}_{11} - {\bf G}_{12} \, , \\ 
{\bf G}_A & = & {\bf G}_{11} - {\bf G}_{21} \, , \\ 
{\bf G}_K &=& (1 + \tilde{f}) {\bf G}_{12} - \tilde{f} {\bf 
G}_{21} \, . 
\label{owani} 
\end{eqnarray} 
For equilibrium systems, ${\bf \Pi}_K = {\bf G}_K = 0$. From Eq. 
(\ref{symm3}), follows the symmetry relations $(U, V, = T, T', L, 
G)$, 
\begin{eqnarray} 
\left( G_{Aj}^{UV} (P) \right)^* & = & \sigma_j^{UV} G_{Rj}^{VU} (P) 
\, , 
\label{2.28d} \\ 
\left( G_{Kj}^{UV} (P) \right)^* & = & - \sigma_j^{UV} G_{Kj}^{VU} 
(P) \, , 
\label{3.32dc} \\ 
G_{Rj}^{UV} (P) & = & G_{Aj}^{VU} (- P) = \sigma_j^{UV} (G_{Rj}^{UV} 
(- P))^* \, , \nonumber \\ 
&& 
\label{asi} \\ 
G_{Kj}^{UV} (P) & = & G_{Kj}^{VU} (- P) \nonumber \\ 
&& + \epsilon (p_0) \left[ N (|p_0|, \vec{p}) - N (|p_0|, - \vec{p}) 
\right] \nonumber \\ 
&& \times \left( G_{Rj}^{VU} (- P) - G_{Aj}^{VU} (- P) \right) \, . 
\label{hikkuri} 
\end{eqnarray} 

Substitution of Eqs. (\ref{bare0g}), (\ref{hossh}), and 
(\ref{hossh1}) into Eq. (\ref{SDg}) yields 
\begin{eqnarray} 
\hat{\bf G}^{(0)} &=& \hat{\bf D}^{(0)} - \hat{\bf D}^{(0)} \hat{\bf 
\Pi} \hat{\bf G}^{(0)} \, , 
\label{2.344} \\ 
{\bf G}_K &=& {\bf D}_K - {\bf D}_R {\bf \Pi}_R {\bf G}_K + {\bf 
D}_R {\bf \Pi}_K {\bf G}_A - {\bf D}_K {\bf \Pi}_A {\bf G}_A \, . 
\nonumber \\ 
&& 
\label{Kne} 
\end{eqnarray} 
Eq. (\ref{2.344}) is formally solved to give 
\begin{equation}
{\bf G}_{R (A)} = \left[ {\bf D}^{- 1} + {\bf \Pi}_{R (A)} 
\right]^{- 1} \, . 
\label{ade} 
\end{equation}
For later convenience, we rewrite Eq. (\ref{ade}) as (cf. Eq. 
(\ref{2.8d})) 
\begin{eqnarray} 
{\bf G}_R &=& \left[ {\bf D}_0^{- 1} + {\bf \Pi}_R' \right]^{- 1} \, 
, 
\label{asita} \\ 
(D_0^{- 1})^{\mu \nu} &=& - \tilde{P}^2 \left[ {\cal P}_T^{\mu \nu} 
+ {\cal P}_L^{\mu \nu} + \frac{1}{\lambda} {\cal P}_G^{\mu \nu} 
\right] \, , \\ 
\Pi_R^{' \mu \nu} &=& \Pi_R^{\mu \nu} - p_0^2 \left( {\cal P}_T^{\mu 
\nu} + {\cal P}_G^{\mu \nu} \right) \nonumber \\ 
&& + p_0 (n^\mu \tilde{P}^\nu + \tilde{P}^\mu n^\nu) \, . 
\label{dash} 
\end{eqnarray} 
The SF for ${\bf G}_{R (A)}$ will be given in the next section. 

As for ${\bf G}_K$, Eq. (\ref{Kne}), through similar procedure as in 
the quark case (cf. Appendix A), we obtain 
\begin{eqnarray}
{\bf G}_K & = & {\bf G}_K^{(1)} + {\bf G}_K^{(2)} + {\bf G}_K^{(3)} 
\, , \nonumber \\ 
{\bf G}_K^{(1)} &=& {\bf G}_R {\bf \Pi}_K {\bf G}_A \, , \nonumber 
\\ 
{\bf G}_K^{(2)} &=& - {\bf G}_R \left[ \tilde{\bf C} {\bf \Pi}_A 
- {\bf \Pi}_R \tilde{\bf C} \right] {\bf G}_A \nonumber \\ 
& \equiv & - {\bf G}_R \tilde{\bf H}_l {\bf G}_A \, , 
\label{Tnyol} \\ 
{\bf G}_K^{(3)} &=& {\bf G}_R \tilde{\bf C} - \tilde{\bf C} {\bf 
G}_A \, , \nonumber 
\end{eqnarray}
where $\tilde{\bf C}$ is as in Eq. (\ref{3.13d}). SF-elements of 
$\tilde{\bf H}_l$ is given in the next section. The SF elements of 
${\bf G}_K$ is obtained by repeatedly using the formulae in Appendix 
C. 
\subsubsection*{Ghost sector}
The self-energy-part ($\hat{\tilde{\Pi}}$) resummed propagator 
$\hat{\tilde{G}}$ obeys 
\begin{equation}
\hat{\tilde{G}} (P) = \hat{\tilde{D}} (P) \left[ 1 + 
\hat{\tilde{\Pi}} (P) \hat{\tilde{G}} (P) \right] = \left[ 1 + 
\hat{\tilde{G}} (P) \hat{\tilde{\Pi}} (P) \right] \hat{\tilde{D}} 
\label{SDghost} \, . 
\end{equation}
Since, $\hat{\tilde{D}}$, Eq. (\ref{ghostp}), is a diagonal $(2 
\times 2)$-matrix, $\hat{\tilde{\Pi}}$ is also diagonal. Then, from 
Eq. (\ref{SDghost}), $\hat{\tilde{G}}$ is diagonal also. Among the 
components of $\hat{\tilde{\Pi}}$ ($\hat{\tilde{G}}$), there is the 
same relation as (\ref{seip}) ((\ref{seip1})). Then we 
have 
\begin{equation}
\hat{\tilde{\Pi}} = \tilde{\Pi} \hat{\tau}_3 \, , 
\mbox{\hspace*{13ex}} \hat{\tilde{G}} = \tilde{G} \hat{\tau}_3 
\label{realyo} 
\end{equation}
with $\tilde{\Pi}$ and $\tilde{G}$ real, and 
\begin{equation}
\tilde{G} (P) = \frac{1}{\tilde{P}^2 - \tilde{\Pi} (P)} 
= - \frac{1}{\vec{p}^{\, 2} + \tilde{\Pi} (P)} \, . 
\label{kotae}
\end{equation}
\subsection{Self-energy-part resummed gluon propagator}
\subsubsection*{Form for $\left( G_{R (A)} \right)_{UV}$}
Through a Slavnov-Taylor identity, $(\hat{G}_1)_{UG}$ $(U = G, T, 
L)$ and $(\hat{G}_2)_{TG}$ (and then also $(\hat{G}_1)_{GU}$ and 
$(\hat{G}_2)_{GT}$ via Eq. (\ref{symm3})) are related to the 
self-energy-part resummed FP-ghost propagator $\hat{\tilde{G}}$ and 
the FP-ghost \lq\lq pre self-energy part''\footnote{$\; 
\hat{\tilde{\Pi}}_\mu$ is evaluated by replacing the vertex factor 
$g C_{a b c} \tilde{P}^\mu$ at the \lq\lq end vertex'' with $g C_{a 
b c}$. Here the \lq\lq end vertex'' is the vertex from which the 
outgoing ghost comes out of the diagram. Then, the ghost self-energy 
part $\hat{\tilde{\Pi}}$ is related to $\hat{\tilde{\Pi}}_\mu$ 
through $\hat{\tilde{\Pi}} = \tilde{P}^\mu \hat{\tilde{\Pi}}_\mu$.} 
$\hat{\tilde{\Pi}}_\mu$. 

The Slavnov-Taylor identity reads \cite{7}: 
\begin{eqnarray} 
\hat{G}_{\mu \nu} \tilde{P}^\nu & = & \lambda \left[ \hat{\tau}_3 
\hat{\tilde{\Pi}}_\mu - P_\mu \right] \hat{\tilde{G}} \nonumber \\ 
& = & \lambda \left[ \hat{\tau}_3 \left( \hat{\tilde{\Pi}}_\mu - 
p_0 n_\mu \hat{\tau}_3 \right) - \tilde{P}_\mu \right] 
\hat{\tilde{G}} \, . 
\label{ST} 
\end{eqnarray}
Here $\hat{\tilde{G}} (P)$ is as in Eq. (\ref{realyo}) with Eq. 
(\ref{kotae}). As in Eq. (\ref{realyo}), $\hat{\tilde{\Pi}}_\mu$ is 
diagonal $(2 \times 2)$-matrix, 
\begin{equation} 
\hat{\tilde{\Pi}}_\mu = \tilde{\Pi}_\mu \hat{\tau}_3 
\mbox{\hspace*{15ex}} (\tilde{\Pi}_\mu^* = \tilde{\Pi}_\mu) \, . 
\label{preg}
\end{equation}

Substitution of the SF for $\hat{G}^{\mu \nu}$ into Eq. (\ref{ST}) 
yields 
\begin{eqnarray}
G_{R1}^{GG} (P) &=& G_{A1}^{GG} (P) = \lambda \frac{1}{\tilde{P}^2} 
\left[ \tilde{\Pi} (P) - \tilde{P}^2 \right] \tilde{G} (P) \nonumber 
\\ 
& = & - \lambda \frac{1}{\tilde{P}^2} \, , 
\label{GP3} \\ 
G_{R1}^{TG} (P) &=& G_{A1}^{TG} (P) = \lambda \frac{1}{\tilde{P}^2 
E_\perp^2} \left( E_\perp^\mu \tilde{\Pi}_\mu (P) \right) \tilde{G} 
(P) \, , \nonumber \\ 
&& 
\label{61} \\ 
G_{R2}^{TG} (P) &=& G_{A2}^{TG} (P) = \lambda \frac{1}{\tilde{P}^2 
\tilde{\zeta}^2} \left( \tilde{\zeta}^\mu \tilde{\Pi}_\mu (P) 
\right) \tilde{G} (P) \, , 
\label{60} \\ 
G_{R1}^{LG} (P) &=& G_{A1}^{LG} (P) = \lambda \frac{1}{\tilde{P}^2 
n^2} \left( n^\mu \tilde{\Pi}_\mu (P) - p_0 \right) \tilde{G} (P) \, 
, \nonumber \\ 
&& \label{GP1} \\ 
G_{K1}^{GG} &=& G_{K1}^{TG} = G_{K2}^{TG} = G_{K1}^{LG} = 0 \, . 
\label{GK1} 
\end{eqnarray}
All the above quantities are real. In deriving Eq. (\ref{GP3}), 
Eq. (\ref{kotae}) has been used. Substituting Eqs. (\ref{GP3}) - 
(\ref{GK1}) into Eq. (\ref{hikkuri}), we obtain $G_{Kj}^{GT} = 
G_{K1}^{GL} = 0$. From the above formulae, we see that 
$\hat{G}^{\mu \nu}_{UG}$ ($U = T, L, G$) and $\hat{G}^{\mu 
\nu}_{GU}$ ($U = T, L$) vanish in the strict Coulomb gauge ($\lambda 
= 0$). 

We are now in a position to obtain $\left( {\bf G}_R \right)_{UV}$ 
($U \neq G, V \neq G$) from Eq. (\ref{asita}). We divide $\left( 
{\bf G}_R \right)_{UV}$ into two pieces, $\left( {\bf G}_R 
\right)_{UV} = \left( {\bf G}_R^{(\lambda = 0)} \right)_{UV} + 
\left( {\bf G}_R^{(\lambda)} \right)_{UV}$, the latter of which 
vanishes in the strict Coulomb gauge ($\lambda =  0$). 

Straightforward manipulation of Eq. (\ref{asita}) using the formulae 
in Appendix C yields, for the SF-elements of ${\bf G}^{(\lambda = 
0)}$ ($\equiv {\bf G}^{(\lambda = 0)}_R$) ($\Pi_j^{UV} \equiv 
\Pi_{Rj}^{UV}$), 
\widetext 
\begin{eqnarray}
{\cal D} G_1^{(\lambda = 0) TT} &=& - \left( \tilde{P}^2 - 
\Pi_1^{'TT} - \tilde{\zeta}^2 \Pi_2^{TT} \right) \left( \tilde{P}^2 
- \Pi_1^{LL} \right) + \tilde{\zeta}^2 n^2 \Pi_1^{TL} \Pi_1^{LT} 
\, , \nonumber \\ 
{\cal D} G_2^{(\lambda = 0) TT} &=& - \left( \tilde{P}^2 - 
\Pi_1^{LL} \right) \Pi_2^{TT} + \tilde{P}^2 n^4 \Pi_2^{TL} 
\Pi_2^{LT} + n^2 \Pi_1^{TL} \Pi_1^{LT} \, , \nonumber \\ 
{\cal D} G_3^{(\lambda = 0) T' T} &=& - \left( \tilde{P}^2 - 
\Pi_1^{LL} \right) \Pi_3^{T' T} - n^2 \Pi_1^{LT} \Pi_2^{TL} \, , 
\nonumber \\ 
{\cal D} G^{(\lambda = 0) LL}_1 &=& - \left( \tilde{P}^2 - 
\Pi_1^{'TT} - \tilde{\zeta}^2 \Pi_2^{TT} \right) \left( \tilde{P}^2 
- \Pi_1^{'TT} \right) + \tilde{P}^2 \tilde{\zeta}^4 n^2 \Pi_3^{TT'} 
\Pi_3^{T'T} \, , \nonumber \\ 
{\cal D} G_2^{(\lambda = 0) TL} &=& - \left( \tilde{P}^2 - 
\Pi_1^{'TT} - \tilde{\zeta}^2 \Pi_2^{TT} \right) \Pi_2^{TL} - 
\tilde{\zeta}^2 \Pi_1^{TL} \Pi_3^{T' T} \, , \nonumber \\ 
{\cal D} G_1^{(\lambda = 0) LT} &=& - \left( \tilde{P}^2 - 
\Pi_1^{'TT} \right) \Pi_1^{LT} - \tilde{P}^2 \tilde{\zeta}^2 n^2 
\Pi_2^{LT} \Pi_3^{T' T} \, , 
\label{zero1} 
\end{eqnarray}
where 
\begin{eqnarray*}
{\cal D} & = & \left[ \left( \tilde{P}^2 - \Pi_1^{'TT} \right) 
\left( \tilde{P}^2 - \Pi_1^{LL} \right) - \tilde{P}^2 
\tilde{\zeta}^2 n^4 \Pi_2^{LT} \Pi_2^{TL} \right] \left( \tilde{P}^2 
- \Pi_1^{'TT} - \tilde{\zeta}^2 \Pi_2^{TT} \right) \nonumber \\ 
&& - \tilde{\zeta}^2 n^2 \Pi_1^{LT} \left[ \left( \tilde{P}^2 - 
\Pi_1^{'TT} \right) \Pi_1^{TL} + \tilde{P}^2 \tilde{\zeta}^2 n^2 
\Pi_3^{TT'} \Pi_2^{TL} \right] \nonumber \\ 
&& + {\vec p}^{\, 2} \tilde{\zeta}^4 \Pi_3^{T' T} \left[ \left( 
\tilde{P}^2 - \Pi_1^{LL} \right) \Pi_3^{TT'} + n^2 \Pi_1^{TL} 
\Pi_2^{LT} \right] \, . 
\end{eqnarray*}
Here, we note that, from Eq. (\ref{dash}), $\tilde{P}^2 - 
\Pi_1^{'TT} = P^2 - \Pi_1^{TT}$ holds. The SF-elements of the 
gauge-parameter dependent part ${\bf G}^{(\lambda)} \equiv {\bf 
G}^{(\lambda)}_R$ reads 
\begin{eqnarray}
{\cal D} G_1^{(\lambda) TT} &=& - \tilde{P}^4 \tilde{\zeta}^2 
n^2 \left[ \left( \tilde{P}^2 - \Pi_1^{'TT} - \tilde{\zeta}^2 
\Pi_2^{TT} \right) \left\{ \left( \tilde{P}^2 - \Pi_1^{LL} 
\right) \Pi_1^{GT} + n^2 \Pi_1^{'GL} \Pi_2^{LT} \right\} 
\right. \nonumber \\ 
&& + \tilde{\zeta}^2 \Pi_2^{GT} \left\{ \left( \tilde{P}^2 - 
\Pi_1^{LL} \right) \Pi_3^{TT'} + n^2 \Pi_1^{TL} \Pi_2^{LT} 
\right\} \nonumber \\ 
&& \left. - \tilde{\zeta}^2 n^2 \Pi_1^{LT} \left\{ \Pi_1^{GT} 
\Pi_1^{TL} - \Pi_1^{'GL} \Pi_3^{TT'} \right\} 
\right] G_1^{TG} \, , 
\nonumber \\ 
{\cal D} G^{(\lambda) TT}_2 &=& - \frac{\tilde{P}^2}{\tilde{\zeta}^2 
\Pi_3^{T' T}} \left[ \left\{ (\tilde{P}^2 - \Pi_1^{LL}) \Pi_2^{GT} + 
n^2 \Pi_1^{'GL} \Pi_1^{LT} \right\} \left( \tilde{P}^2 - \Pi_1^{'TT} 
\right)^2 \right. \nonumber \\ 
&& + \tilde{P}^2 \tilde{\zeta}^2 n^4 \Pi_2^{TL} \left( \tilde{P}^2 
- \Pi_1^{'TT} \right) \left\{ \Pi_1^{LT} \Pi_1^{GT} - \Pi_2^{LT} 
\Pi_2^{GT} \right\} \nonumber \\ 
&& - \tilde{P}^2 \tilde{\zeta}^4 n^2 \Pi_3^{T' T} \left\{ 
(\tilde{P}^2 - \Pi_1^{LL}) \left( \Pi_2^{GT} \Pi_3^{TT'} - 
\Pi_1^{GT} \Pi_2^{TT} \right) \right. \nonumber \\ 
&& \left. \left. + n^2 \Pi_1^{'GL} \left( \Pi_1^{LT} \Pi_3^{TT'} - 
\Pi_2^{TT} \Pi_2^{LT} \right) + n^2 \Pi_1^{TL} \left( \Pi_2^{GT} 
\Pi_2^{LT} - \Pi_1^{GT} \Pi_1^{LT} \right) \right\} \right] G_1^{TG} 
\nonumber \\ 
&& - \frac{n^2}{\tilde{\zeta}^2} \frac{\Pi_2^{TL} 
G_1^{LG}}{\Pi_3^{T'T} G_1^{TG}} G_3^{(\lambda) T' T} {\cal D} + 
\frac{\tilde{P}^2}{\tilde{\zeta}^4} \frac{\Pi_1^{TG}}{\Pi_3^{T' T}} 
G_2^{GT} {\cal D} \, , 
\nonumber \\ 
{\cal D} G_3^{(\lambda) T' T} &=& \tilde{P}^2 \left[ - \tilde{P}^2 
\tilde{\zeta}^2 n^2 \Pi_1^{GT} \left\{ \left( \tilde{P}^2 - 
\Pi_1^{LL} \right) \Pi_3^{T' T} + n^2 \Pi_1^{LT} \Pi_2^{TL} \right\} 
\right. \nonumber \\ 
&& - \Pi_2^{GT} \left\{ \left( \tilde{P}^2 - \Pi_1^{'TT} \right) 
(\tilde{P}^2 - \Pi_1^{LL}) - \tilde{P}^2 \tilde{\zeta}^2 n^4 
\Pi_2^{TL} \Pi_2^{LT} \right\} \nonumber \\ 
&& \left. - n^2 \Pi_1^{'GL} \left\{ \left( \tilde{P}^2 - \Pi_1^{'TT} 
\right) \Pi_1^{LT} + \tilde{P}^2 \tilde{\zeta}^2 n^2 \Pi_2^{LT} 
\Pi_3^{T' T} \right\} \right] G_1^{TG} \, , 
\nonumber \\ 
{\cal D} G^{(\lambda) LL}_1 &=& {\vec p}^{\, 2} \left[ \left( 
\tilde{P}^2 - \Pi_1^{'TT} - \tilde{\zeta}^2 \Pi_2^{TT} \right) 
\left\{ \left( \tilde{P}^2 - \Pi_1^{'TT} \right) \Pi_1^{'GL} + 
\tilde{P}^2 \tilde{\zeta}^2 n^2 \Pi_1^{GT} \Pi_2^{TL} \right\} 
\right. \nonumber \\ 
&& + \tilde{\zeta}^2 \Pi_2^{GT} \left\{ \left( \tilde{P}^2 - 
\Pi_1^{'TT} \right) \Pi_1^{TL} + \tilde{P}^2 \tilde{\zeta}^2 n^2 
\Pi_2^{TL} \Pi_3^{TT'} \right\} \nonumber \\ 
&& \left. - \tilde{P}^2 \tilde{\zeta}^4 n^2 \Pi_3^{T'T} \left\{ 
\Pi_1^{'GL} \Pi_3^{T T'} - \Pi_1^{GT} \Pi_1^{TL} \right\} 
\right] G_1^{LG} \, , \nonumber \\ 
{\cal D} G^{(\lambda) TL}_2 &=& \tilde{P}^2 \left[ - \Pi_1^{'G L} 
\left( \tilde{P}^2 - \Pi_1^{'TT} \right)^2 + \tilde{P}^2 
\tilde{\zeta}^4 n^2 \Pi_1^{GT} \left\{ \Pi_2^{TT} \Pi_2^{TL} - 
\Pi_1^{TL} \Pi_3^{T' T} \right\} \right. \nonumber \\ 
&& + \tilde{\zeta}^2 \left( \tilde{P}^2 - \Pi_1^{'TT} \right) 
\left\{ - \tilde{P}^2 n^2 \Pi_1^{GT} \Pi_2^{TL} + \Pi_2^{TT} 
\Pi_1^{'GL} - \Pi_2^{GT} \Pi_1^{TL} \right\} \nonumber \\ 
&& \left. - \tilde{P}^2 \tilde{\zeta}^4 n^2 \Pi_3^{TT'} \left\{ 
\Pi_2^{GT} \Pi_2^{TL} - \Pi_1^{'GL} \Pi_3^{T' T} \right\} \right] 
G_1^{TG} \, , \nonumber \\ 
G^{(\lambda) LT}_1 &=& G_3^{(\lambda) T' T} 
\frac{G_1^{LG}}{G_1^{TG}} \, . 
\label{GM2} 
\end{eqnarray}
\narrowtext 
\noindent 
Here, $G_1^{TG}$ $\left( \equiv G_{R1}^{TG} \right)$ and $G_1^{LG}$ 
$\left( \equiv G_{R1}^{LG} \right)$ are as in Eqs. (\ref{61}) and 
(\ref{GP1}), respectively. $G_2^{GT}$ $\left( \equiv G_{R2}^{GT} 
\right)$ is obtained from $G_{A2}^{TG}$, Eq. (\ref{60}), with the 
help of Eq. (\ref{2.28d}). 

$\left( {\bf G}_A \right)_{UV}$ is obtained from th above formulae 
with the substitutions $\Pi_j^{UV}$ $(\equiv \Pi_{Rj}^{UV})$ $\to$ 
$\Pi_{Aj}^{UV}$. $G_{R3}^{TT'}$, $G_{R1}^{TL}$, and $G_{R2}^{LT}$ 
are obtained, in respective order, from $G_{A3}^{T'T}$, 
$G_{A1}^{LT}$, and $G_{A2}^{TL}$ with the help of Eq. (\ref{2.28d}) 
or Eq. (\ref{asi}). 
\subsubsection*{Form for $\left( \tilde{H}_l \right)_{UV}$ in Eq. 
(\ref{Tnyol})} 
Straightforward computation using the formulae in Appendix C yields, 
for the SF-elements of $\tilde{H}_l$,  
\begin{eqnarray}
( \tilde{H}_{l} )_1^{TT} &=& 2 i \tilde{\zeta}^2 E_\perp^2 Im \left( 
C^{TT'}_3 \Pi_{R3}^{T'T} \right) \, , \nonumber \\ 
( \tilde{H}_{l} )_2^{TT} &=& 2 i Im \left\{ - C_2^{TT} \left( 
\Pi_{R1}^{TT} + \tilde{\zeta}^2 \Pi_{R2}^{TT} \right) \right. 
\nonumber \\ 
&& \left. + E_\perp^2 C_3^{T'T} \left( \Pi_{R3}^{TT'} + 
\Pi_{A3}^{TT'} \right) \right\} \, , \nonumber \\ 
( \tilde{H}_{l} )_3^{TT'} &=& \tilde{\zeta}^2 C_2^{TT} 
\Pi_{A3}^{TT'} - C_3^{TT'} \left( 2 i Im \Pi_{R1}^{TT} + 
\tilde{\zeta}^2 \Pi_{R2}^{TT} \right) \, , \nonumber \\ 
( \tilde{H}_{l} )_1^{TL} &=& \tilde{\zeta}^2 C_2^{TT} \Pi_{A1}^{TL} 
- E_\perp^2 C_3^{TT'} \Pi_{A2}^{TL} \, , \nonumber \\ 
( \tilde{H}_{l} )_2^{TL} &=& \tilde{\zeta}^2 C_3^{T'T} \Pi_{A1}^{TL} 
\, . 
\label{etti}
\end{eqnarray}
$( \tilde{H}_{l} )_3^{T'T}$, $( \tilde{H}_{l} )_1^{LT}$, and $( 
\tilde{H}_{l} )_2^{LT}$ are obtained using Eqs. (\ref{2.28d}) and 
(\ref{hikkuri}). Other SF-elements than the above ones vanish. 
\setcounter{equation}{0}
\setcounter{section}{3}
\section{Gradient parts of the propagators and the generalized 
Boltzmann equations} 
\def\theequation{\mbox{\arabic{section}.\arabic{equation}}}
Here, we deduce the gradient terms of the quark and gluon 
propagators, and derive generalized Boltzmann equations and their 
relatives. Procedure goes parallel to those in \cite{8,9,10}, 
and then we describe briefly. 
\subsection{Quark sector} 
A configuration-space counterpart of $F (P, X)$ is denoted by 
$\underline{F} (x, y)$: 
\begin{eqnarray*}
\underline{F} (x, y) & = & \int \frac{d^{\, 4} P}{(2 \pi)^4} \, 
e^{- i P \cdot (x - y)} F (P, X) \;\;\; \left( X = \frac{x + y}{2} 
\right) \, , \nonumber \\ 
& \equiv & \left( F (P, X) \right)_{\mbox{\scriptsize{IWT}}} (x, y) 
\, . 
\end{eqnarray*}
If $F (P, X)$ is independent of X, $\underline{F} (x, y) = 
\underline{F} (x - y) \equiv \left( F (P) 
\right)_{\mbox{\scriptsize{IFT}}} (x - y)$. Here \lq\lq IWT'' 
(\lq\lq IFT'') stands for an inverse Wigner (Fourier) transform. 
\subsubsection{Preliminary} 
Configuration-space counterparts of Eqs. (\ref{14-0}) and 
(\ref{P-P}) are, with obvious notation, 
\begin{eqnarray} 
\underline{A} (x, y) &=& \sum_{\rho, \, \sigma = \pm} \left[ 
\underline{\cal P}_\rho \cdot \underline{A}^{\rho \sigma} \cdot 
\underline{\cal P}_\sigma \right] (x, y) \, , 
\label{mm} \\ 
\underline{A}^{\rho \rho} &=& \underline{A}_1^{\rho \rho} + 
\frac{1}{2} \left( \underline{A}_2^{\rho \rho} \cdot (i 
\partial\kern-0.045em\raise0.3ex\llap{/}\kern0.25em\relax) 
+ (i \partial\kern-0.045em\raise0.3ex\llap{/}\kern0.25em\relax) 
\cdot \underline{A}_2^{\rho \rho} \right) \nonumber \\ 
&& + \frac{1}{2} \left( \underline{A}_3^{\rho \rho} \cdot 
\underline{N}\kern-0.153em\raise0.3ex\llap{/}\kern0.177em\relax 
+ \underline{N}\kern-0.153em\raise0.3ex\llap{/}\kern0.177em\relax 
\cdot \underline{A}_3^{\rho \rho} \right) \nonumber \\ 
&& + \frac{1}{2} \left( \underline{A}_4^{\rho \rho} \cdot 
(i \partial\kern-0.045em\raise0.3ex\llap{/}\kern0.25em\relax) \cdot 
\underline{N}\kern-0.153em\raise0.3ex\llap{/}\kern0.177em\relax 
+ (i \partial\kern-0.045em\raise0.3ex\llap{/}\kern0.25em\relax) 
\cdot 
\underline{N}\kern-0.153em\raise0.3ex\llap{/}\kern0.177em\relax 
\cdot \underline{A}_4^{\rho \rho} \right) \, , \nonumber \\ 
\underline{A}^{\rho - \rho} &=& \gamma_5 \left[ 
\underline{A}_1^{\rho - \rho} + \frac{1}{2} \left( 
\underline{A}_2^{\rho - \rho} \cdot (i 
\partial\kern-0.045em\raise0.3ex\llap{/}\kern0.25em\relax) + 
(i \partial\kern-0.045em\raise0.3ex\llap{/}\kern0.25em\relax) 
\cdot \underline{A}_2^{\rho - \rho} \right) \right. \nonumber \\ 
&& + \frac{1}{2} \left( \underline{A}_3^{\rho - \rho} \cdot 
\underline{N}\kern-0.153em\raise0.3ex\llap{/}\kern0.177em\relax + 
\underline{N}\kern-0.153em\raise0.3ex\llap{/}\kern0.177em\relax 
\cdot \underline{A}_3^{\rho - \rho} \right) \nonumber \\ 
&& \left. + \frac{1}{2} \left( \underline{A}_4^{\rho - 
\rho} \cdot (i 
\partial\kern-0.045em\raise0.3ex\llap{/}\kern0.25em\relax) \cdot 
\underline{N}\kern-0.153em\raise0.3ex\llap{/}\kern0.177em\relax 
+ (i \partial\kern-0.045em\raise0.3ex\llap{/}\kern0.25em\relax) 
\cdot 
\underline{N}\kern-0.153em\raise0.3ex\llap{/}\kern0.177em\relax 
\cdot \underline{A}_4^{\rho - \rho} \right) \right] \, . \nonumber 
\end{eqnarray} 
Here we have used the short-hand notation $\underline{F} \cdot 
\underline{G}$, which is a function whose \lq\lq $(x, 
y)$-component'' is 
\begin{equation} 
[\underline{F} \cdot \underline{G}] (x, y) = \int d^{\, 4} z \, 
\underline{F} (x, z) \underline{G} (z, y) \, . 
\label{nn}
\end{equation} 
For later use, we display the Wigner transform of $\underline{F}_G 
= \underline{F} \cdot \underline{G}$: 
\begin{equation} 
F_G (P, X) = F (P, X) G (P,X) - \frac{i}{2} \left\{ F (P, X) , \, \; 
G (P, X) \right\} \, , 
\label{wiech}
\end{equation} 
which is valid to the gradient approximation. The \lq\lq Poisson 
bracket'' in Eq. (\ref{wiech}) is defined as 
\begin{equation} 
\left\{ F , \; \, G \right\} \equiv \frac{\partial F}{\partial 
X^\mu} \frac{\partial G}{\partial P_\mu} - 
\frac{\partial F}{\partial P_\mu} \frac{\partial G}{\partial X^\mu} 
\, . 
\label{Poi} 
\end{equation} 
\subsubsection{Bare propagator and counter-Lagrangian} 
We proceed as in \cite{9}. We start from the expression for the free 
propagator $\underline{\hat{S}} (x, y)$ (cf. Eqs. (\ref{333}) - 
(\ref{335})), 
\begin{eqnarray*} 
\underline{\hat{S}} (x, y) &=& \sum_{\rho = \pm} \underline{{\cal 
P}}_\rho \cdot \underline{\hat{S}}_\rho \cdot \underline{\cal 
P}_\rho + \underline{S}_K \hat{M}_+ \, , \\   
\underline{\hat{S}}_\rho &=& \left( 
\begin{array}{cc} 
\underline{\hat{S}}_R & \;\, 0 \\ 
\underline{\hat{S}}_R - \underline{\hat{S}}_A & \;\, 
- \underline{\hat{S}}_A 
\end{array} 
\right) \nonumber \\ 
&& - \left( \underline{\hat{S}}_R \cdot \underline{f}_\rho - 
\underline{f}_\rho \cdot \underline{\hat{S}}_A \right) \hat{M}_+ \, 
, \\ 
\underline{S}_K &=& \sum_{\rho = \pm} \underline{{\cal P}}_\rho 
\cdot \left( \underline{S}_R \cdot \gamma_5 
\underline{N}\kern-0.153em\raise0.3ex\llap{/}\kern0.177em\relax 
\cdot \underline{C}_{\rho - \rho} \right. \nonumber \\ 
&& \left. - \underline{C}_{\rho - \rho} \cdot \gamma_5 
\underline{N}\kern-0.153em\raise0.3ex\llap{/}\kern0.177em\relax 
\cdot \underline{S}_A \right) \cdot \underline{\cal P}_{- 
\rho} \, . 
\end{eqnarray*} 
For the time being, $\underline{f}_\rho (x, y)$ and 
$\underline{C}_{\rho - \rho} (x, y)$ in the above equations are left 
to be arbitrary. Specification of them will be made in Subsubsec. 5. 
$\hat{S}$ in Eq. (\ref{333}) is the leading part of the derivative 
expansion (DEX) of $\hat{S} (P, X)$ $\left(
\stackrel{\mbox{\scriptsize{WT}}}{\longleftarrow} 
\underline{\hat{S}} (x, y) \right)$. 
Straightforward calculation within the gradient approximation yields 
\widetext 
\begin{eqnarray} 
\underline{\hat{S}}^{- 1} \cdot \underline{\hat{S}} &=& 
\underline{\hat{S}} \cdot \underline{\hat{S}}^{- 1} = 1 \, , 
\label{bareL} \\ 
\underline{\hat{S}}^{- 1} (x, y) & = &  \left( i 
\partial\kern-0.045em\raise0.3ex\llap{/}\kern0.25em\relax_x - m 
\right) \delta^{\, 4} (x - y) \hat{\tau}_3 - \underline{L}_c (x, y) 
\hat{M}_- \, , \nonumber \\ 
\underline{L}_c &=& \underline{L}_{c1} + \underline{L}_{c2} \, , 
\nonumber \\ 
L_{c1} &=& i \sum_{\rho = \pm} \left( 
\partial\kern-0.045em\raise0.3ex\llap{/}\kern0.25em\relax_{\!\!X} 
f_\rho \right) {\cal P}_\rho \nonumber \\ 
&=& i \sum_{\rho = \pm} {\cal P}_\rho \left[ \frac{P \cdot 
\partial_X}{P^2} P\kern-0.1em\raise0.3ex\llap{/}\kern0.15em\relax 
+ \frac{N \cdot \partial_X}{N^2} 
N\kern-0.153em\raise0.3ex\llap{/}\kern0.177em\relax + \frac{\rho 
\epsilon (p_0)}{e_\perp^2} (e_\perp \cdot \partial_X) 
P\kern-0.1em\raise0.3ex\llap{/}\kern0.15em\relax 
N\kern-0.153em\raise0.3ex\llap{/}\kern0.177em\relax \right] f_\rho 
{\cal P}_\rho \, , 
\label{key} \\ 
L_{c2} & = & i \sum_{\rho = \pm} {\cal P}_\rho \gamma_5 \left[ 
\frac{\partial 
N\kern-0.153em\raise0.3ex\llap{/}\kern0.177em\relax}{\partial 
P_\alpha} \frac{\partial C_{\rho - \rho}}{\partial X^\alpha} 
(P\kern-0.1em\raise0.3ex\llap{/}\kern0.15em\relax - m) + \left( 
\partial\kern-0.045em\raise0.3ex\llap{/}\kern0.25em\relax_{\!\!X} 
C_{\rho - \rho} \right) 
N\kern-0.153em\raise0.3ex\llap{/}\kern0.177em\relax \right] {\cal 
P}_{- \rho} \nonumber \\ 
&=& i \sum_{\rho = \pm} {\cal P}_\rho \gamma_5 \left[ \left\{ - 
\frac{\rho \epsilon (p_0)}{N^2} 
N\kern-0.153em\raise0.3ex\llap{/}\kern0.177em\relax e_\perp^\mu 
\frac{\partial N_\mu}{\partial P_\alpha} - \frac{
P\kern-0.1em\raise0.3ex\llap{/}\kern0.15em\relax 
N\kern-0.153em\raise0.3ex\llap{/}\kern0.177em\relax}{2 N^2} 
\frac{\partial N^2}{\partial P_\alpha} \right. \right. \nonumber \\ 
&& \left. - m \left( - \frac{
P\kern-0.1em\raise0.3ex\llap{/}\kern0.15em\relax}{P^2} N^\alpha + 
\frac{N\kern-0.153em\raise0.3ex\llap{/}\kern0.177em\relax}{2 N^2} 
\frac{\partial N^2}{\partial P_\alpha} - \frac{\rho \epsilon 
(p_0)}{e_\perp^2} P\kern-0.1em\raise0.3ex\llap{/}\kern0.15em\relax 
N\kern-0.153em\raise0.3ex\llap{/}\kern0.177em\relax e_\perp^\mu 
\frac{\partial N^\mu}{\partial P_\alpha} \right) \right\} 
\frac{\partial C_{\rho - \rho}}{\partial X^\alpha} \nonumber \\ 
&& \left. + \left\{ \frac{\rho \epsilon (p_0)}{P^2} 
P\kern-0.1em\raise0.3ex\llap{/}\kern0.15em\relax (e_\perp \cdot 
\partial_X) + \frac{P\kern-0.1em\raise0.3ex\llap{/}\kern0.15em\relax 
N\kern-0.153em\raise0.3ex\llap{/}\kern0.177em\relax}{P^2} (P \cdot 
\partial_X) \right\} C_{\rho - \rho} \right] {\cal P}_{- \rho} \, .  
\label{con12} 
\end{eqnarray}
\narrowtext 
\noindent 
Here $1 / P^2 \equiv {\bf P}/ P^2$, $1 / N^2 \equiv {\bf P}/ N^2$, 
and $1 / e_\perp^2 = {\bf P} / e_\perp^2$, with ${\bf P}$ denoting 
to take the principal part. 

Eq. (\ref{bareL}) tells us that the free action of the theory 
\cite{6,9} is 
\begin{eqnarray}
{\cal A}_0 & = & \int d^{\, 4} x \, d^{\, 4} y \, \hat{\bar{\psi}} 
(x) \underline{\hat{S}}^{- 1} (x, y) \hat{\psi} (y) \, , 
\label{4.7d} \\ 
&& \hat{\bar{\psi}} = (\bar{\psi}_1 , \, \bar{\psi}_2) \, , 
\;\;\;\;\;\;\;\;\;\; \hat{\psi} = \left( 
\begin{array}{cc} 
\psi_1 \\ 
\psi_2 
\end{array} 
\right) \, . 
\nonumber
\end{eqnarray}
Since the term with $\underline{L}_c (x, y)$ $\left( \in 
\underline{\hat{S}}^{- 1} \right)$ in ${\cal A}_0$ is absent in the 
original action, we should introduce a counteraction to compensate 
it: 
\begin{equation}
{\cal A}_c = \int d^{\, 4} x \, d^{\, 4} y \, \hat{\bar{\psi}} (x) 
\underline{L}_c (x, y) \hat{M}_- \hat{\psi} (y) \, , 
\label{countl1} 
\end{equation}
which yields a (two-point) vertex factor 
\begin{equation}
i \underline{L}_c (x, y) \hat{M}_- 
\stackrel{\mbox{\scriptsize{WT}}}{\longrightarrow} i \left( L_{c1} 
(P, X) + L_{c2} (P, X) \right) \hat{M}_- \, . 
\label{tuy} 
\end{equation}
Here \lq\lq WT'' indicates to take Wigner transformation. 
\subsubsection{Dyson equation}
Let us start with considering a \lq\lq product'' of $\underline{A}$ 
and $\underline{B}$ of the type (\ref{mm}) (cf. Eq. (\ref{nn})), 
\begin{eqnarray} 
\underline{C} (x, y) &=& \left[ \underline{A} \cdot \underline{B} 
\right] (x, y) \nonumber \\ 
&=& \left[ \sum_{\rho , \, \xi, \, \sigma = \pm} \underline{\cal 
P}_\rho \cdot \underline{A}^{\rho \xi} \cdot \underline{\cal P}_\xi 
\cdot \underline{B}^{\xi \sigma} \cdot \underline{\cal P}_\sigma 
\right] (x, y) \, . \nonumber \\ 
&& 
\label{mama} 
\end{eqnarray} 
Using Eq. (\ref{wiech}), we obtain, for the Wigner transform of 
$\underline{C}$ to the gradient approximation, 
\begin{eqnarray} 
C (P, X) &=& \sum_{\rho , \, \xi, \, \sigma = \pm} \left[ {\cal 
P}_\rho A^{\rho \xi} (P, X) B^{\xi \sigma} (P, X) {\cal P}_\sigma 
\right. \nonumber \\ 
&& \left. + \frac{i}{2} {\cal P}_\rho \left\{ A^{\rho \xi} 
\frac{\partial {\cal P}_\xi}{\partial P_\mu} \frac{\partial B^{\xi 
\sigma}}{\partial X^\mu} \right. \right. \nonumber \\ 
&& \left. \left. - \frac{\partial A^{\rho \xi}}{\partial X^\mu} 
\frac{\partial {\cal P}_\xi}{\partial P_\mu} B^{\xi \sigma} \right\} 
{\cal P}_\sigma + ... \right] \, , 
\end{eqnarray} 
where \lq\lq $\, ... \, $'' stands for other pieces of the gradient 
terms than the second term. Thanks to the relation, 
\[
{\cal P}_{\pm \rho} \frac{\partial {\cal P}_\rho}{\partial P_\mu} 
= \frac{\partial {\cal P}_\rho}{\partial P_\mu} {\cal P}_{\mp 
\rho} \, , 
\]
the second term vanishes, ${\cal P}_\rho \left\{ ... \right\} {\cal 
P}_\sigma = 0$. Then, to the gradient approximation, $\underline{C} 
(x, y)$ in Eq. (\ref{mama}) may be 
written as 
\[ 
\underline{C} (x, y) \simeq \left[ \sum_{\rho , \, \xi, \, \sigma = 
\pm} \underline{\cal P}_\rho \cdot \underline{A}^{\rho \xi} \cdot 
\underline{B}^{\xi \sigma} \cdot \underline{\cal P}_\sigma \right] 
(x, y) \, . 
\] 
Thus, as in Eq. (\ref{SD11}), we can use the $(2 \times 2)$ matrix 
notation in a \lq\lq polarization space'', $\left( \underline{\bf A} 
\right)^{\rho \sigma} = \underline{A}^{\rho \sigma}$ $(\rho, \, 
\sigma = \pm)$. 

The self-energy-part $\left( \underline{\hat{ \bf \Sigma}} (x, y) 
\right)$ resummed propagator $\underline{\hat{\bf G}}$ obeys 
\begin{eqnarray} 
\underline{\hat{\bf G}} &=& \underline{\hat{\bf S}} + 
\underline{\hat{\bf S}} \cdot \underline{\hat{\bf \Sigma}} \cdot 
\underline{\hat{\bf G}} = \underline{\hat{\bf S}} + 
\underline{\hat{\bf G}} \cdot \underline{\hat{\bf \Sigma}} \cdot 
\underline{\hat{\bf S}} \, , 
\label{SDiti} \\ 
\underline{\hat{\bf S}} &=& \left( 
\begin{array}{cc} 
\underline{S}_R & \;\, 0 \\ 
\underline{S}_R - \underline{S}_A & \; \, - \underline{S}_A 
\end{array} 
\right) {\bf 1} \nonumber \\ 
&& - \left( \underline{S}_R \cdot \underline{\bf f} - 
\underline{\bf f} \cdot \underline{S}_A - \underline{\bf S}_K 
\right) \hat{M}_+ \, , \nonumber \\ 
\underline{\bf S}_K &=& \underline{S}_R \cdot \gamma_5 
\underline{N}\kern-0.153em\raise0.3ex\llap{/}\kern0.177em\relax 
\cdot \underline{\bf C} - \underline{\bf C} \cdot \gamma_5 
\underline{N}\kern-0.153em\raise0.3ex\llap{/}\kern0.177em\relax 
\cdot \underline{S}_A \, , \nonumber \\ 
\underline{\bf C} &=& \left( 
\begin{array}{cc} 
0 & \;\, \underline{C}_{+-} \\ 
\underline{C}_{- +} & \; \, 0 
\end{array} 
\right) \, . \nonumber 
\end{eqnarray} 
For $\underline{\hat{\bf G}}$ and $\underline{\hat{\bf \Sigma}}$, we 
have (cf. Eqs. (\ref{gii}) - (\ref{gii3}), (\ref{siguma}) - 
(\ref{sigma3})) 
\begin{eqnarray} 
\underline{\hat{\bf G}} &=& \left( 
\begin{array}{cc} 
\underline{\bf  G}_R & \;\; 0 \\ 
\underline{\bf G}_R - \underline{\bf G}_A & \;\; - 
\underline{\bf G}_A 
\end{array} 
\right) \nonumber \\ 
&& - \left[ \underline{\bf G}_R \cdot \underline{\bf f} - 
\underline{\bf f} \cdot \underline{\bf G}_A - \underline{\bf G}_K 
\right] \hat{M}_+ \, , 
\label{ippan} \\ 
\underline{\hat{\bf \Sigma}} &=& \left( 
\begin{array}{cc} 
\underline{\bf \Sigma}_R & \;\; 0 \\ 
- \underline{\bf \Sigma}_R + \underline{\bf \Sigma}_A & \;\; - 
\underline{\bf \Sigma}_A 
\end{array} 
\right) \nonumber \\ 
&& - \left[ \underline{\bf \Sigma}_R \cdot \underline{\bf f} - 
\underline{\bf f} \cdot \underline{\bf \Sigma}_A + 
\underline{\bf \Sigma}_K \right] \hat{M}_- \, , \nonumber \\ 
\underline{\bf G}_K &=& \underline{\bf G}_R \cdot \underline{\bf f} 
- \underline{\bf f} \cdot \underline{\bf G}_A + \underline{\bf 
G}_{12} \, , \nonumber \\ 
\underline{\bf \Sigma}_K &=& - \underline{\bf \Sigma}_R \cdot 
\underline{\bf f} + \underline{\bf f} \cdot \underline{\bf 
\Sigma}_A + \underline{\hat{\bf \Sigma}}_{12} \, . 
\label{sigmK} 
\end{eqnarray}
Eq. (\ref{SDiti}) may be solved to give 
\begin{eqnarray} 
\underline{\bf G}_{R (A)} (x, y) & = & \left[ {\bf 1} (i 
\partial\kern-0.045em\raise0.3ex\llap{/}\kern0.25em\relax_x - m)^{- 
1} \delta^{\, 4} (x - y) \right. \nonumber \\ 
&& \left. - \underline{\bf \Sigma}_{R (A)} (x, y) \right]^{- 1} \, , 
\label{Ryone} \\ 
\underline{\bf G}_K &=& \underline{\bf G}_K^{[1]} + \underline{\bf 
G}_K^{[2]} + \underline{\bf G}_K^{[3]} \, , \\ 
\underline{\bf G}_K^{[1]} &=& - \underline{\bf G}_R \cdot 
\underline{\bf \Sigma}_K \cdot \underline{\bf G}_A \, , 
\label{GK1yo} \\ 
\underline{\bf G}_K^{[2]} &=& \underline{\bf G}_R \cdot \left[ 
\gamma_5 
\underline{N}\kern-0.153em\raise0.3ex\llap{/}\kern0.177em\relax 
\cdot \underline{\bf C} \cdot \underline{\bf \Sigma}_A \right. 
\nonumber \\ 
&& \left. - \underline{\bf \Sigma}_R \cdot \underline{\bf C} \cdot 
\gamma_5 
\underline{N}\kern-0.153em\raise0.3ex\llap{/}\kern0.177em\relax 
\right] \cdot \underline{\bf G}_A \nonumber \\ 
& \equiv & \underline{\bf G}_R \cdot \underline{\bf H} \cdot 
\underline{\bf G}_A \, , 
\label{GK2yo} \\ 
\underline{\bf G}_K^{[3]} &=& \underline{\bf G}_R \cdot \gamma_5 
\underline{N}\kern-0.153em\raise0.3ex\llap{/}\kern0.177em\relax 
\cdot \underline{\bf C} - \underline{\bf C} \cdot \gamma_5 
\underline{N}\kern-0.153em\raise0.3ex\llap{/}\kern0.177em\relax 
\cdot \underline{\bf G}_A \, . 
\label{GK3yo} 
\end{eqnarray} 
The form for the leading part of the DEX of $\hat{G} (X, P)$ $\left( 
\stackrel{\mbox{\scriptsize{WT}}}{\longleftarrow} \underline{G} (x, 
y) \right)$ is the $\hat{G}$ that is deduced in Sec. II. 
\subsubsection{Gradient piece of the self-energy-part resummed 
propagator} 
\subsubsection*{Form for $G_{R (A)}$} 
From Eq. (\ref{Ryone}), we obtain, for the component 
$\underline{G}_R^{\rho \sigma}$, 
\begin{eqnarray} 
\underline{G}_R^{\rho \rho} (x, y) &=& \left[ \left( i 
\partial\kern-0.045em\raise0.3ex\llap{/}\kern0.25em\relax - m 
\right) \delta^{\, 4} (x - y) - \underline{\Sigma}_R^{\rho \rho} 
\right. \nonumber \\ 
&& \left. - \underline{\Sigma}_R^{\rho - \rho} \cdot 
\underline{G}_R^{\mbox{\scriptsize{(pre)}} - \rho - \rho} \cdot 
\underline{\Sigma}_R^{- \rho \rho} \right]^{- 1} \, , 
\label{Ra} \\ 
\underline{G}_R^{\rho - \rho} (x, y) &=& 
\underline{G}_R^{\mbox{\scriptsize{(pre)}} \rho \rho} \cdot 
\underline{\Sigma}_R^{\rho - \rho} \cdot \underline{G}_R^{- \rho - 
\rho} \nonumber \\ 
& = & \underline{G}_R^{\rho \rho} \cdot \underline{\Sigma}_R^{\rho 
- \rho} \cdot \underline{G}_R^{\mbox{\scriptsize{(pre)}} - \rho - 
\rho} \, , 
\label{Rb} \\ 
\underline{G}_R^{\mbox{\scriptsize{(pre)}} \rho \rho} (x, y) &=& 
\left[ \left( i 
\partial\kern-0.045em\raise0.3ex\llap{/}\kern0.25em\relax - m 
\right) \delta^{\, 4} (x - y) - \underline{\Sigma}_R^{\rho \rho} (x, 
y) \right]^{- 1} \, . \nonumber \\ 
&& 
\label{Rc} 
\end{eqnarray}
Solving Eqs. (\ref{Ra}) and (\ref{Rb}), we write the solutions as 
$G_R^{\rho \sigma} = G_R^{(0) \rho \sigma} + G_R^{(1) \rho 
\sigma}$. Here, $G_R^{(0) \rho \sigma}$ is the leading part of the 
DEX of $G_R^{\rho \sigma}$, whose form has been obtained in Sec. II. 
The gradient part $G_R^{(1) \rho \sigma}$ is obtained as 
\widetext 
\begin{eqnarray} 
G_R^{(1) \rho \rho} &=& \frac{i}{2} G_R^{(0) \rho \rho} \left[ 
\left\{ \left( G_R^{(0) \rho \rho} \right)^{- 1} , \;\, G_R^{(0) 
\rho \rho} \right\} - \left\{\Sigma_R^{\rho - \rho} | 
G_R^{\mbox{\scriptsize{(pre)}}^{- \rho - \rho}} | \Sigma_R^{- \rho 
\rho} \right\} G_R^{(0) \rho \rho} \right. \nonumber \\ 
&& \left. - \Sigma_R^{\rho - \rho} \left\{ 
G_R^{\mbox{\scriptsize{(pre)}} - \rho - \rho} , \;\, \Sigma_R^{- 
\rho \rho} \right\} G_R^{(0) \rho \rho} -  \left\{ \Sigma_R^{\rho - 
\rho} , \;\, G_R^{\mbox{\scriptsize{(pre)}}^{- \rho - \rho}} 
\right\} \Sigma_R^{- \rho \rho} G_R^{(0) \rho \rho} \right] 
\nonumber \\ 
&=& \frac{i}{2} \left[ \left\{ G_R^{(0) \rho \rho} , \;\, \left( 
G_R^{(0) \rho \rho} \right)^{- 1} \right\} - G_R^{(0) \rho \rho} 
\left\{\Sigma_R^{\rho - \rho} | G_R^{\mbox{\scriptsize{(pre)}}^{- 
\rho - \rho}} | \Sigma_R^{- \rho \rho} \right\} \right. \nonumber \\ 
&& \left. - G_R^{(0) \rho \rho} \left\{ \Sigma_R^{\rho - \rho} , 
\;\, G_R^{\mbox{\scriptsize{(pre)}}^{- \rho - \rho}} \right\} 
\Sigma_R^{- \rho \rho} -  G_R^{(0) \rho \rho} \Sigma_R^{\rho - \rho} 
\left\{ G_R^{\mbox{\scriptsize{(pre)}}^{- \rho - \rho}} , \;\, 
\Sigma_R^{- \rho \rho} \right\} \right] G_R^{(0) \rho \rho} \, , 
\label{iika} \\ 
G_R^{(1) \rho - \rho} &=& - \frac{i}{2} \left[ \left\{ 
G_R^{\mbox{\scriptsize{(pre)}} \rho \rho} , \;\, \Sigma_R^{\rho - 
\rho} \right\} G_R^{(0)- \rho - \rho} + \left\{ 
G_R^{\mbox{\scriptsize{(pre)}} \rho \rho} | \Sigma_R^{\rho - \rho} | 
G_R^{(0) - \rho - \rho} \right\} \right. \nonumber \\ 
&& \left. + G_R^{\mbox{\scriptsize{(pre)}} \rho \rho} \left\{ 
\Sigma_R^{\rho - \rho} , \;\, G_R^{(0) - \rho - \rho} \right\} 
\right] \nonumber \\ 
&=& - \frac{i}{2} \left[ \left\{ G_R^{(0) \rho \rho} , \;\, 
\Sigma_R^{\rho - \rho} \right\} G_R^{\mbox{\scriptsize{(pre)}} - 
\rho - \rho} + \left\{ G_R^{(0) \rho \rho} | \Sigma_R^{\rho - \rho} 
| G_R^{\mbox{\scriptsize{(pre)}} - \rho - \rho} \right\} \right. 
\nonumber \\ 
&& \left. + G_R^{(0) \rho \rho} \left\{ \Sigma_R^{\rho - \rho} , 
\;\, G_R^{\mbox{\scriptsize{(pre)}} - \rho - \rho} \right\} \right] 
\, , \nonumber 
\end{eqnarray}
where 
\begin{equation}
\left\{ A | B |C \right\} \equiv \frac{\partial A}{\partial 
X^\mu} B \frac{\partial C}{\partial P_\mu} - \frac{\partial 
A}{\partial P_\mu} B \frac{\partial C}{\partial X^\mu} \, . 
\label{nota} 
\end{equation}
Then, $\underline{G}_R (x, y)$ is written as (cf. Eq. (\ref{mm})) 
\[ 
\underline{G}_R (x, y) = \sum_{\rho, \, \sigma = \pm} \left[ 
\underline{\cal P}_\rho \cdot \left( G_R^{\rho \sigma} (P, X) 
\right)_{\mbox{\scriptsize{IWT}}} \cdot \underline{\cal P}_\sigma 
\right] (x, y) \, , 
\] 

We write the solution to Eq. (\ref{Rc}) as 
$G_R^{\mbox{\scriptsize{(pre)}} \rho \rho} = 
G_R^{\mbox{\scriptsize{(pre) (0)}} \rho \rho} + 
G_R^{\mbox{\scriptsize{(pre) (1)}} \rho \rho}$. The form for 
$G_R^{\mbox{\scriptsize{(pre) (0)}} \rho \rho}$ is given in Sec. II, 
while the form for $G_R^{\mbox{\scriptsize{(pre) (1)}} \rho \rho}$ 
is given by Eq. (\ref{iika}) with the following replacements, 
\[ 
G_R^{(0) \rho \sigma} \to G_R^{\mbox{\scriptsize{(pre)}}(0) \rho 
\sigma} \, , \;\;\; G_R^{(1) \rho \sigma} \to 
G_R^{\mbox{\scriptsize{(pre)}}(1) \rho \sigma} \, , \;\;\; 
G_R^{\mbox{\scriptsize{(pre)}} \mp \rho \mp \rho} \to 0 \, . 
\] 
$G_A (P, X)$ is obtained from $G_R (P, X)$ with $\Sigma_A$'s for 
$\Sigma_R$'s. 
\subsubsection*{Form for $\Sigma_K$, which is involved in 
$G_K^{[1]}$ in Eq. (\ref{GK1yo})} Computation of Eq. (\ref{sigmK}) 
to the gradient approximation yields 
\begin{eqnarray}
\underline{\Sigma}_K &=& \underline{L}_{c1} + \underline{L}_{c2} - 
\sum_{\rho, \, \sigma = \pm} \underline{\cal P}_\rho \cdot \left[ 
\left( \Sigma_R^{\rho \sigma} f_\sigma 
\right)_{\mbox{\scriptsize{IWT}}} - \left( f_\rho \Sigma_A^{\rho 
\sigma} \right)_{\mbox{\scriptsize{IWT}}} - 
\underline{\Sigma}_{12}^{\rho \sigma} \right] \cdot \underline{\cal 
P}_\sigma + \Sigma_K^{[1]} \, , 
\label{ngan} \\ 
\Sigma_K^{[1]} & = & - \frac{i}{2} \sum_{\rho, \, \sigma = \pm} 
{\cal P}_\rho \left[ \left\{ f_\sigma, \;\, \Sigma_R^{\rho 
\sigma} \right\} + \left\{ f_\rho, \;\, \Sigma_A^{\rho \sigma} 
\right\} \right] {\cal P}_\sigma \, . 
\label{SiK1} 
\end{eqnarray}
$\underline{L}_{c1}$ and $\underline{L}_{c2}$ in Eq. (\ref{ngan}) 
come from ${\cal A}_c$ in Eq. (\ref{countl1}) (see Eq. (\ref{tuy})). 
The standard forms for $\Sigma_K^{[1]}$, $H$ (Eq. (\ref{GK2yo})), 
and $\gamma_5 
\underline{N}\kern-0.153em\raise0.3ex\llap{/}\kern0.177em\relax 
\cdot \underline{\bf C}$ and $\underline{\bf C} \cdot \gamma_5 
\underline{N}\kern-0.153em\raise0.3ex\llap{/}\kern0.177em\relax$ in 
Eq.(\ref{GK3yo}) are given in Appendix E. 
\subsubsection{Perturbation theories --- Generalized Boltzmann 
equations and their relatives}
The aim of this subsubsection is to construct perturbation theories. 
We are employing the interaction picture in the sense of \cite{au}. 
Then, the quark-gluon system of our concern is characterized by a 
density matrix $\rho$ at an initial time $X^0 = X^0_i$, from which 
$f_\rho (P, X^0_i, \vec{X})$ and $C_{\rho - \rho} (P, X^0_i, 
\vec{X})$ (cf. Eqs. (\ref{333}) - (\ref{335})) are determined. It 
should be emphasized that there is no information at this stage on 
how $f_\rho (P, X)$ and $C_{\rho - \rho} (P, X)$ evolve in 
spacetime. Then, in the course of construction of a perturbative 
framework, certain evolution equations that describe the spacetime 
evolution for $f_\rho$ and $C_{\rho - \rho}$ should be settled. As a 
matter of fact, one can choose any forms for the evolution 
equations, on the basis of which a perturbative framework is 
constructed. Different frameworks are physically equivalent in the 
sense that they lead to the same result for the physical quantities 
(see below for more details). In the sequel, we construct two kind 
of perturbative frameworks by employing two different forms for the 
evolution equations. 

As seen from Eq. (\ref{ippan}), the propagator $\hat{G}$ is written 
in terms of $G_R$, $G_A$, and $G_K$. $G_R (P, X)$ [$G_A (P, X)$] is 
analytic in an upper [a lower] half complex $p_0$-plane. Then, in 
calculating some quantity, the parts of $\hat{G}$ that are 
proportional to $G_R$ or to $G_A$ yield well-defined contributions. 
Now, we observe that $G_K^{[1]}$ and $G_K^{[2]}$, Eqs. (\ref{GK1yo}) 
and (\ref{GK2yo}), contain $G_R G_A$. Since $G_A$ is essentially 
the complex conjugate of $G_R$, $G_R G_A$ is disastrously large 
on the energy shells\footnote{How to find the solution to Eq. 
(\ref{eshell}) is given in Appendix F.}, $p_0 = \pm \omega_\pm (\pm 
\vec{p}, X)$, on which 
\begin{equation}
Re \left( G_R^{\rho \rho} (P, X) \right)^{- 1} \, 
\rule[-3mm]{.14mm}{8.5mm} \raisebox{-2.85mm}{\scriptsize{$\; p_0 = 
\pm \omega_\pm (\pm \vec{p}, X)$}} = 0 \, . 
\label{eshell}
\end{equation}
As a matter of fact, in the narrow-width approximation, $Im \left( 
G_R^{\rho \rho} \right)^{- 1} \to \epsilon (p_0) 0^+$, $G_R^{\rho 
\rho} G_A^{\rho \rho}$ develops pinch singularities at the energy 
shells in a $p_0$-plane.\footnote{This is a characteristic feature 
of nonequilibrium dynamics\cite{alth}.} Then, $G_K^{[1]}$ and 
$G_K^{[2]}$ yield diverging contribution. In practice, $Im \left( 
G_R^{\rho \rho} \right)^{- 1}$ $(\propto g^2)$ is a small quantity, 
so that the contributions, although not divergent, are large, which 
invalidates the perturbative scheme. These large contributions come 
from the vicinities of the energy shells, on which $Re \left( 
G_R^{\rho \rho} \right)^{- 1} \sim 0$. 

Appropriate use of the first and second equalities of Eq. 
(\ref{final1}) together with Eq. (\ref{hikkuri3}) shows that 
$G_R^{\rho \rho} G_A^{\pm \rho \mp \rho}$ and $G_R^{\rho - \rho} 
G_A^{\pm \rho \mp \rho}$ do not yield large contributions. This is 
because, in general, the energy shells of $G_R^{\rho \rho}$ and 
$G_R^{- \rho - \rho}$, and of $G_R^{\mbox{\scriptsize{(pre)}} \rho 
\rho}$ and $G_R^{\mbox{\scriptsize{(pre)}} - \rho - \rho}$, do not 
coincide. For the case of $f_+ = f_-$, however, this is not the 
case. For $G_R^{\rho \rho} G_A^{\pm \rho \mp \rho}$, a large 
contribution emerges from the region $Re \left( G_R^{\rho \rho} 
\right)^{- 1} \sim 0$, and, for $G_R^{\rho - \rho} G_A^{\pm \rho \mp 
\rho}$, large contributions emerge from the regions $Re \left( 
G_R^{\rho \rho} \right)^{- 1} \sim 0$ and from the region $Re \left( 
G_R^{\mbox{\scriptsize{(pre)}} \rho \rho} \right)^{- 1} \sim 0$. 

From Eqs. (\ref{GK1yo}) and (\ref{GK2yo}) with Eq. (\ref{ngan}), we 
have for the $(\rho \rho)$- and $(\rho - \rho)$-components $(\rho = 
\pm)$ of $\underline{H} - \underline{\Sigma}_K$ $\left( = 
\underline{G}_R^{- 1} \cdot (\underline{G}_K^{[1]} + 
\underline{G}_K^{[2]}) \cdot \underline{G}_A^{- 1} \right)$, 
\begin{eqnarray}
\underline{H}^{\rho \rho} - \underline{\Sigma}_K^{\rho \rho} & 
\stackrel{\mbox{\scriptsize{WT}}}{\longrightarrow} & - i \left[ 
\frac{P \cdot \partial_X}{P^2} 
P\kern-0.1em\raise0.3ex\llap{/}\kern0.15em\relax + \frac{N \cdot 
\partial_X}{N^2} N\kern-0.153em\raise0.3ex\llap{/}\kern0.177em\relax 
+ \frac{\rho \epsilon (p_0)}{e_\perp^2} (e_\perp \cdot \partial_X) 
P\kern-0.1em\raise0.3ex\llap{/}\kern0.15em\relax 
N\kern-0.153em\raise0.3ex\llap{/}\kern0.177em\relax \right] f_\rho 
+ i \tilde{\Gamma}_p^{\rho \rho} - \Sigma_B^{\rho \rho} \, , 
\label{kaku} \\ 
\underline{H}^{\rho - \rho} - \underline{\Sigma}_K^{\rho - \rho} & 
\stackrel{\mbox{\scriptsize{WT}}}{\longrightarrow} & - i \gamma_5 
\left[ \left\{ \frac{m}{P^2} (N \cdot \partial_X) + \frac{\rho 
\epsilon (p_0)}{P^2} (e_\perp \cdot \partial_X) \right\} 
P\kern-0.1em\raise0.3ex\llap{/}\kern0.15em\relax \right. \nonumber 
\\ 
&& - \left\{ \frac{\rho \epsilon (p_0)}{N^2} e_\perp^\mu 
\frac{\partial N_\mu}{\partial P_\alpha} + \frac{m}{2 N^2} 
\frac{\partial N^2}{\partial P_\alpha} \right\} 
N\kern-0.153em\raise0.3ex\llap{/}\kern0.177em\relax \nonumber 
\frac{\partial}{\partial X^\alpha} \\ 
&& \left. - \left\{ \frac{1}{2 N^2} \frac{\partial N^2}{\partial 
P_\alpha} \frac{\partial}{\partial X^\alpha} - m \frac{\rho \epsilon 
(p_0)}{e_\perp^2} e_\perp^\mu \frac{\partial N^\mu}{\partial 
P_\alpha} \frac{\partial}{\partial X^\alpha} - \frac{P \cdot 
\partial_X}{P^2} \right\} 
P\kern-0.1em\raise0.3ex\llap{/}\kern0.15em\relax 
N\kern-0.153em\raise0.3ex\llap{/}\kern0.177em\relax \right] C_{\rho 
- \rho} \nonumber \\ 
&& + i \tilde{\Gamma}_p^{\rho - \rho} - \Sigma_B^{\rho - \rho} \, , 
\label{kaku1} \\ 
\tilde{\Gamma}_p^{\rho \sigma} & = & i \left[ (1 - f_\sigma) 
\Sigma_{12}^{\rho \sigma} + f_\rho \Sigma_{21}^{\rho \sigma} 
+ (f_\rho - f_\sigma) \Sigma_{11}^{\rho \sigma} \right] \, , 
\label{2.26cd} \\ 
\Sigma_B^{\rho \sigma} & = & \Sigma_K^{[1] \rho \sigma} - \left[ 
\left( \gamma_5 
\underline{N}\kern-0.153em\raise0.3ex\llap{/}\kern0.177em\relax 
\cdot \underline{C} \cdot \underline{\bf \Sigma}_A - \underline{\bf 
\Sigma}_R \cdot \underline{C} \cdot \gamma_5 
\underline{N}\kern-0.153em\raise0.3ex\llap{/}\kern0.177em\relax 
\right)^{\rho \sigma} \right]_{\mbox{\scriptsize{WT}}} \, . 
\label{tarou} 
\end{eqnarray}
The first term on the R.H.S. of Eq. (\ref{kaku}) (Eq. (\ref{kaku1})) 
comes from the counter Lagrangian $\underline{L}_{c1}$ 
($\underline{L}_{c2}$), Eq. (\ref{key}) (Eq. (\ref{con12})), in Eq. 
(\ref{ngan}). 

For later reference, we note that the physical number densities, 
$N_\pm^{(\mbox{\scriptsize{ph}})} (P, X)$ and 
$\bar{N}_\pm^{(\mbox{\scriptsize{ph}})} (P, X)$, are obtained 
through computing current density, 
\begin{eqnarray}
\langle j^\mu (x) \rangle & \equiv & \mbox{Tr} \left[ 
\bar{\psi} (x) \gamma^\mu \psi(x) \rho \right] \nonumber \\ 
& = & - \frac{i}{2} \mbox{Tr} \left\{ \gamma^\mu \left[ 
\underline{G}_{21} (x, x) + \underline{G}_{12} (x, x) \right] \rho 
\right\} \, . 
\label{ref1} 
\end{eqnarray}
Similarly, the physical $C_{\pm \mp} (P, X)$, $C_{\pm 
\mp}^{({\mbox{\scriptsize{ph}}})} (P, X)$, is obtained from 
\begin{eqnarray}
\langle j^\mu_5 (x) \rangle & \equiv & \mbox{Tr} \left[ 
\bar{\psi} (x) \gamma_5 \gamma^\mu \psi(x) \rho \right] \nonumber \\ 
& = & - \frac{i}{2} \mbox{Tr} \left\{ \gamma_5 \gamma^\mu \left[ 
\underline{G}_{21} (x, x) + \underline{G}_{12} (x, x) \right] \rho 
\right\} \, . 
\label{ref2} 
\end{eqnarray}
\subsubsection*{Bare-$N$ scheme}
As has been emphasized at the beginning of this subsubsection, 
$f_\rho (P, X)$ in Eq. (\ref{kaku}) and $C_{\rho - \rho} (P, X)$ 
in Eq. (\ref{kaku1}) $(X^0_i < X^0)$ have not been defined so far. 
For the purpose of determining them, we impose here the condition 
that the counter Lagrangian $L_{c}$ is absent, $L_{c} = 0$: 
\begin{eqnarray} 
&& P \cdot \partial_X f_\rho^{(\mbox{\scriptsize{B}})} = N \cdot 
\partial_X f_\rho^{(\mbox{\scriptsize{B}})} = e_\perp \cdot 
\partial_X f_\rho^{(\mbox{\scriptsize{B}})} = 0 \, , 
\label{fB} \\ 
&& \left[ m (N \cdot \partial_X) + \rho \epsilon (p_0) (e_\perp 
\cdot \partial_X) \right] C_{\rho - \rho}^{(\mbox{\scriptsize{B}})} 
= \left[ \rho \epsilon (p_0) e_\perp^\mu \frac{\partial 
N_\mu}{\partial P_\alpha} + \frac{m}{2} \frac{\partial N^2}{\partial 
P_\alpha} \right] \frac{\partial C_{\rho - 
\rho}^{(\mbox{\scriptsize{B}})}}{\partial X^\alpha} \nonumber \\ 
&& \mbox{\hspace*{10ex}} = \left[ \frac{P^2}{2} \frac{\partial 
N^2}{\partial P_\alpha} \frac{\partial}{\partial X^\alpha} + m \rho 
\epsilon (p_0) e_\perp^\mu \frac{\partial N^\mu}{\partial 
P_\alpha} \frac{\partial}{\partial X^\alpha} - N^2 (P \cdot 
\partial_X) \right] C_{\rho - \rho}^{(\mbox{\scriptsize{B}})} = 0 
\, , 
\label{fB1}
\end{eqnarray} 
\narrowtext 
\noindent 
where we have written $f_\rho^{(\mbox{\scriptsize{B}})}$ $(C_{\rho 
- \rho}^{(\mbox{\scriptsize{B}})})$ for $f_\rho$ $(C_{\rho - 
\rho})$. Then, Eq. (\ref{kaku}) (Eq. (\ref{kaku1})), of which the 
first term on the RHS is absent, is to be solved under the given 
initial data $f_\rho (P, X_i^0, \vec{X})$ ($C_{\rho - \rho} (P, 
X_i^0, \vec{X}))$. Eq. (\ref{fB}) is a \lq\lq free Boltzmann 
equation'' and its relatives for the \lq\lq bare'' number densities, 
$N_\rho^{(\mbox{\scriptsize{B}})} (p_0, \vec{p}, X) = \theta (p_0) 
f_\rho^{(\mbox{\scriptsize{B}})} (P, X)$ and 
$\bar{N}_\rho^{(\mbox{\scriptsize{B}})} (|p_0|, \vec{p}, X) = 1 - 
\theta (- p_0) f_\rho^{(\mbox{\scriptsize{B}})} (p_0, - \vec{p}, 
X)$, (cf. Eq. (\ref{bareN})). 

The physical number densities, which are obtained from $\langle 
j^\mu (x) \rangle$ (Eq. (\ref{ref1})), and the physical $C_{\rho - 
\rho}^{(\mbox{\scriptsize{ph}})}$, which is obtained from $\langle 
j^\mu_5 (x) \rangle$ (Eq. (\ref{ref2})), are functionals of 
$f_\sigma^{(\mbox{\scriptsize{B}})}$ and $C_{\sigma - 
\sigma}^{(\mbox{\scriptsize{B}})}$: 
\begin{eqnarray*}
f_\rho^{(\mbox{\scriptsize{ph}})} (P, X) & = & \theta (p_0) 
N_\rho^{(\mbox{\scriptsize{ph})}} (p_0, \vec{p}) \\ 
&& + \theta (- p_0) \left[ 1 - 
\bar{N}_\rho^{(\mbox{\scriptsize{ph}})} (|p_0|, - \vec{p}) \right] 
\nonumber \\ 
& = & {\cal F}_\rho (P, X; [f_\sigma^{(\mbox{\scriptsize{B}})}], 
[C_{\sigma - \sigma}^{(\mbox{\scriptsize{B}})}]) \, , \\ 
C_{\rho - \rho}^{(\mbox{\scriptsize{ph}})} (P, X) & = & 
{\cal G}_\rho (P, X; [f_\sigma^{(\mbox{\scriptsize{B}})}], 
[C_{\sigma - \sigma}^{(\mbox{\scriptsize{B}})}]) \, . 
\end{eqnarray*}
${\cal F}_\rho$ and ${\cal G}_\rho$ here contain large contributions 
mentioned above. Solving these equations for 
$f_\rho^{(\mbox{\scriptsize{B}})}$ and $C_{\rho - 
\rho}^{(\mbox{\scriptsize{B}})}$, , one obtains 
\begin{eqnarray} 
f_\rho^{(\mbox{\scriptsize{B}})} = f_\rho^{(\mbox{\scriptsize{B}})} 
(P, X; [f_\sigma^{(\mbox{\scriptsize{ph}})}], 
[C_{\sigma - \sigma}^{(\mbox{\scriptsize{ph}})}]) \, , 
\label{raberu1} \\   
C_{\rho - \rho}^{(\mbox{\scriptsize{B}})} = C_{\rho - 
\rho}^{(\mbox{\scriptsize{B}})} (P, X; 
[f_\sigma^{(\mbox{\scriptsize{ph}})}], [C_{\sigma - 
\sigma}^{(\mbox{\scriptsize{ph}})}]) \, . 
\label{raberu2} 
\end{eqnarray} 
In the case of scalar theory \cite{8}, the physical number density 
is shown to obey the generalized Boltzmann equation. 

Computation of some physical quantity yields the expression $F 
([f_\rho^{(\mbox{\scriptsize{B}})}], [C_{\rho - 
\rho}^{(\mbox{\scriptsize{B}})}])$, which includes large 
contribution. Substituting the RHS's of Eqs. (\ref{raberu1}) and 
of (\ref{raberu2}) for, in respective order, 
$f_\rho^{(\mbox{\scriptsize{B}})}$ and $C_{\rho - 
\rho}^{(\mbox{\scriptsize{B}})}$ in $F$, one obtains the expression 
$F' ([f_\rho^{(\mbox{\scriptsize{ph}})}], [C_{\rho - 
\rho}^{(\mbox{\scriptsize{ph}})}])$, which does not include large 
contribution. 

The perturbation theory thus constructed is called the \lq\lq 
bare-$N$ scheme'' in \cite{8}. 
\subsubsection*{Physical-$N$ scheme} 
Here we aim at constructing a perturbation theory, on the basis of 
which no large contribution appear. Then, in such a scheme, there 
are no large terms in the relations between 
$(f_\rho^{(\mbox{\scriptsize{ph}})}, \; C_{\rho - 
\rho}^{(\mbox{\scriptsize{ph}})})$ and $(f_\rho, C_{\rho - \rho})$. 
This is achieved if the condition, $\mbox{Eq. (\ref{kaku})} = 
\mbox{Eq. (\ref{kaku1})} = 0$, could be imposed. This is, however, 
not possible. Nevertheless, it is possible to construct the scheme 
that is free from the large contributions. 

For determining so far arbitrary $f_\rho (P, X)$ and $C_{\rho - 
\rho} (P, X)$ $(X^0_i < X^0)$, we impose the conditions, 
\begin{eqnarray} 
&& \mbox{Tr} \left( P\kern-0.1em\raise0.3ex\llap{/}\kern0.15em\relax 
+ \Omega_f (P, X) \right) \left( \mbox{Eq. (\ref{kaku})} \right) = 
\mbox{Tr} N\kern-0.153em\raise0.3ex\llap{/}\kern0.177em\relax \left( 
\mbox{Eq. (\ref{kaku})} \right) \nonumber \\ 
&& \mbox{\hspace*{10ex}} = \mbox{Tr} 
P\kern-0.1em\raise0.3ex\llap{/}\kern0.15em\relax 
N\kern-0.153em\raise0.3ex\llap{/}\kern0.177em\relax \left( \mbox{Eq. 
(\ref{kaku})} \right) = 0 \, , 
\label{51} \\ 
&& \mbox{Tr} \gamma_5 \left( 
P\kern-0.1em\raise0.3ex\llap{/}\kern0.15em\relax - m \right) 
N\kern-0.153em\raise0.3ex\llap{/}\kern0.177em\relax \left( \mbox{Eq. 
(\ref{kaku1})} \right) \nonumber \\ 
&& \mbox{\hspace*{10ex}} = \mbox{Tr} \gamma_5 \left( 
P\kern-0.1em\raise0.3ex\llap{/}\kern0.15em\relax - \Omega_C (P, X) 
\right) \left( \mbox{Eq. 
(\ref{kaku1})} \right) \nonumber \\ 
&& \mbox{\hspace*{10ex}} = 
\mbox{Tr} \gamma_5 
N\kern-0.153em\raise0.3ex\llap{/}\kern0.177em\relax \left( \mbox{Eq. 
(\ref{kaku1})} \right) = 0 \, . 
\label{52d} 
\end{eqnarray} 
Here, $\Omega_f (P, X)$ and $\Omega_C (P, X)$ are arbitrary 
functions with the property, 
\begin{eqnarray}
&& \Omega_f (p_0 = \pm \omega_\pm (\pm \vec{p}, X), \vec{p}, X) 
\nonumber \\ 
&& \mbox{\hspace*{15ex}} = 
\Omega_C (p_0 = \pm \omega_\pm (\pm \vec{p}, X), \vec{p}, X) 
\nonumber \\ 
&& \mbox{\hspace*{15ex}} = \left[ \left( \omega_\pm (\pm \vec{p}, X) 
\right)^2 - \vec{p}^{\, 2} \right]^{1 / 2} \, . 
\label{jyou}
\end{eqnarray}
As has been discussed in \cite{8}, this arbitrariness does not 
matter (see also subsubsection 6 below).  Computation of Eqs. 
(\ref{51}) and (\ref{52d}) yields, in respective order,  
\begin{eqnarray}
P \cdot \partial_X f_\rho & = & \frac{1}{4} \mbox{Tr} \left( 
P\kern-0.1em\raise0.3ex\llap{/}\kern0.15em\relax + \Omega_f (P, X) 
\right) \left[ \tilde{\Gamma}_p^{\rho \rho} + i \Sigma_B^{\rho \rho} 
\right] \, , \nonumber \\ 
&& 
\label{aa} \\ 
N \cdot \partial_X f_\rho & = & \frac{1}{4} \mbox{Tr} 
N\kern-0.153em\raise0.3ex\llap{/}\kern0.177em\relax \left[ 
\tilde{\Gamma}_p^{\rho \rho} + i \Sigma_B^{\rho \rho} \right] \, , 
\label{bi} \\ 
e_\perp \cdot \partial_X f_\rho & = & \frac{1}{4} \rho \epsilon 
(p_0) \mbox{Tr} P\kern-0.1em\raise0.3ex\llap{/}\kern0.15em\relax 
N\kern-0.153em\raise0.3ex\llap{/}\kern0.177em\relax \left[ 
\tilde{\Gamma}_p^{\rho \rho} + i \Sigma_B^{\rho \rho} \right] \, , 
\label{cu} 
\end{eqnarray}
and 
\begin{eqnarray}
&& \left[ N^2 P \cdot \partial_X - \frac{1}{2} (P^2 - m^2) 
\frac{\partial N^2}{\partial P_\alpha} \frac{\partial}{\partial 
X^\alpha} \right] C_{\rho - \rho} \nonumber \\ 
&& \mbox{\hspace*{4ex}} = - \frac{1}{4} \mbox{Tr} \gamma_5 ( 
P\kern-0.1em\raise0.3ex\llap{/}\kern0.15em\relax - m) 
N\kern-0.153em\raise0.3ex\llap{/}\kern0.177em\relax \left[ 
\tilde{\Gamma}_p^{\rho - \rho} + i \Sigma_B^{\rho - \rho} \right] 
\, , 
\label{ia} \\ 
&& \left[ m N \cdot \partial_X + \rho \epsilon (p_0) e_\perp \cdot 
\partial_X \right] C_{\rho - \rho} \nonumber \\ 
&& \mbox{\hspace*{4ex}} = - \frac{1}{4} \mbox{Tr} \gamma_5 \left( 
P\kern-0.1em\raise0.3ex\llap{/}\kern0.15em\relax - \Omega_C (P, X) 
\right) \left[ \tilde{\Gamma}_p^{\rho - \rho} + i \Sigma_B^{\rho - 
\rho} \right] \, , \nonumber \\ 
&& 
\label{meso} \\ 
&& \left[ \rho \epsilon (p_0) e_\perp^\mu \frac{\partial 
N_\mu}{\partial P_\alpha} \frac{\partial}{\partial X^\alpha} + 
\frac{m}{2} \frac{\partial N^2}{\partial P_\alpha} 
\frac{\partial}{\partial X^\alpha} \right] C_{\rho - \rho} \nonumber 
\\ 
&& \mbox{\hspace*{4ex}} = \frac{1}{4} \mbox{Tr} \gamma_5 
N\kern-0.153em\raise0.3ex\llap{/}\kern0.177em\relax \left[ 
\tilde{\Gamma}_p^{\rho - \rho} + i \Sigma_B^{\rho - \rho} \right] 
\, . 
\label{fin1} 
\end{eqnarray}
These equations are the determining equations for $f_\rho$ and 
$C_{\rho - \rho}$, which are to be solved under the given initial 
data, $f_\rho (P, X_i^0, \vec{X})$ and $C_{\rho - \rho} (P, X_i^0, 
\vec{X})$, respectively. 

After imposition of Eqs. (\ref{aa}) - (\ref{fin1}), $H^{\rho \pm 
\rho} - \Sigma_K^{\rho \pm \rho}$, Eqs. (\ref{kaku}) and 
(\ref{kaku1}), turns out to 
\begin{eqnarray}
H^{\rho \rho} - \Sigma_K^{\rho \rho} & = & \frac{i}{4 P^2} \left( 
P^2 - \Omega_f P\kern-0.1em\raise0.3ex\llap{/}\kern0.15em\relax 
\right) \mbox{Tr} \left( \tilde{\Gamma}_p^{\rho \rho} + i 
\Sigma_B^{\rho \rho} \right) \, , \nonumber \\ 
&& \label{ake} \\ 
H^{\rho - \rho} - \Sigma_K^{\rho - \rho} & = & \frac{i}{4 P^2} 
\gamma_5 \left( P^2 - \Omega_C 
P\kern-0.1em\raise0.3ex\llap{/}\kern0.15em\relax \right) \nonumber 
\\ 
&& \times \mbox{Tr} \gamma_5 \left( \tilde{\Gamma}_p^{\rho - \rho} 
+ i \Sigma_B^{\rho - \rho} \right) \, . 
\label{mesomeso}
\end{eqnarray}
On the energy shells $p_0 = \pm \omega_\pm$, these quantities 
vanish, since $(P^2 - \Omega_{f (C)} 
P\kern-0.1em\raise0.3ex\llap{/}\kern0.15em\relax) (P^2 + \Omega_{f 
(C)} P\kern-0.1em\raise0.3ex\llap{/}\kern0.15em\relax) = P^2 (P^2 - 
\Omega^2_{f (C)})$. Then, the above mentioned large contributions, 
which turn out to be diverging contributions in the narrow-width 
approximation, do not appear. Thus, $G_K^{[1]} + G_K^{[2]}$ turns 
out to be a well-behaved function. As a matter of fact, in the 
narrow-width approximation, 
\begin{eqnarray*}
&& G_K^{[2]} + G_K^{[2]} \propto \frac{p_0 \mp \omega_\pm}{(p_0 \mp 
\omega_\pm)^2 + (0^+)^2} = \frac{\bf P}{p_0 \mp \omega_\pm} \\ 
&& \mbox{\hspace*{20ex}} (p_0 \simeq \pm \omega_\pm) \, , 
\end{eqnarray*}
which is a well-defined distribution. In particular, the relations 
between the physical $\left( f_\rho^{\mbox{\scriptsize{(ph)}}}, \; 
C_{\rho - \rho}^{\mbox{\scriptsize{(ph)}}} \right)$ and $\left( 
f_\rho, \; C_{\rho - \rho} \right)$ contain no large term: 
\begin{equation} 
f_\rho^{\mbox{\scriptsize{(ph)}}} = f_\rho + \Delta f_\rho \, , 
\;\;\;\;\; 
C_{\rho - \rho}^{\mbox{\scriptsize{(ph)}}} = C_{\rho - \rho} + 
\Delta C_{\rho - \rho} 
\label{hose} 
\end{equation} 
with $\Delta f_\rho$ and $\Delta C_{\rho - \rho}$ the perturbative 
corrections. 

Proceeding as in \cite{9}, from Eq. (\ref{aa}) on the energy shells, 
one obtains a generalized Boltzmann equation. In fact, the term with 
$\tilde{\Gamma}_p^{\rho \rho}$ on the RHS of Eq. (\ref{aa}) is 
proportional to the net production rate. To avoid complete 
repetition, we do not reproduce it here. 
\subsubsection{Discussion}
Here we like to mention a similarity between the two schemes 
presented here, the bare-$N$ scheme and the physical-$N$ scheme, 
and those in the ultra-violet (UV) renormalization scheme in 
quantum-field theory. For simplicity of presentation, taking a 
complex-scalar theory, we focus on the mass renormalization and 
do not mention on the coupling constant- and wave 
function-renormalizations. 
\subsubsection*{Summary of the UV-renormalization theory}  
{\em \lq\lq Bare'' UV renormalization scheme}: The free Lagrangian 
density reads ${\cal L}_0 = - \phi^\dagger (x) (\partial^2 + 
m^2_B) \phi (x)$ with $m_B$ the bare mass. Computation of the 
physical mass $M_{\mbox{\scriptsize{ph}}}$ yields 
$M_{\mbox{\scriptsize{ph}}} = 
M_{\mbox{\scriptsize{ph}}} (m_B)$, which includes diverging terms. 
Solving this equation for $m_B$, we have $m_B = m_B 
(M_{\mbox{\scriptsize{ph}}})$. 
Perturbative computation of some physical 
quantity $F$ yields the expression $F = F (m_B)$, which contains, in 
general, UV-divergences. Substituting the equation $m_B = 
m_B (M_{\mbox{\scriptsize{ph}}})$ for $m_B$ in $F (m_B)$, one gets 
$F_R (M_{\mbox{\scriptsize{ph}}}) \equiv F (m_B 
(M_{\mbox{\scriptsize{ph}}}))$, which is free from UV-divergence. 

{\em Physical UV renormalization scheme}: One introduces new free 
Lagrangian ${\cal L}_0' = - \phi^\dagger (x) (\partial^2 + m^2) \phi 
(x)$ with $m$ the renormalized mass. Then, the counter Lagrangian 
should be introduced, ${\cal L}_c = {\cal L}_m - {\cal L}_m' = 
\phi^\dagger (x) \left[ m^2 - m_B^2 \right] \phi (x)$. $m^2 - m_B^2$ 
is determined so that the perturbatively computed physical mass 
$M_{\mbox{\scriptsize{ph}}}$ is free from the UV-divergence. Thus, 
no diverging term is involved in the relation 
$M_{\mbox{\scriptsize{ph}}} = M_{\mbox{\scriptsize{ph}}} (m)$. 
However, there is arbitrariness in the definition of the finite part 
of $m$, which is determined by imposing some condition. This 
arbitrariness is called a \lq\lq renormalization scheme dependence'' 
(see, e.g., \cite{ste}). It is well known that, when one computes 
some physical quantity $F$ up to, say, $n$th order of perturbation 
theory, the above arbitrariness affects $F$ at the next to the 
$n$th order. The renormalization scheme, in which $m = 
M^{\mbox{\scriptsize{(ph)}}}$, is convenient for many cases.  
\subsubsection*{Summary of the two schemes presented above} 
{\em Bare-$N$ scheme}: No counter Lagrangian is introduced. 
Computation of the physical number densities (that are related to 
$f_\rho^{\mbox{\scriptsize{(ph)}}}$) and $C_{\rho - 
\rho}^{\mbox{\scriptsize{(ph)}}}$, which are the functionals of 
$f_\rho^{\mbox{\scriptsize{(B)}}}$ and $C_{\rho - 
\rho}^{\mbox{\scriptsize{(B)}}}$, include large contributions. 
Perturbative computation of some quantity yields the expression, 
which is written in terms of $f_\rho^{\mbox{\scriptsize{(B)}}}$ and 
$C_{\rho - \rho}^{\mbox{\scriptsize{(B)}}}$ and includes large 
contributions. Rewriting it in terms of the physical quantities, 
$f_\rho^{\mbox{\scriptsize{(ph)}}}$ and $C_{\rho - 
\rho}^{\mbox{\scriptsize{(ph)}}}$, one obtains the 
large-contribution free form. 

{\em Physical-$N$ scheme}: We introduce a counter Lagrangian 
$\underline{L}_c = \underline{L}_{c1} + \underline{L}_{c2}$, which 
is determined so that the perturbatively computed physical number 
densities and $C_{\rho - \rho}^{\mbox{\scriptsize{(ph)}}}$ do not 
contain large contributions. There is arbitrariness in the 
definition of the \lq\lq finite parts'' of $f_\rho$ and $C_{\rho - 
\rho}$. The arbitrariness in the choice of the functions $\Omega_f$ 
(Eqs. (\ref{aa}) and (\ref{ake})) and $\Omega_C$ (Eqs. (\ref{meso}) 
and (\ref{mesomeso})) is this arbitrariness. It is worth mentioning 
that, if we could choose $\Omega_f$ and $\Omega_C$ so that $f_\rho 
(P, X) = f_\rho^{\mbox{\scriptsize{(ph)}}} (P, X)$, Eq. (\ref{aa}) 
on the energy-shell turns out to a genuine (generalized) Boltzmann 
equation. In the opposite case, the function that obey the 
generalized Boltzmann equation is $f_\rho$ and the physical 
$f_\rho^{\mbox{\scriptsize{(ph)}}}$ is written as in Eq. 
(\ref{hose}). 

Similar comment to the above one at the end of the {\em Physical 
UV renormalization scheme} may be made here. 
\subsubsection*{Correspondence} 
Above observation discloses the correspondence between the two 
schemes presented here and those in the UV-renormalization scheme: 
\[
\begin{array}{l}
\mbox{Bare scheme:} \\ 
\;\;\;\;\; \int d^4 x \, {\cal L}_0 (x) \leftrightarrow 
{\cal A}_0 \mbox{ in Eq. (\ref{4.7d}) with } \underline{L}_c = 0 
\, , 
\\ 
\;\;\;\;\; m_B \leftrightarrow f_\rho^{\mbox{\scriptsize{(B)}}} 
\mbox{ and } C_{\rho - \rho}^{\mbox{\scriptsize{(B)}}} \, , \\ 
\;\;\;\;\; M_{\mbox{\scriptsize{ph}}} \leftrightarrow 
f_\rho^{\mbox{\scriptsize{(ph)}}} \mbox{ and }
C_{\rho - \rho}^{\mbox{\scriptsize{(ph)}}} \, , \\ 
\hspace*{1ex} \\ 
\mbox{Physical scheme:} \\ 
\;\;\;\;\; \int d^4 x \, {\cal L}_0' (x) \leftrightarrow 
{\cal A}_0 \, , \\ 
\;\;\;\;\; \int d^4 x \, {\cal L}_c (x) \leftrightarrow 
{\cal A}_c \mbox{ in Eq. (\ref{countl1})} \, , \\ 
\;\;\;\;\; m \leftrightarrow  f_\rho \mbox{ and } C_{\rho - \rho} 
\, , \\ 
\;\;\;\;\; M_{\mbox{\scriptsize{ph}}} \leftrightarrow  
f_\rho^{\mbox{\scriptsize{(ph)}}} \mbox{ and } C_{\rho - 
\rho}^{\mbox{\scriptsize{(ph)}}} \, , \\ 
\;\;\;\;\; \mbox{Absence of} \\ 
\;\;\;\;\;\;\;\;\;\;\;\;\;\; \mbox{divergence} \leftrightarrow 
\mbox{ large contribution} \, , \\ 
\;\;\;\;\; \mbox{Arbitrariness in } m \leftrightarrow 
\mbox{Arbitrariness in } 
\Omega_f \mbox{ and } \Omega_C \, , \\ 
\;\;\;\;\; \mbox{Scheme with} \\ 
\;\;\;\;\;\;\;\;\;\;\;\;\;\; m = M^{\mbox{\scriptsize{(ph)}}} 
\leftrightarrow f_\rho = f_\rho^{\mbox{\scriptsize{(ph)}}} 
\mbox{ and } C_{\rho - \rho} = C_{\rho - 
\rho}^{\mbox{\scriptsize{(ph)}}} \, . 
\end{array}
\]
\subsection{Gluon sector}
\subsubsection{Preliminary}
Configuration-space counterparts of Eqs. (\ref{decomp1}) and 
(\ref{standardg}) are, with obvious notation, 
\begin{equation}
\underline{A}^{\mu \nu} (x, y) = \sum_{U, V = T, L, G} \; \sum_{j = 
1}^{J_{UV}} \left[ \underline{\cal R}_{Lj}^{UV} \cdot 
\tilde{\underline{A}}_j^{U V} \cdot \underline{\cal R}_{Rj}^{UV} 
\right]^{\mu \nu} (x, y) \, , 
\label{abb}
\end{equation}
where $J_{TT} = 4$, $J_{LL}= J_{GG} = J_{LG} = J_{GL} = 1$ and 
$J_{TL} = J_{LT} =J_{TG} =J_{GT} = 2$,  and 
\widetext 
\begin{eqnarray*} 
\sum_{j=1}^4 \left[ \underline{\cal R}_j^{TT} \cdot 
\tilde{\underline{A}}_j^{TT} \cdot \underline{\cal R}_j^{TT} 
\right]^{\mu \nu} &\equiv& \underline{\cal P}^{\mu \rho}_T \cdot 
\underline{A}_1^{TT} \cdot (\underline{\cal P}_T)_\rho^{\; \, \nu} 
+ \underline{\tilde{\zeta}}^\mu \cdot \underline{A}_2^{TT} \cdot 
\underline{\tilde{\zeta}}^\nu \nonumber \\ 
&& - \underline{\tilde{\zeta}}^\mu \cdot \underline{A}_3^{T T'} 
\cdot \underline{E}_\perp^\nu + \underline{E}_\perp^\mu \cdot 
\underline{A}_3^{T' T} \cdot \underline{\tilde{\zeta}}^\nu 
\, , \nonumber \\ 
\left[ \underline{\cal R}_j^{LL} \cdot \tilde{\underline{A}}_j^{LL} 
\cdot \underline{\cal R}_j^{LL} \right]^{\mu \nu} &\equiv& n^\mu 
\underline{A}_1^{LL} \, n^\nu \, , \nonumber \\ 
\left[ \underline{\cal R}_j^{GG} \cdot \tilde{\underline{A}}_j^{GG} 
\cdot \underline{\cal R}_j^{GG} \right]^{\mu \nu} &\equiv& (i 
\tilde{\partial}^\mu) \underline{A}_1^{GG} (i \tilde{\partial}^\nu) 
\, , \nonumber \\ 
\sum_{j=1}^2 \left[ \underline{\cal R}_j^{TL} \cdot 
\tilde{\underline{A}}_j^{TL} \cdot \underline{\cal R}_j^{TL} 
\right]^{\mu \nu} &\equiv& \underline{\tilde{\zeta}}^\mu \cdot 
\underline{A}_1^{T L} n^\nu + \underline{E}_\perp^\mu \cdot 
\underline{A}_2^{TL} n^\nu \, , \nonumber \\ 
\sum_{j=1}^2 \left[ \underline{\cal R}_j^{LT} \cdot 
\tilde{\underline{A}}_j^{LT} \cdot \underline{\cal R}_j^{LT} 
\right]^{\mu \nu} &\equiv& n^\mu \underline{A}_1^{LT} \cdot 
\underline{\tilde{\zeta}}^\nu - n^\mu \underline{A}_2^{LT} 
\cdot \underline{E}_\perp^\nu \, , \nonumber \\ 
\sum_{j=1}^2 \left[ \underline{\cal R}_j^{TG} \cdot 
\tilde{\underline{A}}_j^{TG} \cdot \underline{\cal R}_j^{TG} 
\right]^{\mu \nu} &\equiv& \underline{E}_\perp^\mu \cdot 
\underline{A}_1^{TG} (i \tilde{\partial}^\nu) + 
\underline{\tilde{\zeta}}^\mu \cdot \underline{A}_2^{TG} (i 
\tilde{\partial}^\nu) \, , \nonumber \\ 
\sum_{j=1}^2 \left[ \underline{\cal R}_j^{GT} \cdot 
\tilde{\underline{A}}_j^{GT} \cdot \underline{\cal R}_j^{GT} 
\right]^{\mu \nu} &\equiv& (i \tilde{\partial}^\mu) 
\underline{A}_1^{GT} \cdot \underline{E}_\perp^\nu - (i 
\tilde{\partial}^\mu) \underline{A}_2^{GT} \cdot 
\underline{\tilde{\zeta}}^\nu \, , \nonumber \\ 
\left[ \underline{\cal R}_j^{LG} \cdot \tilde{\underline{A}}_j^{LG} 
\cdot \underline{\cal R}_j^{LG} \right]^{\mu \nu} &\equiv& n^\mu 
\underline{A}_1^{LG} (i \tilde{\partial}^\nu) \, , \nonumber \\ 
\left[ \underline{\cal R}_j^{GL} \cdot \tilde{\underline{A}}_j^{GL} 
\cdot \underline{\cal R}_j^{GL} \right]^{\mu \nu} &\equiv& - (i 
\tilde{\partial}^\mu) \underline{A}_1^{GL} n^\nu \, , 
\end{eqnarray*}
with $\tilde{\partial}^\mu = \partial^\mu - n^\mu \partial_0$ and 
$\partial_0 = \partial / \partial X_0$.  
\subsubsection{Bare propagator and counter-Lagrangian}
We proceed as in \cite{10}. We start from the expression for the 
bare propagator $\hat{\underline{D}} (x, y)$ (cf. Eqs. 
(\ref{bare0g}) - (\ref{Kdesu})), 
\begin{eqnarray*}
\hat{\underline{D}}^{\mu \nu} &=& \left( \underline{\cal P}_T 
\cdot \hat{\underline{D}}_T \cdot \underline{\cal P}_T 
\right)^{\mu \nu} + \left( \underline{\cal P}_L \cdot 
\hat{\underline{D}}_L \cdot \underline{\cal P}_L 
\right)^{\mu \nu} + i \tilde{\partial}^\mu \hat{\underline{D}}_G 
(i \tilde{\partial}^\nu) \nonumber \\ 
&& + n^\mu \hat{\underline{D}}_{LT} (i \tilde{\partial}^\nu) 
+ i \tilde{\partial}^\mu \hat{\underline{D}}_{TL} n^\nu 
+ \underline{D}_K \hat{M}_+ \, , \nonumber \\ 
\hat{\underline{D}}_T &=& \left( 
\begin{array}{cc} 
- \underline{\Delta}_R & \;\; 0 \\ 
- \underline{\Delta}_R + \underline{\Delta}_A & \;\; 
\underline{\Delta}_A 
\end{array} 
\right) - [ \underline{\Delta}_R \cdot \tilde{\underline{f}} - 
\tilde{\underline{f}} \cdot \underline{\Delta}_A ] 
\hat{M}_+ \, , \\ 
\underline{D}_L &=& - \left[ \frac{1}{\tilde{P}^2} \left( 1 + 
\lambda \frac{p_0^2}{\tilde{P}^2} \right) 
\right]_{\mbox{\scriptsize{IFT}}} \hat{\tau}_3 \nonumber \\ 
&& - \left[ \left\{ \frac{1}{\tilde{P}^2} \left( 1 + \lambda 
\frac{p_0^2}{\tilde{P}^2} \right) \right\}_{\mbox{\scriptsize{IFT}}} 
\cdot \tilde{\underline{f}} - \tilde{\underline{f}} \cdot \left\{ 
\frac{1}{\tilde{P}^2} \left( 1 + \lambda \frac{p_0^2}{\tilde{P}^2} 
\right) \right\}_{\mbox{\scriptsize{IFT}}} \right] \hat{M}_+ \, , \\ 
\underline{\cal D}_G &=& - \lambda \left[ \frac{1}{\tilde{P}^4} 
\right]_{\mbox{\scriptsize{IFT}}} \hat{\tau}_3 - \lambda \left[ 
\left\{ \frac{1}{\tilde{P}^4} \right\}_{\mbox{\scriptsize{IFT}}} 
\cdot \tilde{\underline{f}} - \tilde{\underline{f}} \cdot \left\{ 
\frac{1}{\tilde{P}^4} \right\}_{\mbox{\scriptsize{IFT}}} \right] 
\hat{M}_+ \, , \\ 
\underline{D}_{LT} &=& \underline{\cal D}_{TL} = - \lambda 
\left[ \frac{p_0}{\tilde{P}^4} \right]_{\mbox{\scriptsize{IFT}}} 
\hat{\tau}_3 - \lambda \left[ \left\{ \frac{p_0}{\tilde{P}^4} 
\right\}_{\mbox{\scriptsize{IFT}}} \cdot \tilde{\underline{f}} - 
\tilde{\underline{f}} \cdot \left\{ \frac{p_0}{\tilde{P}^4} 
\right\}_{\mbox{\scriptsize{IFT}}} \right] \hat{M}_+ \, , \\ 
\underline{D}_K &=& - \underline{\tilde{\zeta}}^\mu \cdot 
\left[ \underline{\Delta}_R \cdot \underline{C}_2^{TT} - 
\underline{C}_2^{TT} \cdot \underline{\Delta}_A \right] \cdot 
\underline{\tilde{\zeta}}^\nu + \underline{\tilde{\zeta}}^\mu \cdot 
\left[ \underline{\Delta}_R \cdot \underline{C}_3^{TT'} - 
\underline{C}_3^{TT'} \cdot \underline{\Delta}_A \right] \cdot 
\underline{E}_\perp^\nu \nonumber \\ 
&& - \underline{E}_\perp^\mu \cdot \left[ \underline{\Delta}_R \cdot 
\underline{C}_3^{T'T} - \underline{C}_3^{T'T} \cdot 
\underline{\Delta}_A \right] \cdot \underline{\tilde{\zeta}}^\nu 
\, . 
\end{eqnarray*}
$\hat{D}$ in Eq. (\ref{bare0g}) is the leading part of the DEX of 
$\hat{D} (P, X)$ $\left( 
\stackrel{\mbox{\scriptsize{WT}}}{\longleftarrow} 
\hat{\underline{D}} (x, y) \right)$. Calculation within the gradient 
approximation yields 
\begin{eqnarray}
(\underline{\hat{D}}^{- 1})^{\mu \nu} \cdot 
\underline{\hat{D}}_\nu^{\;\, \rho} &=& g^{\mu \rho} \, , 
\label{bareL1} \\ 
(\underline{\hat{D}}^{- 1})^{\mu \nu} (x, y) &=& \left( g^{\mu \nu} 
\partial^2 - \partial^\mu \partial^\nu + \frac{1}{\lambda} 
\tilde{\partial}^\mu \tilde{\partial}^\nu 
\right)_{\mbox{\scriptsize{IFT}}} \hat{\tau}_3 - 
\underline{\hat{L}}_c^{\mu \nu} (x, y) \nonumber \\ 
&=& \hat{\tau}_3 \left[ {\cal P}_T^{\mu \nu} (i \partial) \partial^2 
+ {\cal P}_L^{\mu \nu} (i \partial) \tilde{\partial}^2 - \partial_0 
\left( \tilde{\partial}^\mu n^\nu + n^\mu \tilde{\partial}^\nu 
\right) + {\cal P}_G^{\mu \nu} (i \partial) \left( 
\frac{\tilde{\partial}^2}{\lambda} + \partial^2_0 \right) 
\right]_{\mbox{\scriptsize{IFT}}} \nonumber \\ 
&& - \underline{\hat{L}}_c^{\mu \nu} (x, y) \, , 
\label{zero2} \\ 
\hat{L}_c^{\mu \nu} &=& L_c^{\mu \nu} \hat{M}_- \nonumber \\ 
& = & 2 i \hat{M}_- \left[ \left( P \cdot \partial \tilde{f} 
\right) {\cal P}_T^{\mu \nu} + \tilde{P} \cdot \tilde{\partial} 
\tilde{f} n^\mu n^\nu \right. \nonumber \\ 
&& + \left\{ \frac{1}{\tilde{P}^2} \left( \frac{p_0^2}{\tilde{P}^2} 
+ \frac{2}{\lambda} \right) \tilde{P} \cdot \tilde{\partial} 
\tilde{f} + \frac{p_0}{\tilde{P}^2} \partial_0 \tilde{f} \right\} 
\tilde{P}^\mu \tilde{P}^\nu \nonumber \\ 
&& - \left\{ \frac{p_0}{\tilde{P}^2} \tilde{P} \cdot 
\tilde{\partial} \tilde{f} + \frac{1}{2} \partial_0 \tilde{f} 
\right\} (n^\mu \tilde{P}^\nu + \tilde{P}^\mu n^\nu) + P \cdot 
\partial C_2^{TT} (P, X) \tilde{\zeta}^\mu \tilde{\zeta}^\nu 
\nonumber \\ 
&& \left. - P \cdot \partial C_3^{TT'} (P, X) \tilde{\zeta}^\mu 
E_\perp^\nu + P \cdot \partial C_3^{T'T} (P, X) E_\perp^\mu 
\tilde{\zeta}^\nu \right] \, . 
\label{usa}
\end{eqnarray}
Here $P \cdot \partial \tilde{f} = P^\mu \partial \tilde{f} (P, X) / 
\partial X^\mu$, etc. 
\narrowtext 

From Eq. (\ref{bareL1}), we see that the free action of the theory 
is 
\begin{eqnarray}
{\cal A}_0 & = & \frac{1}{2} \int d^{\, 4} x \, d^{\, 4} y \, 
\displaystyle{ \raisebox{1.8ex}{\scriptsize{t}}} 
\mbox{\hspace{-0.33ex}} \hat{A}^\mu (x) \left( \hat{D}^{- 1} (x, 
y) \right)_{\mu \nu} \hat{A}^\nu (y) \, , \nonumber \\ 
&& \displaystyle{ \raisebox{1.8ex}{\scriptsize{t}}} 
\mbox{\hspace{-0.33ex}} \hat{A}^\mu = \left(\hat{A}_1^\mu, \, 
\hat{A}_2^\mu \right) \, , 
\label{guru} 
\end{eqnarray}
where the color index is suppressed. Eq. (\ref{guru}) with Eq. 
(\ref{zero2}) tells us that there emerges a counteraction, 
\[ 
{\cal A}_c = \frac{1}{2} \int d^{\, 4} x \, d^{\, 4} y \, 
\displaystyle{ \raisebox{1.8ex}{\scriptsize{t}}} 
\mbox{\hspace{-0.33ex}} \hat{A}^\mu (x) \underline{\hat{L}}_c^{\mu 
\nu} (x, y) \hat{A}^\nu \, , 
\] 
which yields a (two-point) vertex factor 
\[
i \underline{\hat{L}}_c^{\mu \nu} (x, y) \to i L^{\mu \nu}_c (P, X) 
\hat{M}_- \, . 
\]
\subsubsection{Dyson equation} 
As in Sec. III, we use the $(4 \times 4)$ matrix notation in 
Minkowski space. The self-energy-part $\left( \hat{\underline{\bf 
\Pi}} (x, y) \right)$ resummed propagator $\hat{\underline{\bf G}} 
(x, y)$ obeys 
\begin{eqnarray} 
\hat{\underline{\bf G}} &=& \hat{\underline{\bf D}} - 
\hat{\underline{\bf D}} \cdot \underline{\hat{\bf \Pi}} \cdot 
\hat{\underline{\bf G}} \, , 
\label{SDeqn} \\ 
\hat{\underline{\bf D}} &=& \left( 
\begin{array}{cc} 
\underline{\bf D}_R & \;\; 0 \\ 
\underline{\bf D}_R - \underline{\bf D}_A & \;\; - \underline{\bf 
D}_A 
\end{array} 
\right) \nonumber \\ 
&& + \left[ \underline{\bf D}_R \cdot \underline{\tilde{f}} - 
\underline{\tilde{f}} \cdot \underline{\bf D}_A + 
\underline{\bf D}_K \right] \hat{M}_+ \nonumber \, . 
\end{eqnarray}
For $\hat{\underline{\bf G}}$ and $\hat{\underline{\bf \Pi}}$, 
we have (cf. Eqs. (\ref{hossh1}) -(\ref{owani}) and (\ref{hossh}) - 
(\ref{owani1}).) 
\begin{eqnarray} 
\hat{\underline{\bf G}} &=& \left( 
\begin{array}{cc} 
\underline{\bf G}_R & \;\; 0 \\ 
\underline{\bf G}_R - \underline{\bf G}_A & \;\; - \underline{\bf 
G}_A 
\end{array} 
\right) \nonumber \\ 
&& + \left[ \underline{\bf G}_R \cdot \underline{\tilde{f}} - 
\underline{\tilde{f}} \cdot \underline{\bf G}_A + 
\underline{\bf G}_K \right] \hat{M}_+ \, , \nonumber \\ 
\underline{\hat{\bf \Pi}} &=& \left( 
\begin{array}{cc} 
\underline{\bf \Pi}_R & \;\; 0 \\ 
- \underline{\bf \Pi}_R + \underline{\bf \Pi}_A & \;\; - 
\underline{\bf \Pi}_A 
\end{array} 
\right) \nonumber \\ 
&& + \left[ \underline{\bf \Pi}_R \cdot \underline{\tilde{f}} - 
\underline{\tilde{f}} \cdot \underline{\bf \Pi}_A - 
\underline{\bf \Pi}_K \right] \hat{M}_- \, , \nonumber \\ 
\underline{\bf G}_K &=& - \underline{\bf G}_R \cdot 
\underline{\tilde{f}} + \underline{\tilde{f}} \cdot 
\underline{\bf G}_A + \underline{\bf G}_{12} \, , \nonumber \\ 
\underline{\bf \Pi}_K &=& \underline{\bf \Pi}_R \cdot 
\underline{\tilde{f}} - \underline{\tilde{f}} \cdot 
\underline{\bf \Pi}_A + \underline{\bf \Pi}_{12} \, . 
\label{hiro} 
\end{eqnarray}
Eq. (\ref{SDeqn}) may be solved to give 
\begin{eqnarray} 
\underline{\bf G}_{R (A)} & = & \left[ \underline{\bf D}_{R (A)}^{- 
1} + \underline{\bf \Pi}_{R (A)} \right]^{- 1} \, , 
\label{en} \\ 
\underline{\bf G}_K &=& \underline{\bf G}_K^{[1]} 
+ \underline{\bf G}_K^{[2]} + \underline{\bf G}_K^{[3]} \, , 
\nonumber \\ 
\underline{\bf G}_K^{[1]} &=& \underline{\bf G}_R \cdot 
\underline{\bf \Pi}_K \cdot \underline{\bf G}_A \, , 
\label{2424} \\ 
\underline{\bf G}_K^{[2]} &=& \underline{\bf G}_R \cdot \left[ 
\underline{\bf \Pi}_R \cdot \underline{\bf C} - \underline{\bf C} 
\cdot \underline{\bf \Pi}_A \right] \cdot \underline{\bf G}_A 
\nonumber \\ 
& \equiv & - \underline{\bf G}_R \cdot \underline{\tilde{\bf H}} 
\cdot \underline{\bf G}_A \, , 
\label{hika} \\ 
\underline{\bf G}_K^{[3]} &=& 
\underline{\bf G}_R \cdot \underline{\bf C} - \underline{\bf C}
\cdot \underline{\bf G}_A \, , 
\label{4.43} \\ 
\underline{C}^{\mu \nu} &=& \underline{\tilde{\zeta}}^\mu \cdot 
\underline{C}_2^{TT} \cdot \underline{\tilde{\zeta}}^\nu - 
\underline{\tilde{\zeta}}^\mu \cdot \underline{C}_3^{TT'} \cdot 
\underline{E}_\perp^\nu \nonumber \\ 
&& + \underline{E}_\perp^\mu \cdot \underline{C}_3^{T'T} \cdot 
\underline{\tilde{\zeta}}^\nu \nonumber \, . 
\end{eqnarray}
The form for the leading part of the DEX of $\hat{G} (X, P)$, an 
Wigner transform of $\underline{G} (x, y)$, is the $\hat{G}$ that is 
deduced in Sec. III. 
\widetext 
\subsubsection{Gradient piece of the self-energy-part resummed 
propagator} 
\subsubsection*{Form for $G_{R (A)}$} 
We divide the Wigner transform of $\underline{\bf \Pi}_{R (A)} 
\cdot \underline{\bf G}_{R (A)}$ (cf. Eq. (\ref{en})) into two 
pieces (cf. Eq. (\ref{abb})), 
\begin{eqnarray} 
\underline{\bf \Pi}_{R (A)} \cdot \underline{\bf G}_{R (A)} 
& \stackrel{\mbox{\scriptsize{WT}}}{\longrightarrow} & {\bf \Pi}_{R 
(A)} (P, X) {\bf G}_{R (A)} (P, X) + \left( {\bf \Pi}_{R (A)} {\bf 
G}_{R (A)} \right)^{(1)} \, , 
\label{bunkai} \\ 
\left( {\bf \Pi}_{R (A)} {\bf G}_{R (A)} \right)^{(1) \mu \nu} & = & 
\frac{i}{2} \sum_{U, V, V' = T, L, G} \sum_{j = 1}^{J_{UV}} \sum_{j' 
= 1}^{J_{VV'}} \left[ \frac{\partial {\cal R}_{Lj}^{UV}}{\partial 
P_\mu} \frac{\partial \tilde{\Pi}^{UV}_{R (A) j}}{\partial X^\mu} 
{\cal R}_{Rj}^{UV} {\cal R}_{Lj'}^{VV'} \tilde{G}^{VV'}_{R (A) j'} 
{\cal R}_{Rj'}^{VV'} \right. \nonumber \\ 
&& - {\cal R}_{Lj}^{UV} \frac{\partial \tilde{\Pi}^{UV}_{R (A) 
j}}{\partial X^\mu} \frac{\partial {\cal R}_{Rj}^{UV} {\cal 
R}_{Lj'}^{VV'} \tilde{G}^{VV'}_{R (A) j'} {\cal R}_{Rj'}^{VV'}  
}{\partial P_\mu} \nonumber \\ 
&& + \frac{\partial {\cal R}_{Lj}^{UV} \tilde{\Pi}^{UV}_{R (A) j} 
{\cal R}_{Rj}^{UV} {\cal R}_{Lj'}^{VV'}}{\partial P_\mu} 
\frac{\partial \tilde{G}^{VV'}_{R (A) j'} }{\partial X^\mu} 
{\cal R}_{Rj'}^{VV'} \nonumber \\ 
&& \left. - {\cal R}_{Lj}^{UV} \tilde{\Pi}^{UV}_{R (A) j} {\cal 
R}_{Rj}^{UV} {\cal R}_{Lj'}^{VV'} \frac{\partial \tilde{G}^{VV'}_{R 
(A) j'}}{\partial X^\mu} \frac{\partial {\cal 
R}_{Rj'}^{VV'}}{\partial P_\mu} \right]^{\mu \nu} \, . 
\end{eqnarray} 
Here, ${\bf \Pi}_{R (A)} (P, X)$ and ${\bf G}_{R (A)} (P, X)$ are, 
in respective order, ${\bf \Pi}_{R (A)} (P)$ and ${\bf G}_{R (A)} 
(P)$ in Sec III. Using Eq. (\ref{bunkai}) in Eq. (\ref{en}), we 
obtain the solution for ${\bf G}_{R (A)}$ $\left( = {\bf G}_{R 
(A)}^{(0)} + {\bf G}_{R (A)}^{(1)} \right)$. The form for the 
leading part ${\bf G}_{R (A)}^{(0)}$ is given in Sec. III. The 
gradient part is 
\begin{eqnarray*} 
{\bf G}_{R (A)}^{(1)} & = & {\bf G}_{R (A)}^{(0)} \left[ \left\{ 
i g^{\mu \nu} P \cdot \partial - \frac{i}{2} \left( P^\mu 
\partial^\nu + \partial^\mu P^\nu \right) + \frac{i}{2} \lambda 
\left( \tilde{P}^\mu \tilde{\partial}^\nu + \tilde{\partial}^\mu 
\tilde{P}^\nu \right) \right\} {\bf G}_{R (A)}^{(0)} \right. 
\nonumber \\ 
&& \left. - \left( {\bf \Pi}_{R (A)} {\bf G}_{R (A)}^{(0)} 
\right)^{(1)} \right] \, . 
\end{eqnarray*} 
\subsubsection*{Form for $\Pi_K$, which is involved in 
$G_K^{[1]}$ in Eq. (\ref{2424})} 
In the following, we restrict ourselves to the strict Coulomb gauge, 
$\lambda = 0$, which is a physical gauge. Computation of Eq. 
(\ref{hiro}) to the gradient approximation yields 
\begin{eqnarray} 
\underline{\Pi}_K^{\mu \nu} &=& - \underline{L}_c^{\mu \nu} + 
\sum_j \sum_{UV = T, L} \left[ \underline{\cal R}^{UV}_{Lj} \cdot 
\left\{ \left( \tilde{f} \left[ \left( \tilde{\Pi}_{12} 
\right)_j^{UV} - \left( \tilde{\Pi}_{21} \right)_j^{UV} \right] 
\right)_{\mbox{\scriptsize{IWT}}} + 
(\tilde{\underline{\Pi}}_{12})_j^{UV} \right\} \cdot \underline{\cal 
R}_{Rj}^{UV} \right]^{\mu \nu} \nonumber \\ 
&& + \underline{\Pi}_K^{[1] \mu \nu} + \underline{\Pi}_K^{[2] \mu 
\nu} \, . 
\label{mukku1} 
\end{eqnarray}
In $\underline{G}_K^{[1] \mu \nu}$, Eq. (\ref{2424}), 
$\underline{\Pi}_K^{\mu \nu}$, and then $\underline{L}_c^{\mu 
\nu}$ in Eq. (\ref{mukku1}), are sandwiched between $\underline{\bf 
G}_R$ and $\underline{\bf G}_A$. Then, in the case of $\lambda = 0$, 
$\tilde{P}^\mu \tilde{P}^\nu$ and $(n^\mu \tilde{P}^\nu + 
\tilde{P}^\mu n^\nu)$ terms in $\Pi_c^{\mu \nu}$, Eq. (\ref{usa}), 
do not contribute to ${\bf G}_K^{[1]}$ (cf. Sec. III). 

The Standard forms for the gradient terms, $\Pi_K^{[1] \mu \nu}$ and 
$\Pi_K^{[2] \mu \nu}$, and ${\tilde{H}}$ (Eq. (\ref{hika})) are 
given in Appendix G. 
\subsubsection{Generalized Boltzmann equation and its relatives}
Structure of the theory is fully discussed in Sec. IVA5, so that 
we restrict ourselves to giving a brief description of the 
physical-$N$ scheme only. 

Same reasoning as in Subsec. IVA applies here: ${\bf G}_K^{[1]}$ and 
${\bf G}_K^{[2]}$, Eqs. (\ref{2424}) and (\ref{hika}), bring about 
disaster. This disaster would be overcome if the condition 
\begin{equation}
{\bf \Pi}_K - \tilde{\bf H} = 0 
\label{jyo}
\end{equation}
could be imposed. This is, however, not possible. Eq. (\ref{jyo}) 
may be imposed for the ${\cal P}_T^{\mu \nu}$, $n^\mu n^\nu$, 
$\tilde{\zeta}^\mu \tilde{\zeta}^\nu$, and $\tilde{\zeta}^\mu 
E_\perp^\nu$ components, which read, in respective order, 
\begin{eqnarray}
2 P \cdot \partial \tilde{f} &=& - i \left[( 1 +  \tilde{f} ) \left( 
\Pi_{12} \right)_1^{TT} - \tilde{f} \left( \Pi_{21} \right)_1^{TT} 
\right] - i \Pi_{K1}^{TT} - 2 Im \left( \tilde{\zeta}^2 E_\perp^2 
C_3^{TT'} \Pi_{R3}^{T'T} \right) + i H_1^{(1) TT} \, , \nonumber \\ 
&& \label{Tmu} \\ 
2 \tilde{P} \cdot \tilde{\partial} \tilde{f} &=& - i \left[( 1 +  
\tilde{f} ) \left( \Pi_{12} \right)_1^{LL} - \tilde{f} \left( 
\Pi_{21} \right)_1^{LL} \right] - i \Pi_{K1}^{LL} \, , 
\label{Lmu} \\ 
2 P \cdot \partial C_2^{TT} &=& - i \left[( 1 +  \tilde{f} ) \left( 
\Pi_{12} \right)_2^{TT} - \tilde{f} \left( \Pi_{21} \right)_2^{TT} 
\right] - i \Pi_{K2}^{(1) TT} \nonumber \\ 
&& - 2 Im \left[ - C_2^{TT} \left( 
\Pi_{R1}^{TT} + \tilde{\zeta}^2 \Pi_{R2}^{TT} \right) + E_\perp^2 
C_3^{T'T} \left( \Pi_{R3}^{TT'} + \Pi_{A2}^{TT'} \right) \right] + i 
H_2^{(1) TT} \, , 
\label{4.48} \\ 
2 P \cdot \partial C_3^{TT'} &=& - i \left[( 1 +  \tilde{f} ) \left( 
\Pi_{12} \right)_3^{TT'} - \tilde{f} \left( \Pi_{21} \right)_3^{TT'} 
\right] - i \Pi_{K3}^{(1) TT'} \nonumber \\ 
&& + i \left[ \tilde{\zeta}^2 C_2^{TT} \Pi_{A3}^{TT'} + C_3^{TT'} 
\left( \Pi_{A1}^{TT} - \Pi_{R1}^{TT} - \tilde{\zeta}^2 \Pi_{R2}^{TT} 
\right) \right] + i H_3^{(1) TT'} 
\label{4.49} \, . 
\end{eqnarray}
\narrowtext 
\noindent 
Proceeding as in \cite{10}, from Eqs. (\ref{Tmu}) and (\ref{Lmu}) on 
the energy shells, we obtain the generalized Boltzmann equations for 
the transverse and the longitudinal modes, respectively. (As a 
matter of fact, on the energy shells, the first term on the RHS of 
Eq. (\ref{Tmu}) (Eq. (\ref{Lmu})) is proportional to the net 
production rates of the transverse (longitudinal) mode.) We do not 
reproduce them here. It should be remarked that, in the case of $L$ 
mode, Eq. (\ref{Lmu}), the \lq\lq time-derivative term'', 
$\partial_0 \tilde{f}$, in the Boltzmann equation comes from 
$\Pi_{K1}^{LL}$. More precisely, the $\partial_0 \tilde{f}$ term 
comes from $\Pi_K^{[1] \mu \nu}$ (with $UV = LL$), Eq. 
(\ref{meichan}) in Appendix G, which is in $\Pi_{K1}^{LL}$ in Eq. 
(\ref{Lmu}). Eq. (\ref{4.48}) [(\ref{4.49})] determines spacetime 
evolution of $C_2^{TT}$ [$C_3^{TT'}$] along $P$. An evolution 
equation for $C_3^{T'T}$ is obtained from Eqs. (\ref{4.49}) and 
(\ref{mujina}). 

One cannot impose Eq. (\ref{jyo}) for the remaining 
$\tilde{\zeta}^\mu n^\nu$, $n^\mu \tilde{\zeta}^\nu$, $E_\perp^\mu 
n^\nu$, and $n^\mu E_\perp^\nu$ components. This is because, for 
these components, there are no counterpart of $\tilde{f}$, 
$C_2^{TT}$, $C_3^{TT'}$, and $C_3^{T'T}$.  For equilibrium systems, 
these modes are absent. Then, one can expect that, for the 
quasiuniform systems near equilibrium, these modes do not yield 
disastrously large contributions. 
\subsection{Ghost sector} 
The self-energy-part $\left( \underline{\hat{\tilde{\bf \Pi}}} (x, 
y) \right)$ resummed propagator $\underline{\hat{\tilde{\bf G}}} (x, 
y)$ obeys 
\begin{equation}
\underline{\hat{\tilde{\bf G}}} = \underline{\hat{\tilde{\bf D}}} 
\cdot \underline{\hat{\tilde{\bf \Pi}}} \cdot 
\underline{\hat{\tilde{\bf G}}} \, . 
\label{ghom}
\end{equation}
$\underline{\hat{\tilde{\bf D}}}$ is an inverse Fourier transform of 
$\hat{\tilde{\bf D}}$ in Eq. (\ref{ghostp}). As in Eq. 
(\ref{realyo}), $\underline{\hat{\tilde{\bf D}}}$, 
$\underline{\hat{\tilde{\bf \Pi}}}$, and 
$\underline{\hat{\tilde{\bf G}}}$ are diagonal $(2 \times 2)$-matrix 
functions, $\underline{\hat{\tilde{\bf D}}} = \underline{\tilde{\bf 
D}} \hat{\tau}_3= $, etc. Solving Eq. (\ref{ghom}), we see that the 
gradient part of $\tilde{G} (P, X)$ vanishes and 
\[
\hat{\tilde{G}} (P, X) = \left( \hat{\tilde{G}} (P, X) \right)^* 
= \frac{1}{\tilde{P}^2 - \tilde{\Pi} (P, X)} \, . 
\]
\section*{Acknowledgments}
The author thanks the useful discussion at the Workshop on Thermal 
Field Theories and their Applications, held at the Yukawa Institute 
for Theoretical Physics, Kyoto, Japan, 8 - 10 August, 2002. 
\begin{appendix} 
\setcounter{equation}{0}
\setcounter{section}{0}
\def\theequation{\mbox{\Alph{section}.\arabic{equation}}}
\section{$\,\;$ Resummation of the quark propagator} 
Here we derive Eqs. (\ref{ST20}) - (\ref{2.40d}). Formally solving 
Eq. (\ref{mei}), we obtain 
\begin{eqnarray}
{\bf G}_K & = & - {\bf G}_R {\bf \Sigma}_K {\bf G}_A + {\bf G}_K' 
\, , \nonumber \\ 
{\bf G}_K' & = & {\bf G}_R S_R^{- 1} {\bf S}_K \left( 1 + {\bf 
\Sigma}_A {\bf G}_A \right) \, , 
\label{3444} 
\end{eqnarray} 
where use has been made of Eq. (\ref{RRAA}). Since $S_R^{- 1} 
{\bf S}_K \propto (P^2 - m^2) \delta (P^2 - m^2) = 0$, we have ${\bf 
G}_K' = 0$. This means that the piece ${\bf S}_K$ of the bare 
propagator disappears through resummation, which is unnatural. 

A correct ${\bf G}_K'$ is obtained by substituting Eq. (\ref{esu}) 
for ${\bf S}_K$ in Eq. (\ref{3444}) as follows: 
\begin{eqnarray*}
{\bf G}_K' & = & {\bf G}_R S_R^{- 1} \left( S_R - S_A \right) 
\gamma_5 N\kern-0.1em\raise0.3ex\llap{/}\kern0.15em\relax {\bf C} 
(P) \left( 1 + {\bf \Sigma}_A {\bf G}_A \right) \nonumber \\ 
&=& \left[ {\bf G}_R S_R^{- 1} S_R \gamma_5 
N\kern-0.1em\raise0.3ex\llap{/}\kern0.15em\relax - {\bf G}_R S_R^{- 
1} \gamma_5 N\kern-0.1em\raise0.3ex\llap{/}\kern0.15em\relax S_A 
\right] {\bf C} (P) \nonumber \\ 
&& \times \left( 1 + {\bf \Sigma}_A {\bf G}_A \right) \nonumber \\ 
&=& {\bf G}_R \gamma_5 
N\kern-0.1em\raise0.3ex\llap{/}\kern0.15em\relax {\bf C} (P) \left( 
1 + {\bf \Sigma}_A {\bf G}_A \right) \nonumber \\ 
&& - \left( 1 + {\bf G}_R {\bf \Sigma}_R \right) \gamma_5 
N\kern-0.1em\raise0.3ex\llap{/}\kern0.15em\relax {\bf C} (P) {\bf 
G}_A \, , 
\end{eqnarray*} 
where use has been made of ${\bf G}_R S_R^{- 1} = ({\bf 1} + {\bf 
G}_R {\bf \Sigma}_R)$, which follows from Eq. (\ref{RRAA}). 
This is natural in the sense that 
\[ 
{\bf G}_K \stackrel{\hat{\bf \Sigma} \to 0}{\longrightarrow}  
{\bf S}_K \, .  
\]
\setcounter{equation}{0}
\setcounter{section}{1}
\def\theequation{\mbox{\Alph{section}.\arabic{equation}}}
\section{\lq\lq Multiplications'' of the two 
standard forms for the quark part} 
We define \lq\lq multiplications'' of the functions of the type 
(\ref{P-P}) as the products: 
\begin{equation}
\left( A \otimes B \right)^{\rho \sigma} \equiv A^{\rho \rho} \, 
B^{\rho \sigma} \, , \;\;\;\; \;\;\;\; \left[ A \otimes B 
\right]^{\rho \sigma} \equiv A^{\rho - \rho} \, B^{- \rho \sigma} 
\, . 
\label{teigi} 
\end{equation}
Straightforward manipulation yields the SFs (cf. Eq. (\ref{P-P})) of 
$(A \otimes B)^{\rho \sigma}$, 
\begin{eqnarray}
(A \otimes B)_1^{\rho \pm \rho} & = & A_1^{\rho \rho} B_1^{\rho \pm 
\rho} \pm P^2 A_2^{\rho \rho} B_2^{\rho \pm \rho} \pm N^2 A_3^{\rho 
\rho} B_3^{\rho \pm \rho} \nonumber \\ 
&& - P^2 N^2 A_4^{\rho \rho} B_4^{\rho \pm 
\rho} \, , \nonumber \\ 
(A \otimes B)_2^{\rho \pm \rho} & = & A_1^{\rho \rho} B_2^{\rho \pm 
\rho} \pm A_2^{\rho \rho} B_1^{\rho \pm \rho} \mp N^2 A_3^{\rho 
\rho} B_4^{\rho \pm \rho} \nonumber \\ 
&& + N^2 A_4^{\rho \rho} B_3^{\rho \pm \rho} 
\, , \nonumber \\ 
(A \otimes B)_3^{\rho \pm \rho} & = & A_1^{\rho \rho} B_3^{\rho \pm 
\rho} \pm P^2 A_2^{\rho \rho} B_4^{\rho \pm \rho} \pm A_3^{\rho 
\rho} B_1^{\rho \pm \rho} \nonumber \\ 
&& - P^2 A_4^{\rho \rho} B_2^{\rho \pm \rho} 
\, , \nonumber \\ 
(A \otimes B)_4^{\rho \pm \rho} & = & A_1^{\rho \rho} B_4^{\rho \pm 
\rho} \pm A_2^{\rho \rho} B_3^{\rho \pm \rho} \mp A_3^{\rho \rho} 
B_2^{\rho \pm \rho} \nonumber \\ 
&& + A_4^{\rho \rho} B_1^{\rho \pm \rho} \, . 
\label{decompp}
\end{eqnarray}
$\left[ A \otimes B \right]^{\rho - \rho}_j$ ($j = 1 - 4$) 
is given by $\left( A \otimes B \right)^{\rho \rho}_j$ in Eq. 
(\ref{decompp}) with $B^{- \rho - \rho}_j$ for $B^{\rho \rho}_j$, 
and $\left[ A \otimes B \right]^{\rho \rho}_j$ is given by 
$\left( A \otimes B \right)^{\rho - \rho}_j$ in Eq. 
(\ref{decompp}) with $B^{- \rho \rho}_j$ for $B^{\rho - \rho}_j$. 
\setcounter{equation}{0}
\setcounter{section}{2}
\section{\lq\lq Multiplication'' of the two standard forms for the 
gluon part} 
We define a \lq\lq multiplication'' of the functions, $A^{\mu 
\nu}$ and $B^{\mu \nu}$, of the type, (\ref{decomp1}) with 
(\ref{standardg}), by $C^{\mu \nu} = A^{\mu \rho} B_\rho^{\;\; 
\nu}$. Straightforward computation yields the SF for $C^{\mu \nu}$: 
\widetext 
\begin{eqnarray*}
C_1^{TT} &=& A_1^{TT} B_1^{TT} + \tilde{P}^2 \tilde{\zeta}^4 
n^2 A_3^{T'T} B_3^{TT'} + \tilde{P}^2 \tilde{\zeta}^2 n^4 
A_2^{TL} B_2^{LT} - \tilde{P}^4 \tilde{\zeta}^2 n^2 A_1^{TG} 
B_1^{GT} \, , \\ 
C_2^{TT} &=& A_1^{TT} B_2^{TT} + A_2^{TT} B_1^{TT} + \tilde{\zeta}^2 
A_2^{TT} B_2^{TT} + \tilde{P}^2 \tilde{\zeta}^2 n^2 A_3^{TT'} 
B_3^{T'T} \nonumber \\ 
&& - \tilde{P}^2 \tilde{\zeta}^2 n^2 A_3^{T'T} B_3^{TT'} + 
n^2 A_1^{TL} B_1^{LT} - \tilde{P}^2 n^4 A_2^{TL} B_2^{LT} + 
\tilde{P}^4 n^2 A_1^{TG} B_1^{GT} \nonumber \\ 
&& - \tilde{P}^2 A_2^{TG} B_2^{GT} \, , \\ 
C_3^{TT'} &=& A_1^{TT} B_3^{TT'} + \tilde{\zeta}^2 A_2^{TT} 
B_3^{TT'} + A_3^{TT'} B_1^{TT} + n^2  A_1^{TL} B_2^{LT} - 
\tilde{P}^2 A_2^{TG} B_1^{GT} \, , \nonumber \\ 
C_3^{T'T} &=& A_1^{TT} B_3^{T'T} + A_3^{T'T} B_1^{TT} + 
\tilde{\zeta}^2 A_3^{T'T} B_2^{TT} + n^2 A_2^{TL} B_1^{LT} - 
\tilde{P}^2 A_1^{TG} B_2^{GT} \, , \\ 
C_1^{LL} &=& \tilde{\zeta}^2 n^2 A_1^{LT} B_1^{TL} + 
\tilde{P}^2 \tilde{\zeta}^2 n^4 A_2^{LT} B_2^{TL} + A_1^{LL} 
B_1^{LL} - \tilde{P}^2 n^2 A_1^{LG} B_1^{GL} \, , \\ 
C_1^{GG} &=& - \tilde{P}^4 \tilde{\zeta}^2 n^2 A_1^{GT} 
B_1^{TG} - \tilde{P}^2 \tilde{\zeta}^2 A_2^{GT} B_2^{TG} - 
\tilde{P}^2 n^2 A_1^{GL} B_1^{LG} + A_1^{GG} B_1^{GG} \, , \\  
C_1^{TL} &=& A_1^{TT} B_1^{TL} + \tilde{\zeta}^2 A_2^{TT} 
B_1^{TL} + \tilde{P}^2 \tilde{\zeta}^2 n^2 A_3^{TT'} B_2^{TL} 
+ A_1^{TL} B_1^{LL} - P^2 A_2^{TG} B_1^{GL} \, , \\ 
C_2^{TL} &=& A_1^{TT} B_2^{TL} + \tilde{\zeta}^2 A_3^{T'T} 
B_1^{TL} + A_2^{TL} B_1^{LL} - \tilde{P}^2 A_1^{TG} B_1^{GL} \, , 
\\ 
C_1^{LT} &=& A_1^{LT} B_1^{TT} + \tilde{\zeta}^2 A_1^{LT} 
B_2^{TT} + \tilde{P}^2 \tilde{\zeta}^2 n^2 A_2^{LT} B_3^{T'T} 
+ A_1^{LL} B_1^{LT} - \tilde{P}^2 A_1^{LG} B_2^{GT} 
\, , \\ 
C_2^{LT} &=& \tilde{\zeta}^2 A_1^{LT} B_3^{TT'} + A_2^{LT} 
B_1^{TT} + A_1^{LL} B_2^{LT} - \tilde{P}^2 A_1^{LG} B_1^{GT} \, , 
\\ 
C_1^{TG} &=& A_1^{TT} B_1^{TG} + \tilde{\zeta}^2 A_3^{T'T} 
B_2^{TG} + n^2 A_2^{TL} B_1^{LG} + A_1^{TG} B_1^{GG} \, , \\ 
C_2^{TG} &=& A_1^{TT} B_2^{TG} + \tilde{\zeta}^2 A_2^{TT} 
B_2^{TG} + \tilde{P}^2 \tilde{\zeta}^2 n^2 A_3^{TT'} B_1^{TG} 
+ n^2 A_1^{TL} B_1^{LG} + A_2^{TG} B_1^{GG} \, , \\ 
C_1^{GT} &=& A_1^{GT} B_1^{TT} + \tilde{\zeta}^2 A_2^{GT} 
B_3^{TT'} + n^2 A_1^{GL} B_2^{LT} + A_1^{GG} B_1^{GT} \, , \\ 
C_2^{GT} &=& \tilde{P}^2 \tilde{\zeta}^2 n^2 A_1^{GT} 
B_3^{TT'} + A_2^{GT} B_1^{TT} + \tilde{\zeta}^2 A_2^{GT} 
B_2^{TT} + \tilde{n}^2 A_1^{GL} B_1^{LT} \, , \\ 
C_1^{LG} &=& \tilde{\zeta}^2 A_1^{LT} B_2^{TG} + \tilde{P}^2 
\tilde{\zeta}^2 n^2 A_2^{LT} B_1^{TG} + A_1^{LL} B_1^{LG} 
+ A_1^{LG} B_1^{GG} \, , \\ 
C_1^{GL} &=& \tilde{P}^2 \tilde{\zeta}^2 n^2 A_1^{GT} 
B_2^{TL} + \tilde{\zeta}^2 A_2^{GT} B_1^{TL} + A_1^{GL} B_1^{LL} 
+ A_1^{GG} B_1^{GL} \, . 
\end{eqnarray*}
\narrowtext 
\setcounter{equation}{0}
\setcounter{section}{3}
\section{Gluon propagator in a covariant gauge} 
\def\theequation{\mbox{\Alph{section}.\arabic{equation}}}
Here we present a \lq\lq translation table'' to get the expressions 
for the gluon propagator in a covariant-gauge from the Coulomb-gauge 
counterparts given in Sec. III. 

An orthogonal basis in Minkowski space is given by Eq. 
(\ref{orth2}) with the replacements\footnote{It should be noted 
that $E_\perp^\mu = \epsilon^{\mu \nu \rho \sigma} P_\nu 
\tilde{\zeta}_\rho \tilde{n}_\sigma = \epsilon^{\mu \nu \rho \sigma} 
\tilde{P}_\nu \tilde{\zeta}_\rho n_\sigma$.}, 
\begin{eqnarray} 
\left( \tilde{P}^\mu, \tilde{\zeta}^\mu, n^\mu, E_\perp^\mu \right) 
\;\; &\Rightarrow& \;\; \left( P^\mu, \tilde{\zeta}^\mu, 
\tilde{n}^\mu, E_\perp^\mu \right) \, , 
\label{oki} \\ 
\tilde{n}^\mu &\equiv& n^\mu - \frac{n \cdot P}{P^2} \, P^\mu = 
n^\mu - \frac{p_0}{P^2} \, P^\mu \nonumber \\ 
&& \mbox{\hspace*{10ex}} (\tilde{n}^2 = - \frac{\vec{p}^{\, 
2}}{P^2}) \nonumber \, . 
\end{eqnarray} 
Then, among the projection operators, Eqs. (\ref{Pro1}) - 
(\ref{Pro3}), ${\cal P}_L^{\mu \nu}$ and ${\cal P}_G^{\mu \nu}$ are 
replaced as 
\begin{eqnarray*}
{\cal P}_L^{\mu \nu} (P) = \frac{n^\mu n^\nu}{n^2} & \Rightarrow & 
{\cal P}_L^{\mu \nu} (P) = \frac{\tilde{n}^\mu 
\tilde{n}^\nu}{\tilde{n}^2} \, , 
\\ 
{\cal P}_G^{\mu \nu} (P) = \frac{\tilde{P}^\mu 
\tilde{P}^\nu}{\tilde{P}^2} & \Rightarrow & {\cal P}_G^{\mu \nu} 
(P) = \frac{P^\mu P^\nu}{P^2} \, . 
\end{eqnarray*}
${\cal P}_T$ is the same as in Eq. (\ref{Pro1}). 

Eq. (\ref{2.8d}) is replaced with 
\[ 
(\hat{D}^{-1} (P))^{\mu \nu} = - P^2 \left[ {\cal P}_T^{\mu \nu} + 
{\cal P}_L^{\mu \nu} + \frac{1}{\lambda} {\cal P}_G^{\mu \nu} 
\right] \hat{\tau}_3 \, , 
\] 
which is already in SF. 

The SF-elements of ${\bf D}_K$ in Eq. (\ref{Kdesu}) are replaced by 
\begin{eqnarray*}
D^{TT}_{K2} (P) & = & 2 \pi i C_2^{TT} (P) \epsilon (p_0) \delta 
(P^2) \, , \\ 
D^{TT'}_{K3} (P) & = & 2 \pi i C_3^{TT'} (P) \epsilon (p_0) \delta 
(P^2) \, , \\ 
D^{TL}_{K1} (P) & = & 2 \pi i C_1^{TL} (P) \epsilon (p_0) \delta 
(P^2) \, , \\ 
D^{TL}_{K2} (P) & = & 2 \pi i C_2^{TL} (P) \epsilon (p_0) \delta 
(P^2) \, , \\ 
D^{UV}_{Kj} (P) &=& 0 \mbox{\hspace*{10ex} (otherwise)} \, . 
\end{eqnarray*} 
In obtaining these, we have used the fact that $\left( \hat{D}_{UG} 
\right)^{\mu \nu} = \left( \hat{D}_{GU} \right)^{\mu \nu} = 0$ 
($U = T, L$), which is verified from the \lq\lq bare counterparts'' 
of Eq. (\ref{awa}), below. $D_R^{\mu \nu}$ in Eq. (\ref{Ryodesu}) is 
replaced with 
\[
D_R^{\mu \nu} = \left( D_A^{\mu \nu} \right)^* = - \Delta_R {\cal 
P}_T^{\mu \nu} + \frac{d \Delta_R}{d P^2} P^2 \left( {\cal P}_L^{\mu 
\nu} + \lambda {\cal P}_G^{\mu \nu} \right) \, . 
\]

Eq. (\ref{mujina}) is replaced with 
\[
\begin{array}{ll} 
\left( C_2^{TT} \right)^* = C_2^{TT} \, , & \;\;\;\;\;\; \left( 
C_3^{TT'} \right)^* = - C_3^{T'T} \, , \\ 
\left( C_1^{TL} \right)^* = C_1^{LT} \, , & \;\;\;\;\;\; \left( 
C_2^{TL} \right)^* = - C_2^{LT} \, . 
\end{array} 
\]

Eq. (\ref{ghostp}) is replaced by 
\[
\tilde{\hat{D}} (P) = \left( 
\begin{array}{cc} 
\Delta_R & \;\; 0 \\ 
\Delta_R - \Delta_A & \;\; - \Delta_A 
\end{array} 
\right) + \tilde{f} (\Delta_R - \Delta_A) \hat{M}_+ \, . 
\]

Introduction of ${\bf \Pi}_R'$, Eq. (\ref{asita}), is not necessary, 
${\bf D}_0^{- 1} = {\bf D}^{- 1}$ and ${\bf \Pi}_R' = {\bf \Pi}_R$, 
and, Eqs. (\ref{asita}) - (\ref{dash}) are deleted. 

Description after Eq. (\ref{SDghost}) is replaced with the following 
one: Solving Eq. (\ref{SDghost}), we obtain 
\[
\hat{\tilde{G}} = \left( 
\begin{array}{cc}
\tilde{G}_R & \;\; 0 \\ 
\tilde{G}_R - \tilde{G}_A & \;\; - \tilde{G}_A 
\end{array}
\right) + \left[ \tilde{f} (\tilde{G}_R - \tilde{G}_A) + \tilde{G}_K 
\right] \hat{M}_+ \, , 
\]
where 
\begin{eqnarray*}
\tilde{G}_R (P) &=& \tilde{G}_A^* (P) = \frac{1}{P^2 - \tilde{\Pi}_R 
(P)} \, , \\ 
\tilde{G}_K (P) &=& - \tilde{G}_R (P) \tilde{\Pi}_K (P)\tilde{G}_A 
(P) \, , \\ 
\tilde{\Pi}_R & = & \tilde{\Pi}_A^* = \tilde{\Pi}_{11} + 
\tilde{\Pi}_{12} = - \tilde{\Pi}_{22} - \tilde{\Pi}_{21} \, , \\ 
\tilde{\Pi}_K &=& (1 + \tilde{f}) \tilde{\Pi}_{11} - \tilde{f} 
\tilde{\Pi}_{21} \, . 
\end{eqnarray*}

Eq. (\ref{ST}) is replaced with 
\begin{equation} 
\hat{G}_{\mu \nu} P^\nu = \lambda \left[ \hat{\tau}_3 
\hat{\tilde{\Pi}}_\mu - P_\mu \right] \hat{\tilde{G}} \, . 
\label{awa} 
\end{equation} 
Eq. (\ref{preg}) is deleted. 

Eqs. (\ref{GP3}) - (\ref{GK1}) are replaced by 
\widetext 
\begin{eqnarray*} 
G_{R1}^{GG} (P) &=& = \lambda \frac{1}{P^2 + i p_0 0^+} \left[ 
\tilde{\Pi} (P) - P^2 \right] \tilde{G}_R (P) = - \lambda 
\frac{1}{P^2 + i p_0 0^+} \, , \\ 
G_{R1}^{TG} (P) &=& \lambda \frac{1}{(P^2 + i p_0 0^+) E_\perp^2} 
\left( E_\perp^\mu \tilde{\Pi}_{R \mu} (P) \right) \tilde{G}_R (P) 
\, , \\ 
G_{R2}^{TG} (P) &=& \lambda \frac{1}{(P^2 + i p_0 0^+) 
\tilde{\zeta}^2} \left( \tilde{\zeta}^\mu \tilde{\Pi}_{R \mu} (P) 
\right) \tilde{G}_R (P) \, , \\ 
G_{R1}^{LG} (P) &=& \lambda \frac{1}{(P^2 + i p_0 0^+) \tilde{n}^2} 
\left( \tilde{n}^\mu \tilde{\Pi}_{R \mu} (P) \right) \tilde{G}_R (P) 
\, , \\ 
G_{K1}^{GG} (P) &=& - \lambda \left[ \frac{P^2 \left( P^2 - 
\tilde{\Pi}_R \right) \tilde{G}_K}{(P^2 + i 0^+) (P^2 - i 
0^+)} + \frac{ \tilde{\Pi}_K \tilde{G}_A}{P^2 - i p_0 0^+} 
\right] = 0 \, , \\ 
G_{K1}^{TG} (P) &=& \lambda \frac{1}{E_\perp^2} \left[ 
\frac{E_\perp^\mu \tilde{\Pi}_{R \mu} \, P^2 \tilde{G}_K}{(P^2 
+ i 0^+) (P^2 - i 0^+)} - \frac{ E_\perp^\mu \tilde{\Pi}_{K \mu} 
\, \tilde{G}_A}{P^2 - i p_0 0^+} \right] \, , \\ 
G_{K2}^{TG} (P) &=& \lambda \frac{1}{\tilde{\zeta}^2} \left[ 
\frac{\tilde{\zeta}^\mu \tilde{\Pi}_{R \mu} \, P^2 \tilde{G}_K}{(P^2 
+ i 0^+) (P^2 - i 0^+)} - \frac{\tilde{\zeta}^\mu \tilde{\Pi}_{K 
\mu} \, \tilde{G}_A}{P^2 - i p_0 0^+} \right] \, , \\ 
G_{K1}^{LG} (P) &=& \lambda \frac{1}{\tilde{n}^2} \left[ 
\frac{\tilde{n}^\mu \tilde{\Pi}_{R \mu} \, P^2 \tilde{G}_K}{(P^2 + i 
0^+) (P^2 - i 0^+)} - \frac{\tilde{n}^\mu \tilde{\Pi}_{K \mu} \, 
\tilde{G}_A}{P^2 - i p_0 0^+} \right] \, . 
\end{eqnarray*}

In Eqs. (\ref{zero1}) - (\ref{GM2}), the replacements (\ref{oki}) 
and $\left( \Pi_j' \right)_{UV}$'s $\to$ $\left( \Pi_j 
\right)_{UV}$'s are made, and, in the formulae in Appendix C, the 
replacement (\ref{oki}) is made. 
\setcounter{equation}{0}
\setcounter{section}{4}
\section{Standard forms for the quantities in Sec. IV A} 
\def\theequation{\mbox{\Alph{section}.\arabic{equation}}}
\subsubsection*{Standard form for $\Sigma_K^{[1]}$} 
From Eq. (\ref{SiK1}), we obtain, after some algebra, 
\begin{eqnarray*}
\Sigma_K^{[1]} &=& - \frac{i}{2} \sum_{\rho = \pm} {\cal P}_\rho 
\left[ 2 \left\{ f_\rho , \;\, Re \Sigma_R^{\rho \rho} \right\}' 
\cdot {\cal P}_\rho + \left\{ f_{- \rho} , \;\, \Sigma_R^{\rho - 
\rho} \right\}' \cdot {\cal P}_{- \rho} + \left\{ f_\rho , \;\, 
\Sigma_A^{\rho - \rho} \right\}' {\cal P}_{- \rho} \right. \nonumber 
\\ 
&& + 2 \frac{\partial f_\rho}{\partial X^\alpha} Re \left\{ 
\frac{P\kern-0.1em\raise0.3ex\llap{/}\kern0.15em\relax}{P^2} \left( 
P^\alpha \Sigma_{R2}^{\rho \rho} - N^\alpha \Sigma_{R3}^{\rho \rho} 
- \rho \epsilon (p_0) e_\perp^\alpha \Sigma_{R4}^{\rho \rho} \right) 
\right. \nonumber \\ 
&& + \frac{N\kern-0.153em\raise0.3ex\llap{/}\kern0.177em\relax}{N^2} 
\left( N^\alpha \Sigma_{R2}^{\rho \rho} + \frac{1}{2} \frac{\partial 
N^2}{\partial P_\alpha} \Sigma_{R3}^{\rho \rho} - \rho \epsilon 
(p_0) e_\perp^\mu \frac{\partial N^\mu}{\partial P_\alpha} 
\Sigma_{R4}^{\rho \rho} \right) \nonumber \\ 
&& \left. + \frac{P\kern-0.1em\raise0.3ex\llap{/}\kern0.15em\relax 
N\kern-0.153em\raise0.3ex\llap{/}\kern0.177em\relax}{P^2 N^2} \rho 
\epsilon (p_0) \left( - e_\perp^\alpha \Sigma_{R2}^{\rho \rho} - 
e_\perp^\mu \frac{\partial N_\mu}{\partial P_\alpha} 
\Sigma_{R3}^{\rho \rho} + \rho \epsilon (p_0) N^2 \hat{P}^\alpha 
\Sigma_{R4}^{\rho \rho} \right) \right\} \nonumber \\ 
&& + \gamma_5 \left\{ 
\frac{P\kern-0.1em\raise0.3ex\llap{/}\kern0.15em\relax}{P^2} \left( 
P^\alpha (\Lambda_2^{\rho - \rho})_\alpha - N^\alpha 
(\Lambda_3^{\rho - \rho})_\alpha + \rho \epsilon (p_0) 
e_\perp^\alpha (\Lambda_4^{\rho - \rho})_\alpha \right) \right. 
\nonumber \\ 
&& + \frac{N\kern-0.153em\raise0.3ex\llap{/}\kern0.177em\relax}{N^2} 
\left( N^\alpha (\Lambda_2^{\rho - \rho})_\alpha + \frac{1}{2} 
\frac{\partial N^2}{\partial P_\alpha} (\Lambda_3^{\rho - 
\rho})_\alpha + \rho \epsilon (p_0) e_\perp^\mu \frac{\partial 
N^\mu}{\partial P_\alpha} (\Lambda_4^{\rho - \rho})_\alpha \right) 
\nonumber \\ 
&& \left. \left. + 
\frac{P\kern-0.1em\raise0.3ex\llap{/}\kern0.15em\relax 
N\kern-0.153em\raise0.3ex\llap{/}\kern0.177em\relax}{P^2 N^2} \rho 
\epsilon (p_0) \left( e_\perp^\alpha (\Lambda_2^{\rho - 
\rho})_\alpha + e_\perp^\mu \frac{\partial N_\mu}{\partial P_\alpha} 
(\Lambda_3^{\rho - \rho})_\alpha + \rho \epsilon (p_0) N^2 
\hat{P}^\alpha (\Lambda_4^{\rho - \rho})_\alpha \right) \right\} 
\right] \, , 
\end{eqnarray*} 
where 
\begin{eqnarray*} 
\hat{P}^\alpha &\equiv & P^\alpha + \frac{P^2}{2 N^2} 
\frac{\partial N^2}{\partial P_\alpha} \, , \\ 
(\Lambda_j^{\rho - \rho})_\alpha & \equiv & \Sigma_{Rj}^{\rho - 
\rho} \frac{\partial f^{- \rho}}{\partial X^\alpha} + 
\Sigma_{Aj}^{\rho - \rho} \frac{\partial f^\rho}{\partial 
X^\alpha} \;\;\;\;\;\;\;\; (j = 2, 3, 4) \, , \\ 
\left\{ f_{- \rho}, \; \, \Sigma_R^{\rho - \rho} \right\}' & \equiv 
& \gamma_5 \left[ \left\{ f _{- \rho} , \;\, \Sigma_{R1}^{\rho - 
\rho} \right\} + \left\{ f _{- \rho} , \;\, \Sigma_{R2}^{\rho - 
\rho} \right\} P\kern-0.1em\raise0.3ex\llap{/}\kern0.15em\relax 
\right. \nonumber \\ 
&& \left. + \left\{ f _{- \rho} , \;\, \Sigma_{R3}^{\rho - \rho} 
\right\} N\kern-0.153em\raise0.3ex\llap{/}\kern0.177em\relax 
+ \left\{ f _{- \rho} , \;\, \Sigma_{R4}^{\rho - \rho} \right\} 
P\kern-0.1em\raise0.3ex\llap{/}\kern0.15em\relax 
N\kern-0.153em\raise0.3ex\llap{/}\kern0.177em\relax \right] \, , 
\end{eqnarray*} 
etc. Here, $\left\{ ...,  \;\, ... \right\}$ is as in Eq. 
(\ref{Poi}). 
\subsubsection*{Standard form for $\underline{H}$ in $G_K^{[2]}$ in 
Eq. (\ref{GK2yo})} 
Straightforward manipulation of Eq. (\ref{GK2yo}) yields 
\begin{eqnarray}
\underline{H} &=& \sum_{\rho, \, \sigma = \pm} \underline{\cal 
P}_\rho \cdot \left[ H_l^{\rho \sigma} 
\right]_{\mbox{\scriptsize{IWT}}} \cdot \underline{\cal P}_\sigma + 
\underline{H}^{(1)} \, , \nonumber \\ 
H^{(1)} &=& \frac{i}{2} \sum_{\rho, \, \sigma = \pm} {\cal P}_\rho 
\gamma_5 \left( \frac{\partial 
N\kern-0.153em\raise0.3ex\llap{/}\kern0.177em\relax}{\partial 
P_\alpha} \frac{\partial C_{\rho - \rho} \Sigma_A^{- \rho 
\sigma}}{\partial X^\alpha} - 
N\kern-0.153em\raise0.3ex\llap{/}\kern0.177em\relax \left\{ 
C_{\rho - \rho} , \;\, \Sigma_A^{- \rho \sigma} \right\} \right) 
{\cal P}_\sigma \nonumber \\ 
&& + \frac{i}{2} \sum_{\rho, \, \sigma = \pm} {\cal P}_\rho \left( - 
\frac{\partial \Sigma_R^{\rho \sigma} C_{\sigma - \sigma}}{\partial 
X^\alpha} \frac{\partial 
N\kern-0.153em\raise0.3ex\llap{/}\kern0.177em\relax}{\partial 
P_\alpha} + \left\{ C_{\sigma - \sigma}, \;\, \Sigma_R^{\rho \sigma} 
\right\} N\kern-0.153em\raise0.3ex\llap{/}\kern0.177em\relax \right) 
\gamma_5 {\cal P}_{- \sigma} \, , 
\label{nagai2} 
\end{eqnarray} 
where $H_l^{\rho \sigma}$ is as in Eq. (\ref{Hlead}). The SF for 
each term on the RHS of Eq. (\ref{nagai2}) reads 
\begin{eqnarray*}
{\cal P}_\rho \gamma_5 \frac{\partial 
N\kern-0.153em\raise0.3ex\llap{/}\kern0.177em\relax}{\partial 
P_\alpha} \frac{\partial C_{\rho - \rho} \Sigma^{- \rho 
\rho}_A}{\partial X^\alpha} {\cal P}_\rho &=& {\cal P}_\rho 
\frac{\partial N_\mu}{\partial P_\alpha} \frac{\partial}{\partial 
X^\alpha} \left[ C_{\rho - \rho} \left\{- P^\mu \Sigma_{A2}^{- \rho 
\rho} - N^\mu \Sigma_{A3}^{- \rho \rho} - \rho \epsilon (p_0) 
e_\perp^\mu \Sigma_{A4}^{- \rho \rho} \right. \right. \nonumber \\ 
&& + \frac{P\kern-0.1em\raise0.3ex\llap{/}\kern0.15em\relax}{P^2} 
\left( - P^\mu \Sigma_{A1}^{- \rho \rho} + P^2 N^\mu \Sigma_{A4}^{- 
\rho \rho} + \rho \epsilon (p_0) e_\perp^\mu \Sigma_{A3}^{- \rho 
\rho} \right) \nonumber \\ 
&& + \frac{N\kern-0.153em\raise0.3ex\llap{/}\kern0.177em\relax}{N^2} 
\left( - N^2 P^\mu \Sigma_{A4}^{- \rho \rho} - N^\mu \Sigma_{A1}^{- 
\rho \rho} - \rho \epsilon (p_0) e_\perp^\mu \Sigma_{A2}^{- \rho 
\rho} \right) \nonumber \\ 
&& \left. \left. + 
\frac{P\kern-0.1em\raise0.3ex\llap{/}\kern0.15em\relax 
N\kern-0.153em\raise0.3ex\llap{/}\kern0.177em\relax}{P^2 N^2} \left( 
- N^2 P^\mu \Sigma_{A3}^{- \rho \rho} + P^2 N^\mu \Sigma_{A2}^{- 
\rho \rho} + \rho \epsilon (p_0) e_\perp^\mu \Sigma_{A1}^{- \rho 
\rho} \right) \right\} \right] \, , 
\end{eqnarray*} 
\begin{eqnarray*}
{\cal P}_\rho \gamma_5 \frac{\partial 
N\kern-0.153em\raise0.3ex\llap{/}\kern0.177em\relax}{\partial 
P_\alpha} \frac{\partial C_{\rho - \rho} \Sigma^{- \rho - 
\rho}_A}{\partial X^\alpha} {\cal P}_{- \rho} &=& {\cal P}_\rho 
\frac{\partial N_\mu}{\partial P_\alpha} \gamma_5 
\frac{\partial}{\partial X^\alpha} \left[ C_{\rho - \rho} \left\{ 
P^\mu \Sigma_{A2}^{- \rho - \rho} + N^\mu \Sigma_{A3}^{- \rho - 
\rho} - \rho \epsilon (p_0) e_\perp^\mu \Sigma_{A4}^{- \rho - \rho} 
\right. \right. \nonumber \\ 
&& + \frac{P\kern-0.1em\raise0.3ex\llap{/}\kern0.15em\relax}{P^2} 
\left( P^\mu \Sigma_{A1}^{- \rho - \rho} - P^2 N^\mu \Sigma_{A4}^{- 
\rho - \rho} + \rho \epsilon (p_0) e_\perp^\mu \Sigma_{A3}^{- \rho - 
\rho} \right) \nonumber \\ 
&& + \frac{N\kern-0.153em\raise0.3ex\llap{/}\kern0.177em\relax}{N^2} 
\left( N^2 P^\mu \Sigma_{A4}^{- \rho - \rho} + N^\mu \Sigma_{A1}^{- 
\rho - \rho} - \rho \epsilon (p_0) e_\perp^\mu \Sigma_{A2}^{- \rho - 
\rho} \right) \nonumber \\ 
&& \left. \left. + 
\frac{P\kern-0.1em\raise0.3ex\llap{/}\kern0.15em\relax 
N\kern-0.153em\raise0.3ex\llap{/}\kern0.177em\relax}{P^2 N^2} \left( 
N^2 P^\mu \Sigma_{A3}^{- \rho - \rho} -  P^2 N^\mu \Sigma_{A2}^{- 
\rho - \rho} + \rho \epsilon (p_0) e_\perp^\mu \Sigma_{A1}^{- \rho - 
\rho} \right) \right\} \right] \, , 
\end{eqnarray*} 
\begin{eqnarray*}
{\cal P}_\rho \frac{\partial \Sigma^{\rho \rho}_R C_{\rho - 
\rho}}{\partial X^\alpha} \frac{\partial 
N\kern-0.153em\raise0.3ex\llap{/}\kern0.177em\relax}{\partial 
P_\alpha} \gamma_5 {\cal P}_{- \rho} &=& - {\cal P}_\rho 
\frac{\partial N_\mu}{\partial P_\alpha} \gamma_5 
\frac{\partial}{\partial X^\alpha} \left[ \left\{ - P^\mu 
\Sigma_{R2}^{\rho \rho} - N^\mu \Sigma_{R3}^{\rho \rho} - \rho 
\epsilon (p_0) e_\perp^\mu \Sigma_{R4}^{\rho \rho} \right. \right. 
\nonumber \\ 
&& + \frac{P\kern-0.1em\raise0.3ex\llap{/}\kern0.15em\relax}{P^2} 
\left( P^\mu \Sigma_{R1}^{\rho \rho} + \rho \epsilon (p_0) 
e_\perp^\mu \Sigma_{R3}^{\rho \rho} + P^2 N^\mu \Sigma_{R4}^{\rho 
\rho} \right) \nonumber \\ 
&& + \frac{N\kern-0.153em\raise0.3ex\llap{/}\kern0.177em\relax}{N^2} 
\left( N^\mu \Sigma_{R1}^{\rho \rho} - \rho \epsilon (p_0) 
e_\perp^\mu \Sigma_{R2}^{\rho \rho} - N^2 P^\mu \Sigma_{R4}^{\rho 
\rho} \right) \nonumber \\ 
&& \left. \left. + 
\frac{P\kern-0.1em\raise0.3ex\llap{/}\kern0.15em\relax 
N\kern-0.153em\raise0.3ex\llap{/}\kern0.177em\relax}{P^2 N^2} \left( 
\rho \epsilon (p_0) e_\perp^\mu \Sigma_{R1}^{\rho \rho} - P^2 N^\mu 
\Sigma_{R2}^{\rho \rho} + N^2 P^\mu \Sigma_{R3}^{\rho \rho} \right) 
\right\} C_{\rho - \rho} \right] \, , 
\end{eqnarray*} 
\begin{eqnarray*}
{\cal P}_\rho \frac{\partial \Sigma^{\rho - \rho}_R C_{- \rho 
\rho}}{\partial X^\alpha} \frac{\partial 
N\kern-0.153em\raise0.3ex\llap{/}\kern0.177em\relax}{\partial 
P_\alpha} \gamma_5 {\cal P}_\rho &=& - {\cal P}_\rho \frac{\partial 
N_\mu}{\partial P_\alpha} \frac{\partial}{\partial X^\alpha} \left[ 
\left\{ - P^\mu \Sigma_{R2}^{\rho - \rho} - N^\mu \Sigma_{R3}^{\rho 
- \rho} + \rho \epsilon (p_0) e_\perp^\mu \Sigma_{R4}^{\rho - \rho} 
\right. \right. \nonumber \\ 
&& + \frac{P\kern-0.1em\raise0.3ex\llap{/}\kern0.15em\relax}{P^2} 
\left( P^\mu \Sigma_{R1}^{\rho - \rho} - \rho \epsilon (p_0) 
e_\perp^\mu \Sigma_{R3}^{\rho - \rho} + P^2 N^\mu \Sigma_{R4}^{\rho 
- \rho} \right) \nonumber \\ 
&& + \frac{N\kern-0.153em\raise0.3ex\llap{/}\kern0.177em\relax}{N^2} 
\left( N^\mu \Sigma_{R1}^{\rho - \rho} + \rho \epsilon (p_0) 
e_\perp^\mu \Sigma_{R2}^{\rho - \rho} - N^2 P^\mu \Sigma_{R4}^{\rho 
- \rho} \right) \nonumber \\ 
&& \left. \left. + 
\frac{P\kern-0.1em\raise0.3ex\llap{/}\kern0.15em\relax 
N\kern-0.153em\raise0.3ex\llap{/}\kern0.177em\relax}{P^2 N^2} \left( 
- \rho \epsilon (p_0) e_\perp^\mu \Sigma_{R1}^{\rho - \rho} - P^2 
N^\mu \Sigma_{R2}^{\rho - \rho} + N^2 P^\mu \Sigma_{R3}^{\rho - 
\rho} \right) \right\} C_{- \rho \rho} \right] \, , 
\end{eqnarray*} 
\begin{eqnarray*}
{\cal P}_\rho \gamma_5 
N\kern-0.153em\raise0.3ex\llap{/}\kern0.177em\relax \left\{ C_{\rho 
- \rho} , \, \Sigma^{- \rho \rho}_A \right\} {\cal P}_\rho &=& - 
{\cal P}_\rho \left[ N^2 \left\{ C_{\rho - \rho} , \;\, \Sigma^{- 
\rho \rho}_{A3} \right\} - N^2 
P\kern-0.1em\raise0.3ex\llap{/}\kern0.15em\relax \left\{ C_{\rho - 
\rho} , \;\, \Sigma^{- \rho \rho}_{A4} \right\} \right. \nonumber \\ 
&& \left. + N\kern-0.153em\raise0.3ex\llap{/}\kern0.177em\relax 
\left\{ C_{\rho - \rho} , \;\, \Sigma^{- \rho \rho}_{A1} \right\} - 
P\kern-0.1em\raise0.3ex\llap{/}\kern0.15em\relax 
N\kern-0.153em\raise0.3ex\llap{/}\kern0.177em\relax 
\left\{ C_{\rho - \rho} , \;\, \Sigma^{- \rho \rho}_{A2} \right\} 
\right] {\cal P}_\rho \nonumber \\ 
&& - \frac{\partial C_{\rho - \rho}}{\partial X^\alpha} {\cal 
P}_\rho \left[ N^\alpha \Sigma_{A2}^{- \rho \rho}  + \frac{1}{2} 
\frac{\partial N^2}{\partial P_\alpha} \Sigma_{A3}^{- \rho \rho} - 
\rho \epsilon (p_0) e_\perp^\mu \frac{\partial N_\mu}{\partial 
P_\alpha} \Sigma_{A4}^{- \rho \rho} \right. \nonumber \\ 
&& + \frac{P\kern-0.1em\raise0.3ex\llap{/}\kern0.15em\relax}{P^2} 
\left( \rho \epsilon (p_0) e_\perp^\alpha \Sigma_{A2}^{- \rho \rho} 
+ \rho \epsilon (p_0) e_\perp^\mu \frac{\partial N_\mu}{\partial 
P_\alpha} \Sigma_{A3}^{- \rho \rho} - N^2 \hat{P}^\alpha 
\Sigma_{A4}^{ - \rho \rho} \right) \nonumber \\ 
&& \left. + \frac{P\kern-0.1em\raise0.3ex\llap{/}\kern0.15em\relax 
N\kern-0.153em\raise0.3ex\llap{/}\kern0.177em\relax}{P^2} \left( - 
P^\alpha \Sigma_{A2}^{- \rho \rho} + N^\alpha \Sigma_{A3}^{- \rho 
\rho} + \rho \epsilon (p_0) e_\perp^\alpha \Sigma_{A4}^{- \rho \rho} 
\right) \right] {\cal P}_\rho \, , 
\end{eqnarray*} 
\begin{eqnarray*}
{\cal P}_\rho \gamma_5 
N\kern-0.153em\raise0.3ex\llap{/}\kern0.177em\relax \left\{ C_{\rho 
- \rho} , \, \Sigma^{- \rho - \rho}_A \right\} {\cal P}_{- \rho} 
&=& {\cal P}_\rho \gamma_5 \left[ N^2 \left\{ C_{\rho - \rho} , \;\, 
\Sigma^{- \rho - \rho}_{A3} \right\} - N^2 
P\kern-0.1em\raise0.3ex\llap{/}\kern0.15em\relax \left\{ C_{\rho - 
\rho} , \;\, \Sigma^{- \rho - \rho}_{A4} \right\} \right. \nonumber 
\\ 
&& \left. + N\kern-0.153em\raise0.3ex\llap{/}\kern0.177em\relax 
\left\{ C_{\rho - \rho} , \;\, \Sigma^{- \rho - \rho}_{A1} \right\} 
- P\kern-0.1em\raise0.3ex\llap{/}\kern0.15em\relax 
N\kern-0.153em\raise0.3ex\llap{/}\kern0.177em\relax 
\left\{ C_{\rho - \rho} , \;\, \Sigma^{- \rho - \rho}_{A2} \right\} 
\right] {\cal P}_{- \rho} \nonumber \\ 
&& + \frac{\partial C_{\rho - \rho}}{\partial X^\alpha} {\cal 
P}_\rho \gamma_5 \left[ N^\alpha \Sigma_{A2}^{- \rho - \rho} + 
\frac{1}{2} \frac{\partial N^2}{\partial P_\alpha} \Sigma_{A3}^{- 
\rho - \rho} + \rho \epsilon (p_0) e_\perp^\mu \frac{\partial 
N_\mu}{\partial P_\alpha} \Sigma_{A4}^{- \rho - \rho} \right. 
\nonumber \\ 
&& + \frac{P\kern-0.1em\raise0.3ex\llap{/}\kern0.15em\relax}{P^2} 
\left( - \rho \epsilon (p_0) e_\perp^\alpha \Sigma_{A2}^{- \rho - 
\rho} - \rho \epsilon (p_0) e_\perp^\mu \frac{\partial 
N_\mu}{\partial P_\alpha} \Sigma_{A3}^{- \rho - \rho} - N^2 
\hat{P}^\alpha \Sigma_{A4}^{- \rho - \rho} \right) \nonumber \\ 
&& \left. + \frac{P\kern-0.1em\raise0.3ex\llap{/}\kern0.15em\relax 
N\kern-0.153em\raise0.3ex\llap{/}\kern0.177em\relax}{P^2} \left( - 
P^\alpha \Sigma_{A2}^{- \rho - \rho} + N^\alpha \Sigma_{A3}^{- \rho 
- \rho} - \rho \epsilon (p_0) e_\perp^\alpha \Sigma_{A4}^{- \rho - 
\rho} \right) \right] {\cal P}_{- \rho} \, , 
\end{eqnarray*} 
\begin{eqnarray*}
{\cal P}_\rho \left\{ C_{\rho - \rho} , \, \Sigma^{\rho \rho}_R 
\right\} N\kern-0.153em\raise0.3ex\llap{/}\kern0.177em\relax 
\gamma_5 {\cal P}_{- \rho} &=& {\cal P}_\rho \gamma_5 \left[ 
N^2 \left\{ C_{\rho - \rho} , \;\, \Sigma^{\rho \rho}_{R3} \right\} 
- N^2 P\kern-0.1em\raise0.3ex\llap{/}\kern0.15em\relax \left\{ 
C_{\rho - \rho} , \;\, \Sigma^{\rho \rho}_{R4} \right\} \right. 
\nonumber \\ 
&& \left. - N\kern-0.153em\raise0.3ex\llap{/}\kern0.177em\relax 
\left\{ C_{\rho - \rho} , \;\, \Sigma^{\rho \rho}_{R1} \right\} + 
P\kern-0.1em\raise0.3ex\llap{/}\kern0.15em\relax 
N\kern-0.153em\raise0.3ex\llap{/}\kern0.177em\relax 
\left\{ C_{\rho - \rho} , \;\, \Sigma^{\rho \rho}_{R2} \right\} 
\right] {\cal P}_{- \rho} \nonumber \\ 
&& + \frac{\partial C_{\rho - \rho}}{\partial X^\alpha} {\cal 
P}_\rho \gamma_5 \left[ N^\alpha \Sigma_{R2}^{\rho \rho} + 
\frac{1}{2} \frac{\partial N^2}{\partial P_\alpha} \Sigma_{R3}^{\rho 
\rho} - \rho \epsilon (p_0) e_\perp^\mu \frac{\partial 
N_\mu}{\partial P_\alpha} \Sigma_{R4}^{\rho \rho} \right. \nonumber 
\\ 
&& + \frac{P\kern-0.1em\raise0.3ex\llap{/}\kern0.15em\relax}{P^2} 
\left( \rho \epsilon (p_0) e_\perp^\alpha \Sigma_{R2}^{\rho \rho} + 
\rho \epsilon (p_0) e_\perp^\mu \frac{\partial N_\mu}{\partial 
P_\alpha} \Sigma_{R3}^{\rho \rho} - N^2 \hat{P}^\alpha 
\Sigma_{R4}^{\rho \rho} \right) \nonumber \\ 
&& \left. + \frac{P\kern-0.1em\raise0.3ex\llap{/}\kern0.15em\relax 
N\kern-0.153em\raise0.3ex\llap{/}\kern0.177em\relax}{P^2} \left( 
P^\alpha \Sigma_{R2}^{\rho \rho} - N^\alpha \Sigma_{R3}^{\rho \rho} 
- \rho \epsilon (p_0) e_\perp^\alpha \Sigma_{R4}^{\rho \rho} \right) 
\right] {\cal P}_{- \rho} \, , 
\end{eqnarray*} 
\begin{eqnarray*}
{\cal P}_\rho \left\{ C_{- \rho \rho} , \, \Sigma^{\rho - \rho}_R 
\right\} N\kern-0.153em\raise0.3ex\llap{/}\kern0.177em\relax 
\gamma_5 {\cal P}_\rho &=& {\cal P}_\rho \left[ N^2 \left\{ C_{- 
\rho \rho} , \;\, \Sigma^{\rho - \rho}_{R3} \right\} - N^2 
P\kern-0.1em\raise0.3ex\llap{/}\kern0.15em\relax \left\{ C_{\rho - 
\rho} , \;\, \Sigma^{\rho - \rho}_{R4} \right\} \right. \nonumber \\ 
&& \left. - N\kern-0.153em\raise0.3ex\llap{/}\kern0.177em\relax 
\left\{ C_{\rho - \rho} , \;\, \Sigma^{\rho - \rho}_{R1} \right\} + 
P\kern-0.1em\raise0.3ex\llap{/}\kern0.15em\relax 
N\kern-0.153em\raise0.3ex\llap{/}\kern0.177em\relax \left\{ C_{\rho 
- \rho} , \;\, \Sigma^{\rho - \rho}_{R2} \right\} \right] {\cal 
P}_\rho \nonumber \\ 
&& + \frac{\partial C_{- \rho \rho}}{\partial X^\alpha} {\cal 
P}_\rho \left[ N^\alpha \Sigma_{R2}^{\rho - \rho}  + \frac{1}{2} 
\frac{\partial N^2}{\partial P_\alpha} \Sigma_{R3}^{\rho - \rho} + 
\rho \epsilon (p_0) e_\perp^\mu \frac{\partial N_\mu}{\partial 
P_\alpha} \Sigma_{R4}^{\rho - \rho} \right. \nonumber \\ 
&& + \frac{P\kern-0.1em\raise0.3ex\llap{/}\kern0.15em\relax}{P^2} 
\left( - \rho \epsilon (p_0) e_\perp^\alpha \Sigma_{R2}^{\rho - 
\rho} - \rho \epsilon (p_0) e_\perp^\mu \frac{\partial 
N_\mu}{\partial P_\alpha} \Sigma_{R3}^{\rho - \rho} - N^2 
\hat{P}^\alpha \Sigma_{R4}^{\rho - \rho} \right) \nonumber \\ 
&& \left. + \frac{P\kern-0.1em\raise0.3ex\llap{/}\kern0.15em\relax 
N\kern-0.153em\raise0.3ex\llap{/}\kern0.177em\relax}{P^2} \left( 
P^\alpha \Sigma_{R2}^{\rho - \rho} - N^\alpha \Sigma_{R3}^{\rho - 
\rho} + \rho \epsilon (p_0) e_\perp^\alpha \Sigma_{R4}^{\rho - \rho} 
\right) \right] {\cal P}_\rho \, . 
\end{eqnarray*} 
\narrowtext 
\subsubsection*{Standard forms for $\gamma_5 
\underline{N}\kern-0.153em\raise0.3ex\llap{/}\kern0.177em\relax 
\cdot \underline{\bf C}$ and $\underline{\bf C} \cdot \gamma_5 
\underline{N}\kern-0.153em\raise0.3ex\llap{/}\kern0.177em\relax$ in 
$G_K^{[3]}$ in Eq. (\ref{GK3yo})} Form for $\gamma_5 
\underline{N}\kern-0.153em\raise0.3ex\llap{/}\kern0.177em\relax 
\cdot \underline{\bf C}$ in ${\bf G}_K^{[3]}$ in Eq. (\ref{GK3yo}) 
is given by Eq. (\ref{nagai2}) with 
\begin{eqnarray*}
\Sigma_{A1}^{\rho \sigma} \to \delta^{\rho \sigma} \, , 
\;\;\;\;\;\;\;\; 
&& \Sigma_{Aj}^{\rho \sigma} \to 0 \;\;\;\;\; (j = 2 - 4) \, , \\ 
&& \Sigma_{Rj}^{\rho \sigma} \to 0 \;\;\;\;\; (j = 1 - 4) \, . 
\end{eqnarray*}
$\underline{\bf C} \cdot \gamma_5 
\underline{N}\kern-0.153em\raise0.3ex\llap{/}\kern0.177em\relax$ in 
Eq. (\ref{GK3yo}) is given by Eq. (\ref{nagai2}) with 
\begin{eqnarray*}
\Sigma_{R1}^{\rho \sigma} \to - \delta^{\rho \sigma} \, , 
\;\;\;\;\;\;\;\; 
&& \Sigma_{Rj}^{\rho \sigma} \to 0 \;\;\;\;\; (j = 2 - 4) \, , \\ 
&& \Sigma_{Aj}^{\rho \sigma} \to 0 \;\;\;\;\; (j = 1 - 4) \, . 
\end{eqnarray*}
\widetext 
\setcounter{equation}{0}
\setcounter{section}{5}
\section{Energy shells of $G_R^{\rho \rho} (P, X)$} 
\def\theequation{\mbox{\Alph{section}.\arabic{equation}}}
To find the energy shells of $G_R^{\rho \rho}$, we need $\left( 
G_R^{\rho \rho} (P, X) \right)^{- 1}$, the inverse of $G_R^{\rho 
\rho} (P, X)$ (cf. Eq. \ref{eshell})). To the gradient 
approximation, we have 
\begin{eqnarray}
\left( G_R^{\rho \rho} \right)^{- 1} &=& \left( G_R^{(0) \rho \rho} 
+ G_R^{(1) \rho \rho} \right)^{- 1} \nonumber \\ 
& \simeq & \left( G_R^{(0) \rho \rho} \right)^{- 1} -  
\left( G_R^{(0) \rho \rho} \right)^{- 1} G_R^{(1) \rho \rho} 
\left( G_R^{(0) \rho \rho} \right)^{- 1} \, . 
\label{F1} 
\end{eqnarray}
Here $\left( G_R^{(0) \rho \rho} \right)^{- 1}$ is the 
$(11)$-element of Eq. (\ref{nagai}) and $G_R^{(1) \rho \rho}$ is as 
in Eq. (\ref{iika}). If we ignore the gradient term in Eq. 
(\ref{F1}), the energy shells are obtained through 
\[
Re \left( G_R^{(0) \rho \rho} (P, X) \right)^{- 1} 
\rule[-3mm]{.14mm}{8.5mm} \raisebox{-2.85mm}{\scriptsize{$\; p_0 = 
\pm \omega_\pm^{(0)} (\pm \vec{p}, X)$}} \propto {\cal D}^{\rho 
\rho} \, \rule[-3mm]{.14mm}{8.5mm} 
\raisebox{-2.85mm}{\scriptsize{$\; p_0 = \pm \omega_\pm^{(0)} (\pm 
\vec{p}, X)$}} = 0 \, , 
\]
where ${\cal D}^{\rho \rho}$ is given by Eq. (\ref{1.54ddd}) 
with the substitutions (\ref{1.57d1}) being made. Then, the true 
energy shells, $p_0 = \pm \omega_\pm (\pm \vec{p}, X)$, are obtained 
from Eq. (\ref{F1}), 
\begin{eqnarray*} 
&& \pm \frac{\partial Re \left( G_R^{(0) \rho \rho} (P, X) 
\right)^{- 1}}{\partial p_0} \, \rule[-3mm]{.14mm}{8.5mm} 
\raisebox{-2.85mm}{\scriptsize{$\; p_0 = \pm \omega_\pm^{(0)} (\pm 
\vec{p}, X)$}} \left( \omega_\pm (\pm \vec{p}, X) - \omega_\pm^{(0)} 
(\pm \vec{p}, X) \right) \\ 
&& \mbox{\hspace*{10ex}} = Re \left[ \left( G_R^{(0) \rho \rho} (P, 
X) \right)^{- 1} G_R^{(1) \rho \rho} (P, X) \left( G_R^{(0) \rho 
\rho} (P, X) \right)^{- 1} \right] \, \rule[-3mm]{.14mm}{8.5mm} 
\raisebox{-2.85mm}{\scriptsize{$\; p_0 = \pm \omega_\pm^{(0)} (\pm 
\vec{p}, X)$}} \, . 
\end{eqnarray*} 
\setcounter{equation}{0}
\setcounter{section}{6}
\section{Standard forms for the quantities in Sec. IVB} 
\def\theequation{\mbox{\Alph{section}.\arabic{equation}}}
\subsubsection*{Standard forms for $\Pi_K^{[1] \mu \nu}$ and 
$\Pi_K^{[2] \mu \nu}$ in Eq. (\ref{mukku1})} 
From Eq. (\ref{hiro}) with Eq. (\ref{mukku1}), we obtain 
\begin{eqnarray} 
\Pi_K^{[1] \mu \nu} &=& \frac{i}{2} \sum_j \sum_{UV = T, L} \left[ 
{\cal R}_{Lj}^{UV} \left\{ \tilde{f} , \;\, \tilde{\Pi}^{UV}_{Rj} 
+ \tilde{\Pi}^{UV}_{Aj} \right\} {\cal R}_{Rj}^{UV} \right]^{\mu 
\nu} \, , 
\label{meichan}  \\ 
\Pi_K^{[2] \mu \nu} &=& 2 i {\cal P}_T^{\mu \nu} Re \left[ 
\Pi_{R1}^{TG} - \frac{\zeta \cdot \tilde{P}}{\tilde{P}^2} 
\Pi_{R3}^{T'T} \right] (E_\perp \cdot \partial) \tilde{f} + 2 i 
\frac{\tilde{\zeta}^\mu \tilde{\zeta}^\nu}{\tilde{\zeta}^2} Re 
\left[ \left( - \frac{\zeta \cdot \tilde{P}}{\tilde{P}^2} 
\Pi_{R2}^{TT} + \Pi_{R2}^{TG} \right) (\tilde{\zeta} \cdot \partial) 
\tilde{f} \right. \nonumber \\ 
&& \left. + \left\{ \frac{\zeta \cdot \tilde{P}}{\tilde{P}^2} 
\left( \Pi_{R3}^{T'T} - \Pi_{R3}^{TT'} \right) - \Pi_{R1}^{TG} 
\right\} (E_\perp \cdot \partial) \tilde{f} \right] \nonumber \\ 
&& + i \tilde{\zeta}^\mu E_\perp^\nu \left[ - \frac{1}{\tilde{P}^2} 
\Pi_{R3}^{TT'} (\tilde{P} \cdot \tilde{\partial}) \tilde{f} + 
\frac{1}{E_\perp^2} \left( \Pi_{R2}^{TG} - \frac{\zeta \cdot 
\tilde{P}}{\tilde{P}^2} \Pi_{R2}^{TT} \right) (E_\perp \cdot 
\partial) \tilde{f} \right. \nonumber \\ 
&& \left. + \frac{1}{\tilde{\zeta}^2} \left( \Pi_{A1}^{GT} + 
\frac{\zeta \cdot \tilde{P}}{\tilde{P}^2} \left( \Pi_{R3}^{TT'} + 
\Pi_{A3}^{TT'} \right) \right) (\tilde{\zeta} \cdot \partial) 
\tilde{f} \right] + 2 i n^\mu n^\nu Re \Pi_{R1}^{LG} \partial_0 
\tilde{f} \nonumber \\ 
&& + i \tilde{\zeta}^\mu n^\nu \left[ \Pi_{R2}^{TG} \partial_0 
\tilde{f} - \frac{1}{\tilde{\zeta}^2} \left( \Pi_{A1}^{GL} + 
\frac{\zeta \cdot \tilde{P}}{\tilde{P}^2} \Pi_{A1}^{TL} \right) 
(\tilde{\zeta} \cdot \partial) \tilde{f} + \frac{\zeta \cdot 
\tilde{P}}{\tilde{P}^2 \tilde{\zeta}^2} \Pi_{A2}^{TL} (E_\perp \cdot 
\partial) \tilde{f} \right] \nonumber \\ 
&& + i E_\perp^\mu n^\nu \left[ \frac{1}{\tilde{P}^2} \Pi_{A2}^{TL} 
(\tilde{P} \cdot \tilde{\partial}) \tilde{f} + \Pi_{R1}^{TG} 
\partial_0 \tilde{f} - \frac{\zeta \cdot \tilde{P}}{\tilde{P}^2 
\tilde{\zeta}^2} \Pi_{A2}^{TL} (\tilde{\zeta} \cdot \partial) 
\tilde{f} \right. \nonumber \\ 
&& \left. - \frac{1}{E_\perp^2} \left( \Pi_{A1}^{GL} + \frac{\zeta 
\cdot \tilde{P}}{\tilde{P}^2} \Pi_{A1}^{TL} \right) (E_\perp \cdot 
\partial) \tilde{f} \right] \nonumber \\ 
&& + i E_\perp^\mu \tilde{\zeta}^\nu [...] + i n^\mu 
\tilde{\zeta}^\nu [...] + i n^\mu E_\perp^\nu [...] \, . 
\end{eqnarray}
$[...]$'s are obtained using Eq. (\ref{maki}). 
\subsubsection*{Standard dorm for $\tilde{H}$ in Eq. (\ref{hika})} 
We write $\underline{\tilde{H}}^{\mu \nu} = 
\underline{\tilde{H}}^{(0) \mu \nu} + \underline{\tilde{H}}^{(1) \mu 
\nu}$, with $\underline{\tilde{H}}^{(0) \mu \nu}$ the leading term 
and $\underline{\tilde{H}}^{(1) \mu \nu}$ the gradient term of the 
DEX of $\tilde{H}^{\mu \nu} (P, X)$. Straightforward manipulation of 
Eq. (\ref{hika}) yields 
\begin{eqnarray} 
\underline{\tilde{H}}^{(0) \mu \nu} &=& 
\underline{\cal P}_T^{\mu \rho} \cdot \left[ \left( \tilde{H}_l 
\right)_1^{TT} \right]_{\mbox{\scriptsize{IWT}}} \cdot \left( 
\underline{\cal P}_T \right)_\rho^{\; \, \nu} + 
\underline{\tilde{\zeta}}^\mu \cdot \left[ \left( \tilde{H}_l 
\right)_2^{TT} \right]_{\mbox{\scriptsize{IWT}}} \cdot 
\underline{\tilde{\zeta}}^\nu - \underline{\tilde{\zeta}}^\mu \cdot 
\left[ \left( \tilde{H}_l \right)_3^{TT'} 
\right]_{\mbox{\scriptsize{IWT}}} \cdot \underline{E}_\perp^\nu 
\nonumber \\ 
&& + \underline{E}_\perp^\mu \cdot \left[ \left( \tilde{H}_l 
\right)_3^{T'T} \right]_{\mbox{\scriptsize{IWT}}} \cdot 
\tilde{\zeta}^\nu + \underline{\tilde{\zeta}}^\mu \cdot \left[ 
\left( \tilde{H}_l \right)_1^{TL} \right]_{\mbox{\scriptsize{IWT}}} 
n^\nu + \underline{E}_\perp^\mu \cdot \left[ \left( \tilde{H}_l 
\right)_2^{TL} \right]_{\mbox{\scriptsize{IWT}}} n^\nu \nonumber \\ 
&& + n^\mu \left[ \left( \tilde{H}_l \right)_1^{LT} 
\right]_{\mbox{\scriptsize{IWT}}} \cdot 
\underline{\tilde{\zeta}}^\nu - n^\mu \left[ \left( \tilde{H}_l 
\right)_2^{LT} \right]_{\mbox{\scriptsize{IWT}}} \cdot 
\underline{E}_\perp^\nu \, , 
\nonumber 
\end{eqnarray}
with $\left( \tilde{H}_l \right)_j^{UV}$ as in Eq. (\ref{etti}), and 
\begin{eqnarray}
\tilde{H}^{(1) \mu \nu} &=& i {\cal P}_T^{\mu \nu} Re \left[ 
\tilde{\zeta}^2 E_\perp^2 \left\{ C^{TT'}_3 , \; \, \Pi_{R3}^{T'T} 
\right\} \right. \nonumber \\ 
&& \left. - 2 (\zeta \cdot \tilde{P}) \left\{ \tilde{\zeta}^2 
C_3^{TT'} (\tilde{\zeta} \cdot \stackrel{\leftrightarrow}{\partial}) 
\Pi_{R3}^{T'T} - \frac{1}{\tilde{P}^2} C_3^{TT'} (E_\perp 
\cdot \partial) \Pi_{R1}^{TT} \right\} \right] \nonumber \\ 
&& + i \tilde{\zeta}^\mu \tilde{\zeta}^\nu Re \left[ - \left\{ 
C_2^{TT} , \; \, \Pi_{R1}^{TT} \right\} - \tilde{\zeta}^2 \left\{ 
C_2^{TT} , \; \, \Pi_{R2}^{TT} \right\} + E_\perp^2 \left\{ 
C_3^{T'T} , \; \, \Pi_{R3}^{TT'} - \Pi_{A3}^{TT'} \right\} \right. 
\nonumber \\ 
&& - 2 \frac{\zeta \cdot \tilde{P}}{\tilde{P}^2} C_2^{TT} 
(\tilde{\zeta} \cdot \stackrel{\leftrightarrow}{\partial}) 
\Pi_{R2}^{TT} + 2 \tilde{\zeta}^2 C_3^{T'T} (\tilde{P} \cdot 
\stackrel{\leftrightarrow}{\partial}) \Pi_{R3}^{TT'} \nonumber \\ 
&& + 2 (\zeta \cdot \tilde{P}) \left( C_3^{TT'} (\tilde{\zeta} 
\cdot \stackrel{\leftrightarrow}{\partial}) \Pi_{R3}^{T'T} - 
C_3^{T'T} (\tilde{\zeta} \cdot \stackrel{\leftrightarrow}{\partial}) 
\Pi_{R3}^{TT'} \right) \nonumber \\ 
&& \left. - 2 \frac{\zeta \cdot \tilde{P}}{\tilde{P}^2 
\tilde{\zeta}^2} \left( C_2^{TT} (\tilde{\zeta} \cdot \partial) + 2 
Re C_3^{TT'} ( E_\perp \cdot \partial) \right) \Pi_{R1}^{TT} 
\right] \nonumber \\ 
&& + \frac{i}{2} \frac{\tilde{\zeta}^\mu E_\perp^\nu}{\tilde{P}^2} 
\left[ \tilde{P}^2 \tilde{\zeta}^2 \left( \left\{ C_2^{TT} , \; \, 
\Pi_{A3}^{TT'} \right\} + \left\{ C_3^{TT'} , \; \, \Pi_{R2}^{TT} 
\right\} \right) + 2 \tilde{P}^2 \left\{ C_3^{TT'} , \; \, Re 
\Pi_{R1}^{TT} \right\} \right. \nonumber \\ 
&& - 2 C_3^{TT'} (\tilde{P} \cdot \tilde{\partial}) \Pi_{A1}^{TT} + 
2 (\zeta \cdot \tilde{P}) \left\{ - C_2^{TT} \frac{E_\perp \cdot 
\partial}{E_\perp^2} \Pi_{A1}^{TT} + C_2^{TT} (\tilde{\zeta} \cdot 
\stackrel{\leftrightarrow}{\partial}) \Pi_{A3}^{TT'} \right. 
\nonumber \\ 
&& \left. \left. + C_3^{TT'} (\tilde{\zeta} \cdot 
\stackrel{\leftrightarrow}{\partial}) \Pi_{R2}^{TT} + 2 i (E_\perp 
\cdot \partial) Im \left( C_3^{TT'} \Pi_{R3}^{T'T} \right) + 
\frac{2}{\tilde{\zeta}^2} C_3^{TT'} (\tilde{\zeta} \cdot \partial) 
Re \Pi_{R1}^{TT} \right\} \right] \nonumber \\ 
&& + \frac{i}{2} \tilde{\zeta}^\mu n^\nu \left[ - \tilde{\zeta}^2 
\left\{ C_2^{TT} , \;\, \Pi_{A1}^{TL} \right\} + E_\perp^2 \left\{ 
C_3^{TT'} , \;\, \Pi_{A2}^{TL} \right\} \right. \nonumber \\ 
&& \left. - \frac{2 \zeta \cdot \tilde{P}}{\tilde{P}^2} C_2^{TT} 
(\tilde{\zeta} \cdot \stackrel{\leftrightarrow}{\partial}) 
\Pi_{A1}^{TL} + 2 C_3^{TT'} \left( \tilde{\zeta}^2 (\tilde{P} \cdot 
\stackrel{\leftrightarrow}{\partial}) - (\zeta \cdot \tilde{P}) 
(\tilde{\zeta} \cdot \stackrel{\leftrightarrow}{\partial}) \right) 
\Pi_{A2}^{TL} \right] \nonumber \\ 
&& - i E_\perp^\mu n^\nu \left[ \frac{\tilde{\zeta}^2}{2} \left\{ 
C_3^{T'T} , \;\, \Pi_{A1}^{TL} \right\} + \frac{\zeta \cdot 
\tilde{P}}{\tilde{P}^2} C_3^{T'T} (\tilde{\zeta} \cdot 
\stackrel{\leftrightarrow}{\partial}) \Pi_{A1}^{TL} \right] 
\nonumber \\ 
&& + i E_\perp^\mu \tilde{\zeta}^\nu [...] + i n^\mu 
\tilde{\zeta}^\nu [...] - i n^\mu E_\perp^\nu [...] \nonumber \, , 
\end{eqnarray}
where $A \stackrel{\leftrightarrow}{\partial} B \equiv A \partial B 
-  A \stackrel{\leftarrow}{\partial} B$, and $[...]$'s are obtained 
using Eqs. (\ref{2.28d}) and (\ref{hikkuri}). 
\end{appendix} 
\narrowtext 

\end{document}